\documentclass[aps,amsmath,twocolumn,amssymb,floatfix,showpacs,superscriptaddress,longbibliography]{revtex4-2}

\usepackage{amsmath}
\usepackage{mathtools}
\usepackage{braket}
\usepackage{bm}
\usepackage{graphicx}
\usepackage{amssymb}
\usepackage[dvipsnames]{xcolor}
\usepackage{multirow}
\usepackage{xfrac}
\usepackage{textcomp}
\usepackage{float}
\usepackage{enumerate}
\usepackage{subfigure}

\usepackage[colorlinks=true,linktoc=page,linkcolor=ForestGreen,citecolor=blue,urlcolor=purple]{hyperref}
\usepackage[toc,page]{appendix}

\usepackage{graphicx} 
\graphicspath{{./Figs/}{./}}

\usepackage{float}

\usepackage{cleveref}
\usepackage{epstopdf}
\usepackage{epsfig}

\usepackage{multirow}
\usepackage{diagbox}
\usepackage{dcolumn}
\usepackage{bm}

\usepackage[mathlines]{lineno}

\usepackage[english]{babel}

\def\nn{\nonumber}

\newcommand{\beq}{\begin{eqnarray}}
\newcommand{\eeq}{\end{eqnarray}}
\newcommand{\bea}{\begin{eqnarray}}
\newcommand{\eea}{\end{eqnarray}}

\newcommand{\Eq}[1]{{\textcolor{blue}{Eq.}}~\!\!(\ref{#1})} 
\newcommand{\Fig}[1] {{\textcolor{blue}{Fig.}}~\!\!\ref{#1}}

\newcommand{\rmrk}[1]{#1} 

\def\<{\langle}
\def\>{\rangle}

\def\nn{\nonumber}

\usepackage[english]{babel}

\usepackage[normalem]{ulem}
\usepackage{sidecap,tikz}
\DeclareRobustCommand{\orcidicon}{\hspace{-1.0mm}
	\begin{tikzpicture}
	\draw[lime, fill=lime] (0.0,0.0) 
	circle [radius=0.15] 
	node[white] {{\fontfamily{qag}\selectfont \tiny \,ID}};
	\draw[white, fill=white] (-0.0525,0.095) 
	circle [radius=0.007];
	\end{tikzpicture}
	\hspace{-3.0mm}
}
\foreach \x in {A, ..., Z}{\expandafter\xdef\csname orcid\x\endcsname{\noexpand\href{https://orcid.org/\csname orcidauthor\x\endcsname}
		{\noexpand\orcidicon}}
}

\begin{document}

\title{Characterization of hybrid quantum eigenstates \\ in systems with mixed classical phasespace}

\author{Anant Vijay Varma\orcidA{}}
\email{varma@post.bgu.ac.il}
\affiliation{Department of Chemistry, Ben-Gurion University of the Negev, Beer Sheva 84105, Israel}

\author{Amichay Vardi\orcidB{}}
\email{avardi@bgu.ac.il}
\affiliation{Department of Chemistry, Ben-Gurion University of the Negev, Beer Sheva 84105, Israel}
\affiliation{ITAMP, Harvard-Smithsonian Center for Astrophysics, Cambridge, MA 02138, USA}

\author{Doron Cohen\orcidC{}}
\email{dcohen@bgu.ac.il}
\affiliation{Department of Physics, Ben-Gurion University of the Negev, Beer Sheva, Israel}


\begin{abstract}
Generic low-dimensional Hamiltonian systems feature a structured, mixed classical phase-space. The traditional Percival classification of quantum spectra into regular states supported by quasi-integrable regions and irregular states supported by quasi-chaotic regions turns out to be insufficient to capture the richness of the Hilbert space. Berry's conjecture and the eigenstate thermalization hypothesis are not applicable and quantum effects such as tunneling, scarring, and localization, do not obey the standard paradigms. We demonstrate these statements for a prototype Bose-Hubbard model. We highlight the hybridization of chaotic and regular regions from opposing perspectives of ergodicity and localization.  
\end{abstract}

\maketitle

\section{Introduction}

In a seminal paper, Percival suggested a classification of quantum mechanical spectra into 'regular' and 'irregular'  eigenstates \cite{ICPercival_1973}, 
supported respectively by quasi-regular islands and chaotic seas within the classical phasespace \cite{Stechel}. Whereas regular states are restricted to invariant tori, irregular states are RMT-ergodic in the chaotic sea \cite{Berry_1977}. For large systems this implies a thermal expectation value for local observables, an observation known as the 'Eigenstate Thermalization Hypothesis' (ETH) \cite{PhysRevE.50.888,doi:10.1080/00018732.2016.1198134, Deutsch_2018,Huse_2015}. The binary classification of quantum eigenstates is a widely accepted paradigm in studies of quantum chaos.

In reality, the picture is more complicated. Sharp distinction between regular and chaotic states is not generally practical, and spectral analysis is not sufficiently revealing \cite{PhysRevE.107.024210}.
In this work, we consider a prototype mixed-phasespace system.  We find that the statistical properties of chaotic eigenstates that dwell in phasespace with mixed regular and chaotic motion, are substantially different from those of eigenstates supported by a globally connected chaotic sea. We also identify special non-ergodic states that are dynamically localized \cite{PhysRevA.35.1360,Prange,Casati} in chaotic regions. The existence of the latter is related to slow dynamics near unstable stationary points, and can be viewed as an extreme type of scarring \cite{LKaplan_1999,Turner2018,PhysRevLett.125.134101,Serbyn2021}. 

Our study demonstrates that quantization of the mixed classical phasespace is a double-edged sword.  On the one hand, quantum tunneling can connect classically separated chaotic and regular regions, resulting in hybrid quantum eigenstates that do not adhere to the standard classification. On the other hand, there is also an opposite effect -- due to dynamical localization there are states that are not fully ergodic despite the prevailing chaos. The underlying mechanism of this localization is the slow dynamics in the vicinity of an unstable hyperbolic point embedded in chaos. As such, it is different from previous studies of dynamical localization in the peripheral regions of the chaotic sea \cite{PhysRevE.97.022127,PhysRevA.101.043603} and from localization by remnants of KAM tori 
\cite{Heller,10.1063/5.0130682,PhysRevLett.132.047201}.




\section{Bose-Hubbard Trimer Hamiltonian}

The Bose-Hubbard (BH) model, see \cite{Dutta_2015} and references within, is a paradigm for quantum chaos studies. Of particular interest is the 3-site (trimer) model, whose mixed phase-space is of interest e.g. in the context of superflow stability \cite{Geva} and phase separation \cite{Penna}. The Hamiltonian for $N$ Bosons is written in terms of three second-quantized modes:
\begin{equation}
H= V \hat{n}_{2} +  \frac{U}{2} \ \sum_{i=1}^{3} \hat{n}_{i}^{2} -  \frac{\Omega}{2} \  ( \hat{a}^{\dagger}_{2} \hat{a}_{1} + \hat{a}^{\dagger}_{3} \hat{a}_{2} + H.c.),
\label{E1}
\end{equation}
where $\hat{a}^{\dagger}_{i}$ and $\hat{a}_{i}$ are Bosonic creation and annihilation operators in the $i$-th local mode. $U$ is the interaction strength, $\Omega$ is the hopping parameter, and $V$ is the middle site bias. The dimensionless parameters of the model are,
\begin{equation}
u = \frac{NU}{\Omega}, 
\ \ \ \ \ v=\frac{V}{\Omega}
\end{equation} 
Throughout the manuscript, we use units of time such that $\Omega=1$ and set $v=0.1$. 

In the classical limit, the field operators $\hat{a}_{i}$ can be replaced by complex numbers  $a_{i} = \sqrt{n_{i}} e^{i \phi_{i} }$. Thus, the classical motion has three degrees of freedom, with $\{{n_{i},\phi_{i}}\}$,  $i=1,2,3$ serving as conjugate action-angle variables. Owing to the U(1) symmetry, the classical phase space can be further reduced to two degrees of freedom. Throughout this paper, our choice of canonical variables is $p_{1} = n_{1}/N$, $p_{2} = n_{2}/N$, $q_{1} = \phi_{1}-\phi_{3}$, and $q_{2}=\phi_{2}-\phi_{3}$, resulting in the classical Hamiltonian:
\begin{eqnarray}
\dfrac{H_{cl}}{N} = && V \ p_{2}+  \frac{NU}{2} \ (p_{1}^{2} + p_{2}^{2} + (1{-}p_{1}{-}p_{2})^{2} )
\label{E6} \\ \nn
&& -  \Omega \left( \sqrt{p_1 p_2} 
\cos (q_{1}{-}q_{2}) +\sqrt{p_{2} (1{-}p_{1}{-}p_{2})} \cos q_{2} \right)
\end{eqnarray}

The quantum Hilbert space of the $N$-particle system is spanned by the Fock basis $\ket{\mathbf{n}}= \ket{n_{1},n_{2}}$, with $n_{3} = N-n_{1}-n_{2}$.  Its dimension is thus,
\begin{equation}
\mathcal{N} \ \ = \ \ \frac{1}{2} (N+1)(N+2)~.
\end{equation}

\section{Eigenstate Characterization}

Diagonalizing $H$ in the Fock basis, we obtain the quantum spectrum $H|E_\nu\rangle=E_\nu |E_\nu\rangle$. The purity of each eigenstate $\ket{E_{\nu}}$ is $S = {\rm Tr}\left[({\rho^{\rm (sp)}})^{2}\right]$  where 
\beq
\rho^{\rm (sp)}=(1/N)\left(\langle a_i^\dag a_j \rangle\right)_{i,j=1,2,3} 
\eeq
is the one-particle probability matrix. Eigenstates may be visualized via their Husimi phasespace distribution,
\begin{equation}
Q_\nu(\alpha) = |\langle \alpha | E_{\nu}\rangle|^{2}
\label{E3}
\end{equation}
where $|\alpha\rangle$ are coherent states  localized at $\alpha=(q_1,q_2,p_1,p_2)$. 
Alternatively, the eigenstates can be represented by their Fock-space distribution,
\begin{equation} \label{eX}
X_{\nu,n}= |\langle \mathbf{n}|   E_{\nu}\rangle|^2~. 
\end{equation}
In order to characterize this distribution for the various eigenstates, we calculate the following measures:
\begin{eqnarray}
R_{q} &=& \sum_{n=1}^{\mathcal{N}} X_{\nu,n}^{q},
\label{E1a}
\\ 
M_{q}  &=&  R_{q}^{-\frac{1}{q-1}}
\end{eqnarray}
The quantities $R_{2}$ and $M_2$ are respectively, the inverse participation ratio (IPR) and the participation number (PN) in the Fock (computational) basis. Higher $q>2$ moments provide more information on the shape of the Fock space distribution. In the limit ${q \rightarrow 1}$ one obtains 
$M_1=\exp(\tilde{R}_{1})$, with $\tilde{R}_{1} = -\sum_{n=1}^{\mathcal{N}} X_{\nu,n} \ln X_{\nu,n}$, aka Shanon's entropy 

Averaging over all eigenstates within a narrow energy window around $E$, we define the mean intensities $X_{n}$. The effective dimension of  this energy shell is thus, 
\begin{equation}
\mathcal{N}_{\text{eff}} = \left[ \sum_n |X_{n}|^2 \right]^{-1}  
\end{equation}
Rescaling the intensities as, 
\begin{equation}
x_{\nu,n} = \mathcal{N}_{\text{eff}}  X_{\nu,n}
\end{equation}
ensures that their average is roughly unity. For GOE chaotic states we expect Porter-Thomas intensity statistics \cite{PhysRev.104.483,PhysRevLett.74.62},
\begin{equation} \label{ePT}
P(x) = \frac{1}{\sqrt{x}} e^{-x/2}~,
\end{equation}
resulting in, 
\begin{equation}
M_q^{\rm GOE} = 
\left[ \frac{2^q}{\sqrt{\pi}}\Gamma(q+1/2) \right]^{-\frac{1}{q-1}} \mathcal{N}_{\text{eff}} 
%
\label{mqgoe}
\end{equation}
It is easily verified that for $q=2$ one obtains the well-known participation number  ${\rm PN}_{\rm GOE}=\mathcal{N}/3$.   For $q=10$ the expected Fock-basis moment for a chaotic eigenstate is $M_{10}^{\rm GOE}=\mathcal{N}/9.54$.

\begin{figure}
\includegraphics[clip=true,width=8.5cm, height=4.5cm]{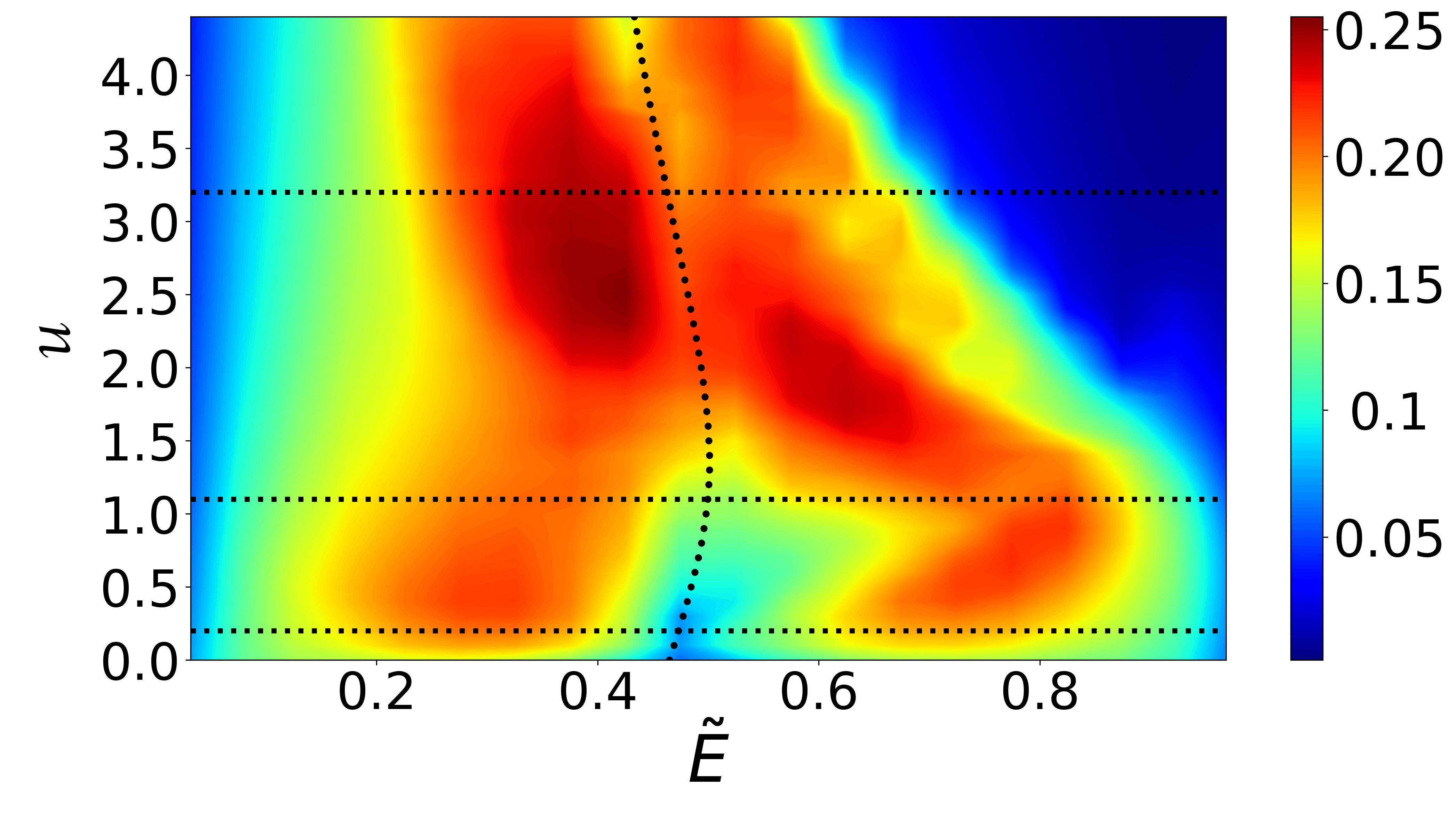} \\
\includegraphics[clip=true,width=8.5cm, height=4.5cm]{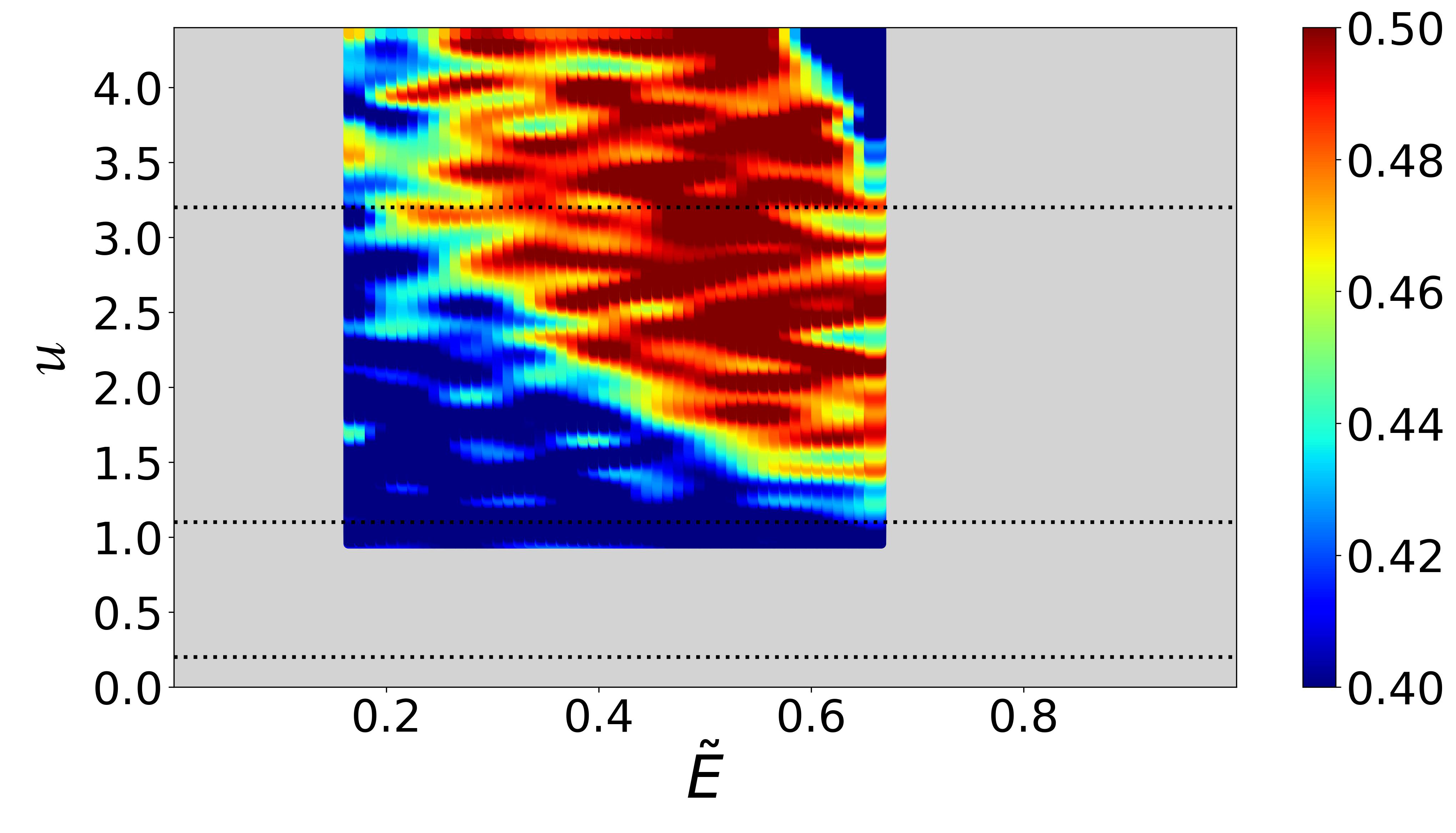}
\caption{
{\bf Parametric evolution of the trimer spectrum.}
Upper panel: mean participation. The trimer spectrum at each given value of $u$ is distributed into 100 equal-width energy bins and the participation ratio $M_2/\mathcal{N}$ is averaged over all eigenstates in each bin. High participation ratios indicate chaos. The total number of particles is $N=150$ and the detuning is $v= 0.1$. The black dotted line marks the energy $\tilde{E}_{SP}$, while the horizontal dotted lines indicate (in order of increasing $u$) the loss of dynamical stability, the onset of chaos, and the restoration of integrability as discussed in Section~\ref{sV}.
\rmrk{Lower panel: The $r$ level-spacing statistics. Each bin is color-coded by the the average $r$ value. Regions of bad statistics are grey. GOE statistics (${r \approx 0.53}$) indicates underlying chaos.}    
} 
\label{fUE}
\end{figure}

\begin{figure}
\centering
\includegraphics[width=8.5cm,height=4cm]{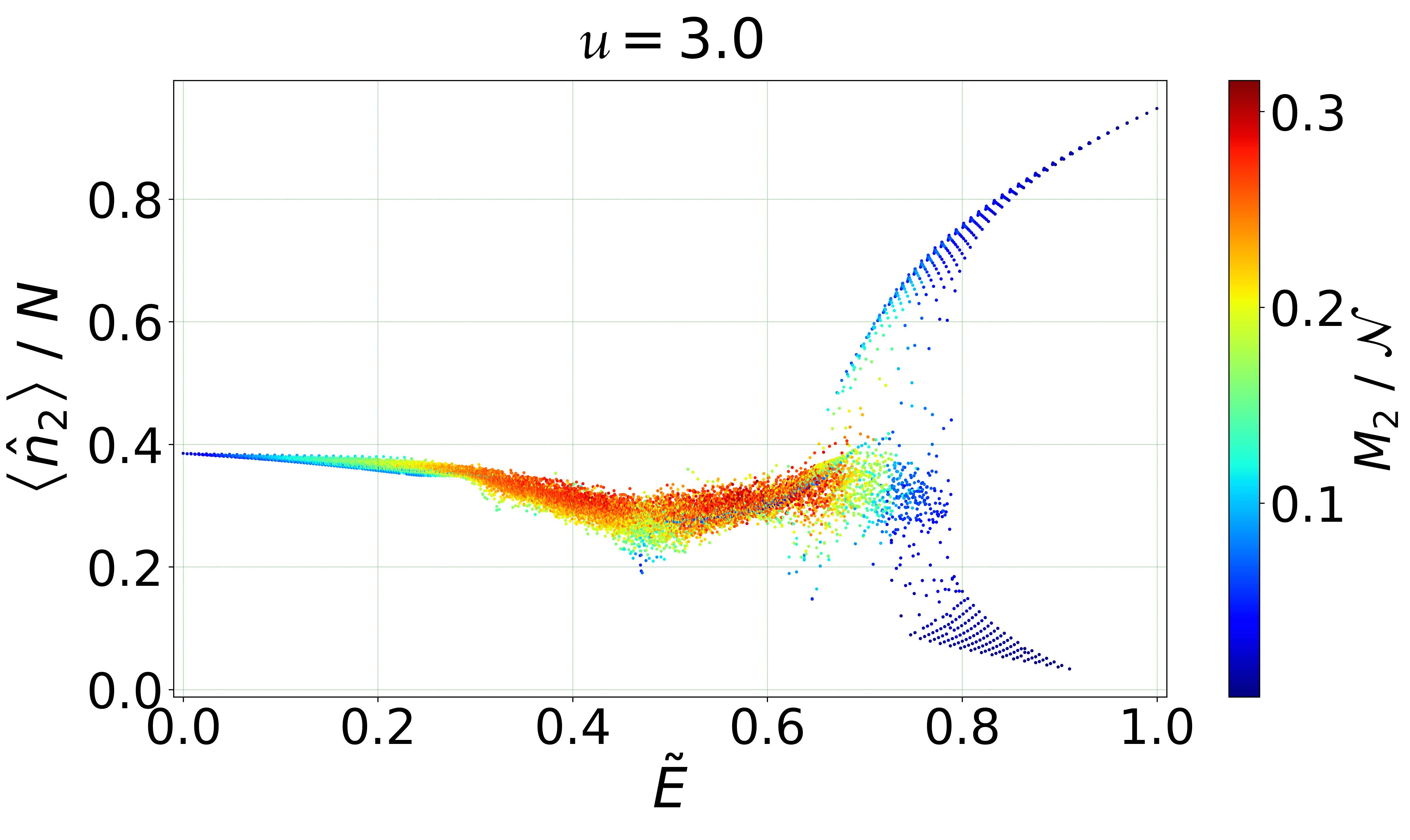} \\
\includegraphics[width=8.5cm, height=4cm]{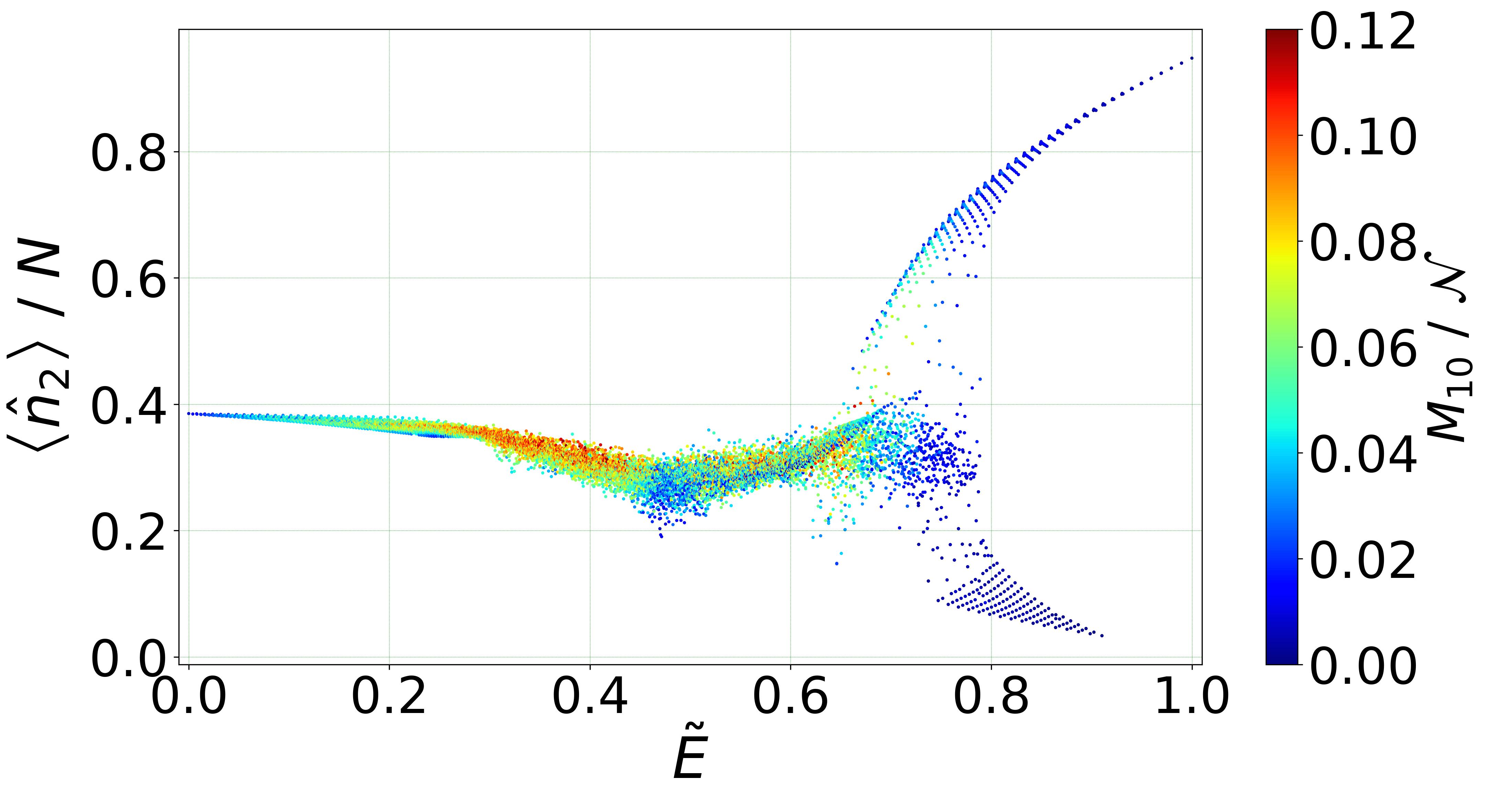}
\includegraphics[width=8.5cm, height=4cm]{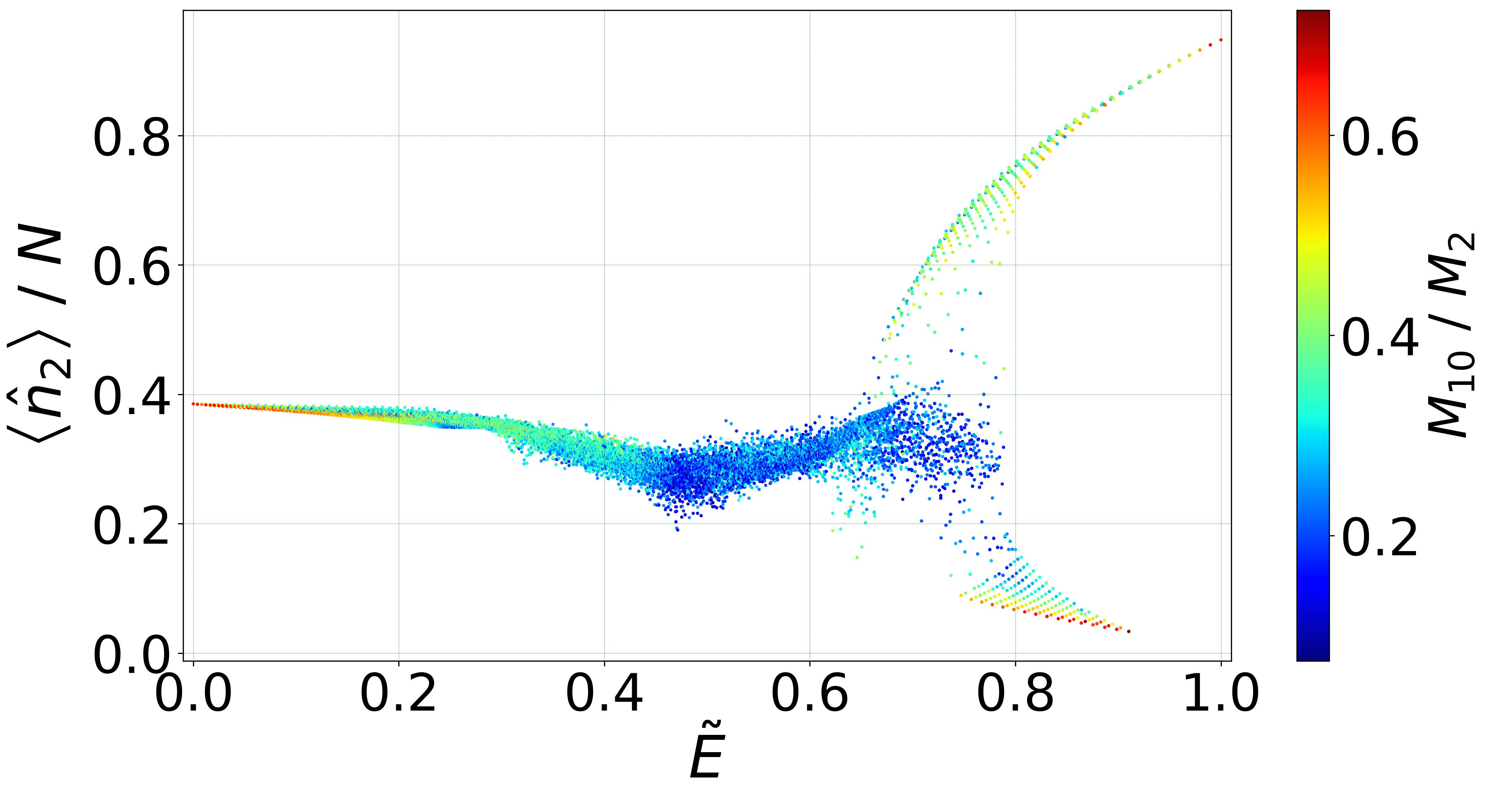}
\caption{
{\bf Tomography of the spectrum.}
Detailed view of the trimer spectrum at  $u=3$. 
Each point corresponds to one many-body eigenstate,  classified according to its rescaled energy  $\tilde{E}_{\nu}$ and central site occupation $\langle\hat{n}_{2}\rangle/N$, 
and color-coded according to $M_2/\mathcal{N}$ (top panel), $M_{10}/{\mathcal N}$ (middle panel) and $M_{10}/M_{2}$ (bottom panel). Parameters are the same as in \Fig{fUE}
} 
\label{fig2_1}
\end{figure}

\begin{figure}
\centering
\includegraphics[width=8.5cm, height=4.0cm]{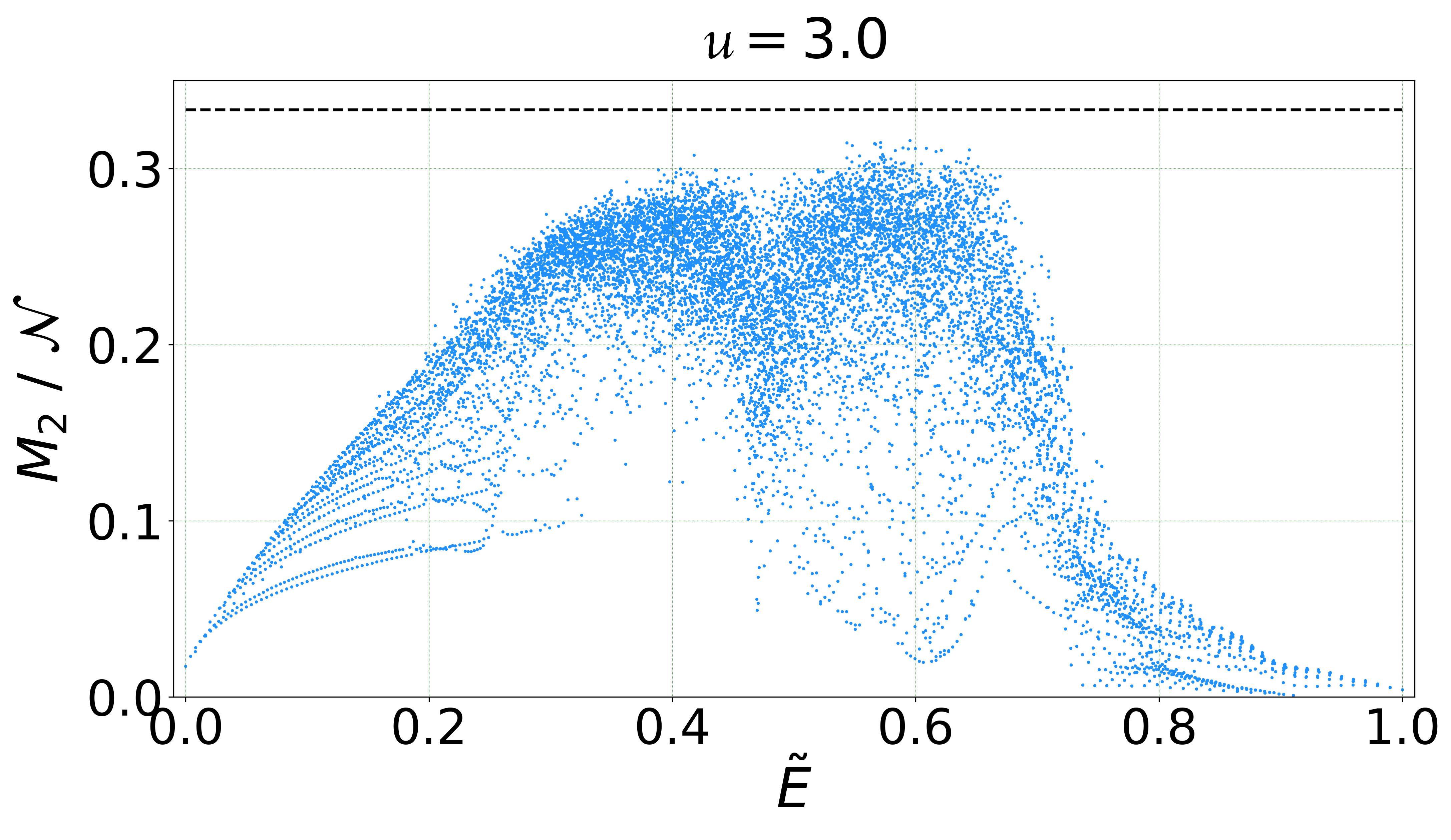} \\
\includegraphics[width=8.5cm, height=4.0cm]{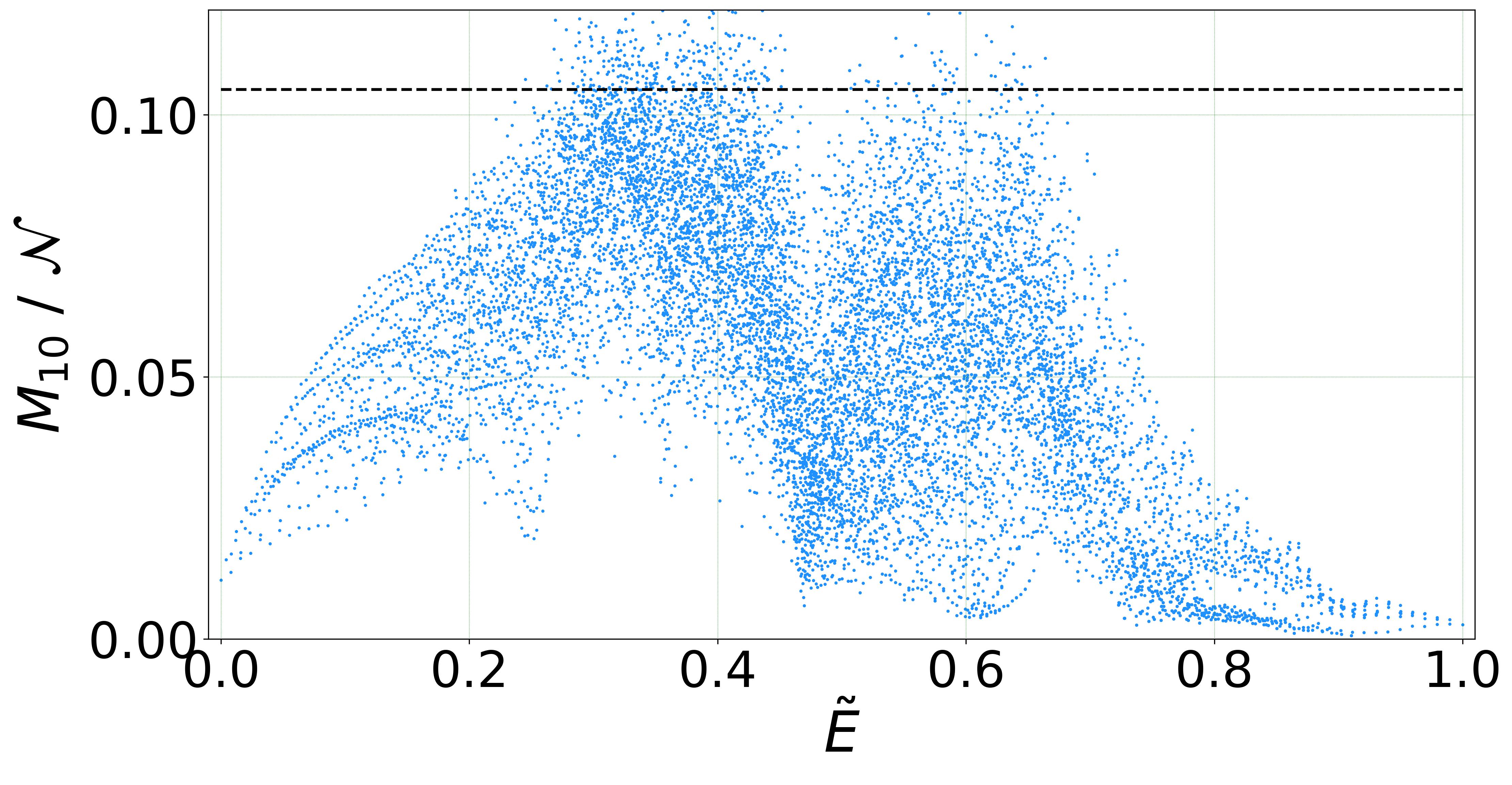} \\
\includegraphics[width=8.5cm, height=4cm]{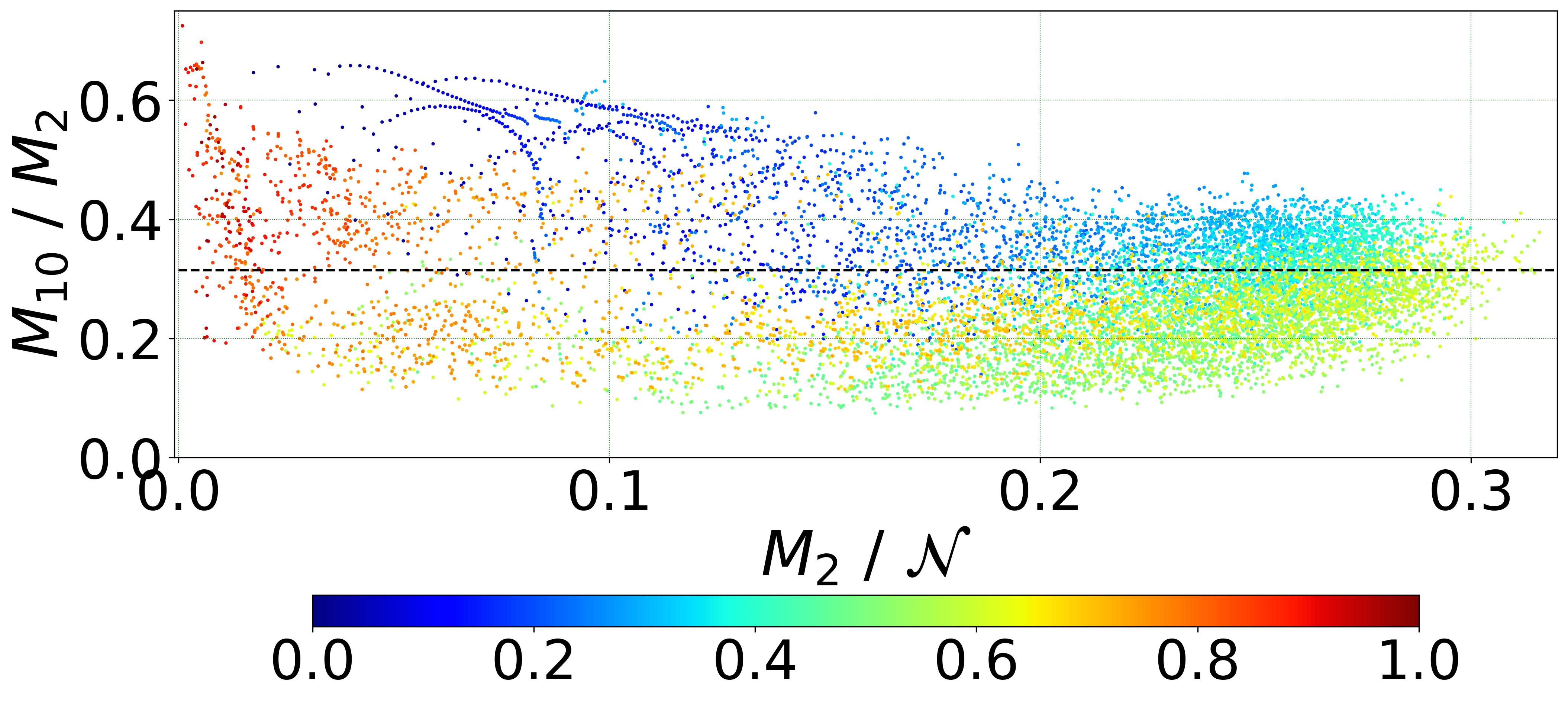}
\caption{
{\bf Moments of the Fock space distribution of the eigenstates.}
Scatter plots of the data displayed in \Fig{fig2_1}: 
(a)~$M_2/\mathcal{N}$ vs $\tilde{E}$; 
(b)~$M_{10}/{\cal N}$ vs $\tilde{E}$;
(3)~$M_{10}/M_{2}$ vs $M_2/\mathcal{N}$, color coded with $\tilde{E}$.  Horizontal dashed lines mark the expected values according to \Eq{mqgoe} for GOE ergodic states. }
\label{2_1a}
\end{figure}

\section{Mapping the trimer spectrum}

This section provides a global view of the Bose-Hubbard trimer's spectrum.  Rescaling the eigenenergies as 
$
\tilde{E}_{\nu}= (E_{\nu}-E_{min})/(E_{max}-E_{min})\in [0,1]$, we plot in \Fig{fUE} the mean Fock participation number $M_2$ of energy eigenstates lying within energy bins around $\tilde E$ at different values of the interaction parameter $u$.  The spectrum evolves parameterically in a non-trivial way, with large PN that indicates underlying chaos observed around ${u\sim 3}$.  Subsequent analysis focuses on this regime. \rmrk{In the lower panel of \Fig{fUE} we show the level statistics. The definition of the spacing ratio $r$ around $E_{\nu}$, and details on the statistical analysis, are provided in Appendix~B of \cite{PhysRevA.101.043603}. An average value ${r \approx 0.53}$ that is based on the Gaussian orthogonal ensemble (GOE), as opposed to the Poissonian ${r \approx 0.386}$, is commonly regarded as an indication for an  underlying chaos.  While the $r$ measure better identifies this chaos, the $M_q$ measure, that can be large also for quasi-integrable states, is more sensitive to variations in quantum ergodicity.}
   
A detailed view of the spectrum for ${u=3.0}$ is provided in \Fig{fig2_1}. 
Each point represents an eigenstate of the Hamiltonian \Eq{E1}. Due to mirror symmetry, specification of the middle site's normalized population $\langle\hat{n}_{2}\rangle/N$ is sufficient to determine the population in the other sites. 
In the absence of bias and interactions $u=v=0$, the system's eigenstates would be orbital-Fock states, i.e. symmetrized direct products of one-particle orbitals. In this case, the spectrum would be degenerate in~$n_2$, because moving pairs of particles from the Dark State orbital  $({ \ket{1}-\ket{3} })/\sqrt{2}$ into the other two orbitals does not change the energy. Introducing a small bias ${v\ne0}$, the spectrum of \Fig{fig2_1} is stretched in the vertical direction, making it easier to understand its structure.  The introduction of finite interaction ($u\ne0$) further deforms the displayed spectrum.   The value ${u = 3}$ is chosen because chaotic regions in the classical phasespace are relatively large.    

The colormaps on the three panels of \Fig{fig2_1} display information on 
the normalized participation number $M_2/\mathcal{N}$, the shape-sensitive moment $M_{10}/\mathcal{N}$, and the ratio between them $M_{10}/M_{2}$.  The same information is displayed in \Fig{2_1a}. 
The low energy range of the spectrum exhibits small PNs. The same applies in the high energy range, where the interaction induces self-trapping either in the middle or in the outer sites. Chaos prevails at intermediate energies in the range ${0.3<\tilde E<0.7}$. 

Interestingly, the ergodicity measures exhibit non-trivial dependence on~$E$.  Both $M_2$ and $M_{10}$ roughly agree with the GOE prediction throughout the chaotic energy range, but show a pronounced narrow dip at the energy $E= E_{\text{SP}}$. This is the energy of an underlying stationary point (SP) that supports the dark state, see further discussion in the following section. Its energy for different values of $u$ is indicated by the black dotted line in \Fig{fUE}. 

More importantly, the $E$ dependence of the $M_{10}/M_{2}$ ratio reveals information that eludes the $M_2$ measure alone. The ratio is significantly lower in the  $E>E_{\text{SP}}$ range compared with the $E<E_{\text{SP}}$ range. As shown in Section~\ref{mbc}, this drop reflects an underlying 
classical transition from {\em hard chaos}, where the motion on the pertinent energy surface is fully chaotic, to {\em mixed chaos}, where the energy surface contains quasi-integrable regions of non-negligible measure.

\begin{figure}
\centering
\includegraphics[width=8.25cm, height=4cm]{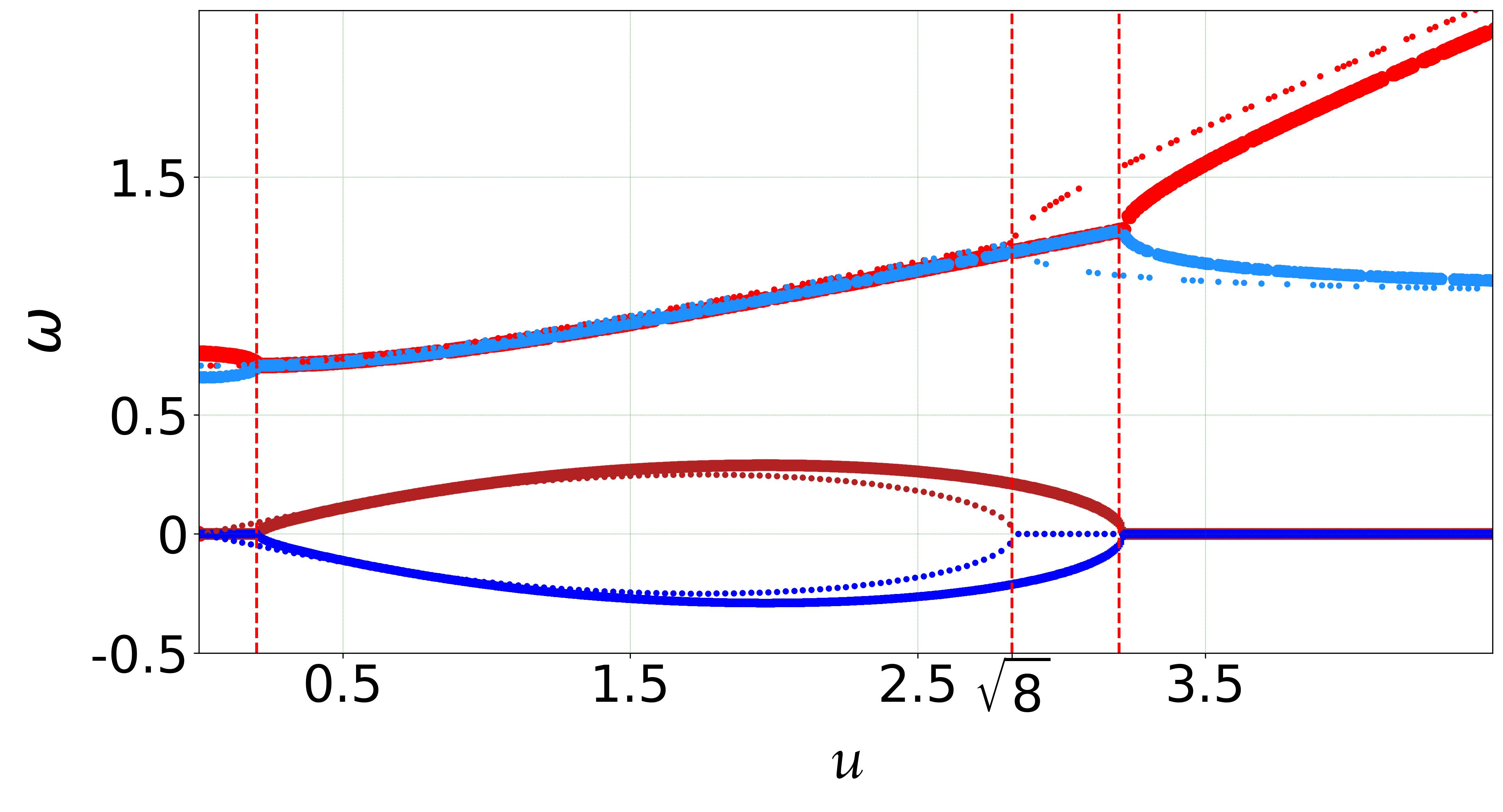}
\includegraphics[width=8.5cm, height=4cm]{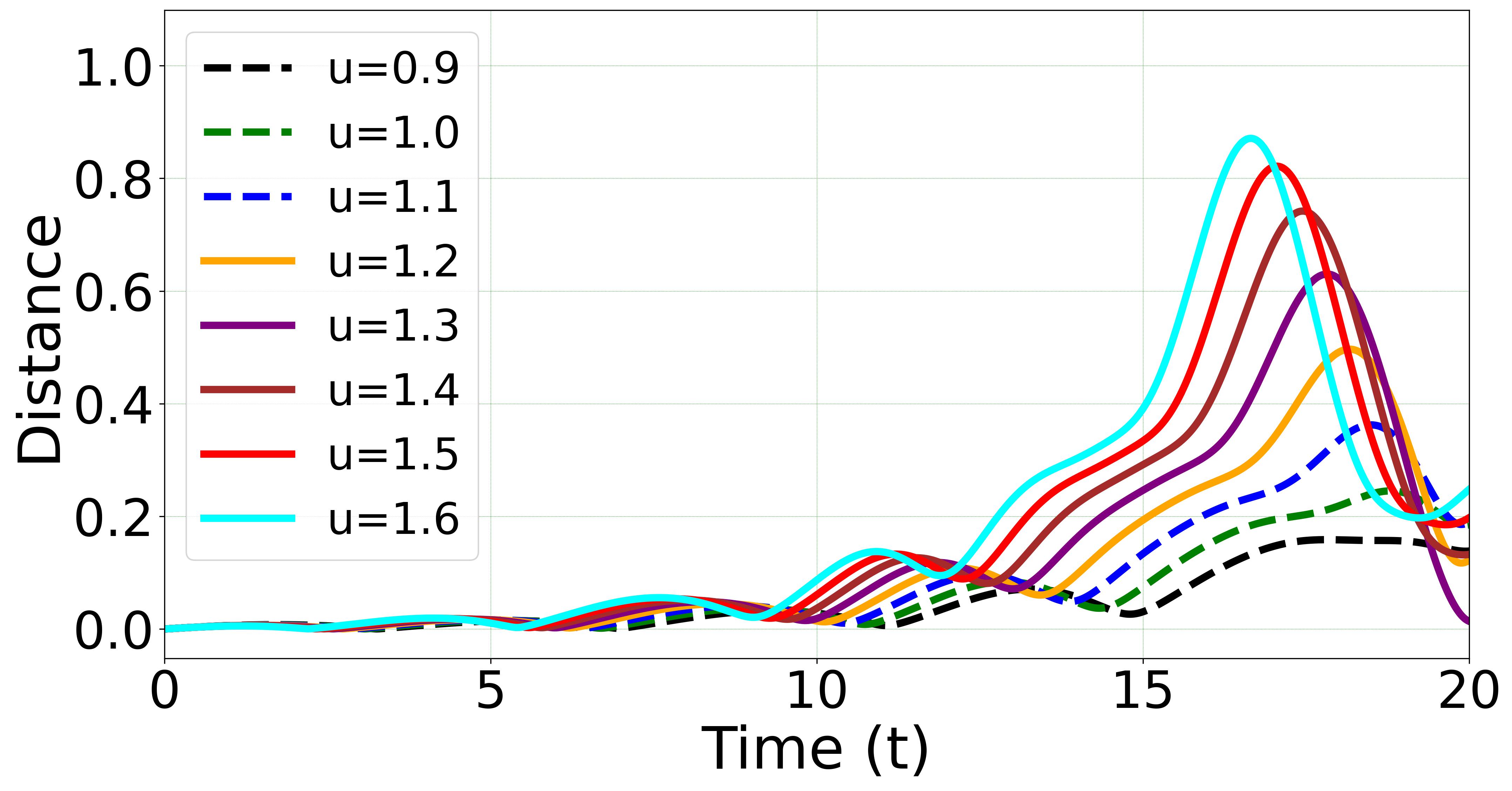}
\caption{{\bf Stability analysis of the SP.}
{\em Upper Panel:}
Upper (lower) lines are the real (imaginary) part of the Bogoliubov frequencies, plotted  against $u$. 
Dotted thin lines and thick solid lines are for detuning $v=0,0.1$ respectively. Dynamical instability is indicated by the non-vanishing imaginary component of the frequencies  in the range indicated by vertical dotted lines. 
{\em Lower Panel:}
Distance from the SP as a function of time for an individual trajectory launched very close to the SP. The dependence on $u$ is illustrated and reflects the instability.} 
\label{fig4_1}
\end{figure}

\begin{figure}
\centering
\includegraphics[width=8.5cm, height=4cm]{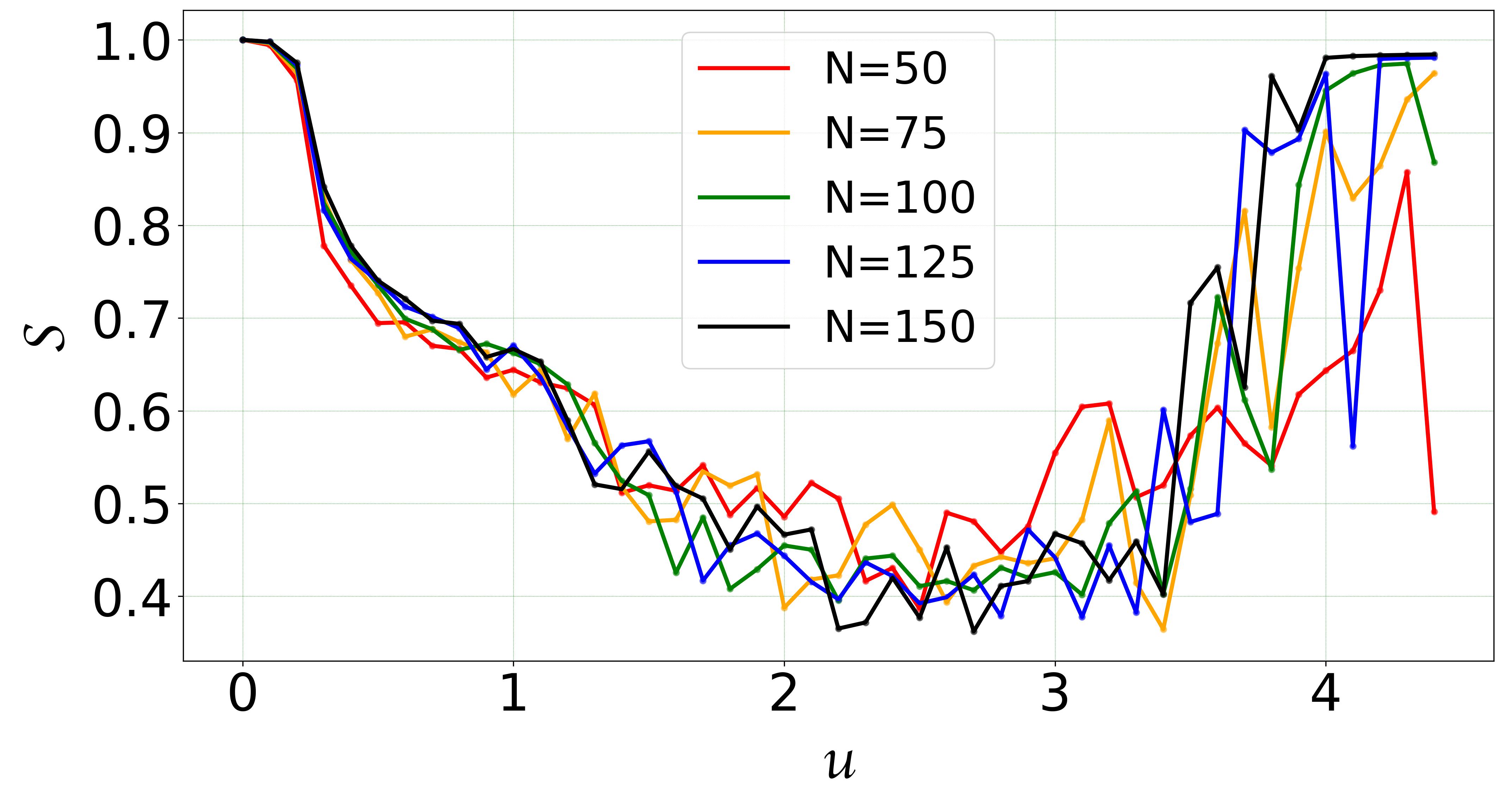}
\includegraphics[width=8.5cm, height=4cm]{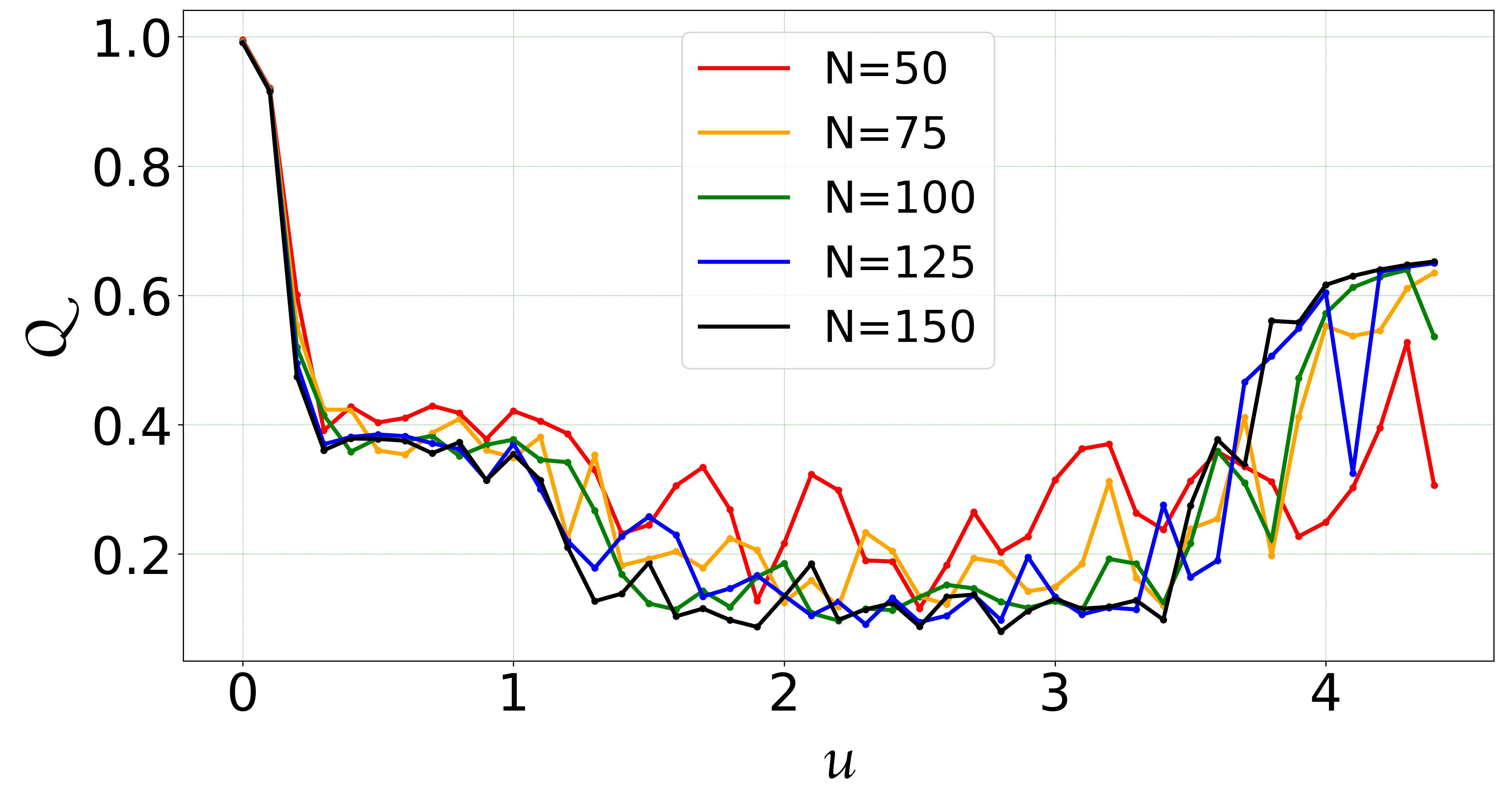}
\includegraphics[width=8.5cm, height=4cm]{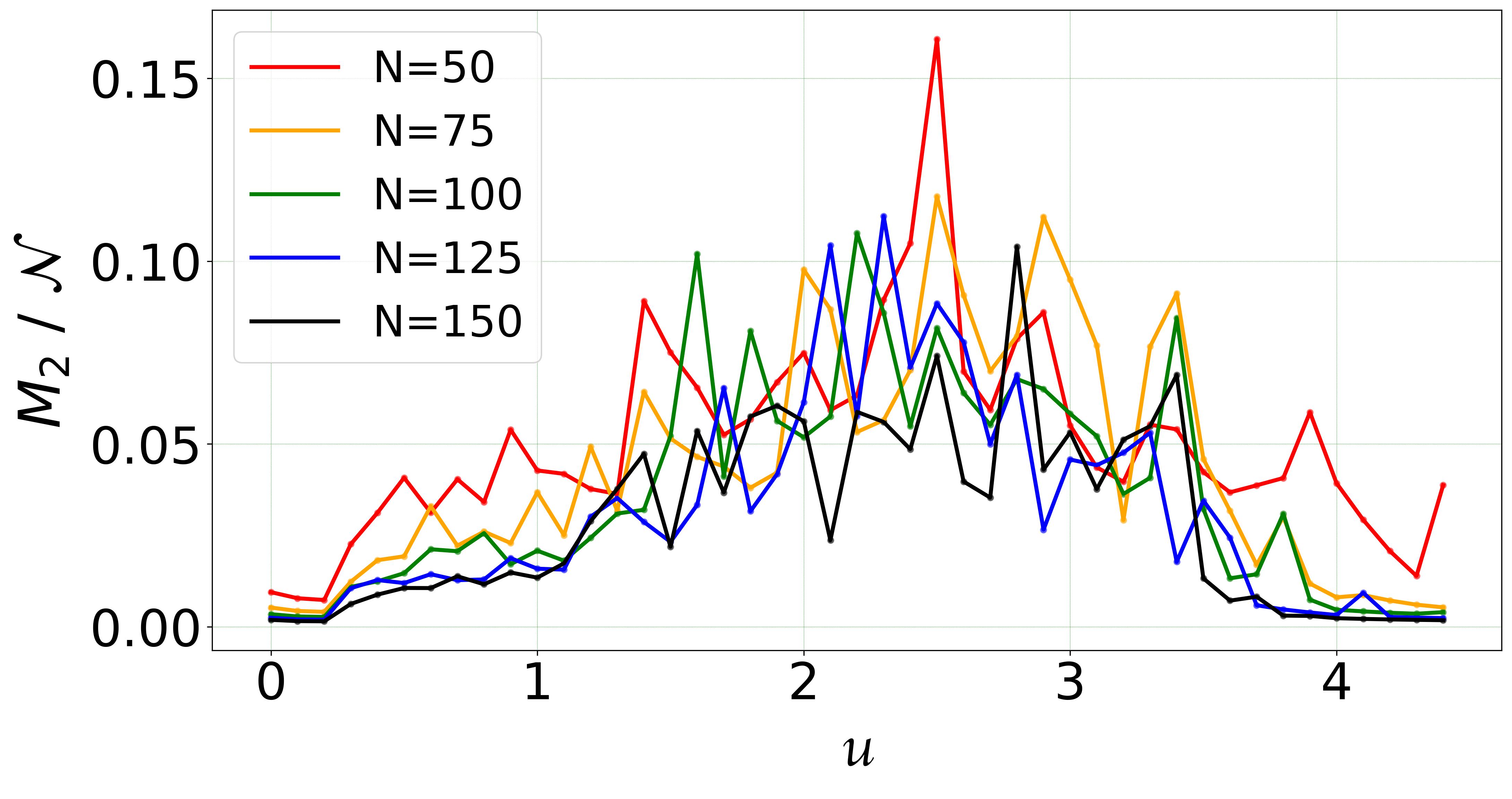}
\caption{{\bf Characterization of the SP-supported state.} 
The purity $S$ (top), the overlap with the dark state  $Q$ (middle), and the participation ratio $M_2/{\mathcal N}$ (bottom) 
of the SP-supported state are plotted versus the interaction strength  $u$ for different values of the total particle number $N$.
} 
\label{fig4_3a}
\end{figure}

\begin{figure*}
\centering
 
\includegraphics[width=3.5cm, height=3.20cm]{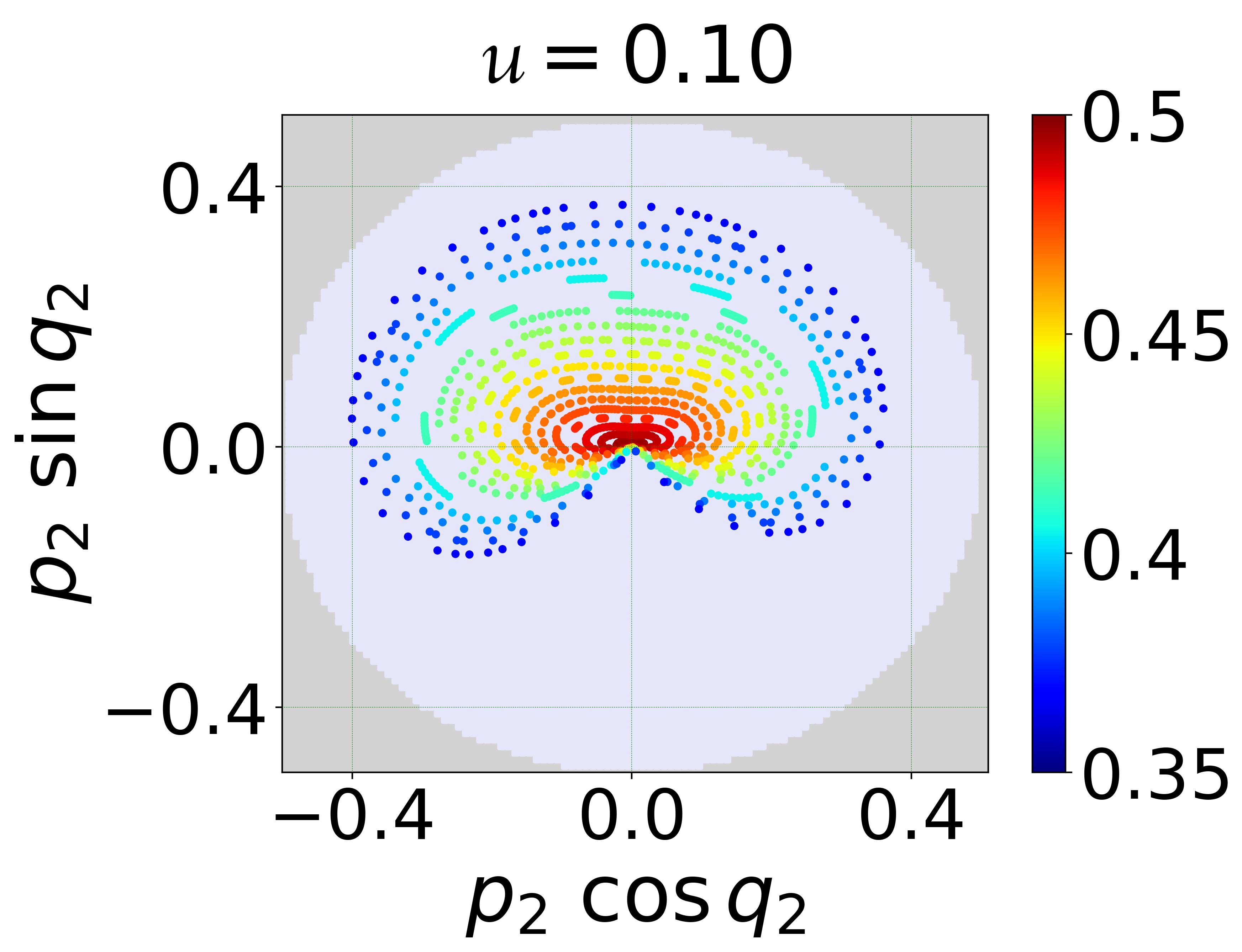}
\includegraphics[width=3.5cm, height=3.20cm]{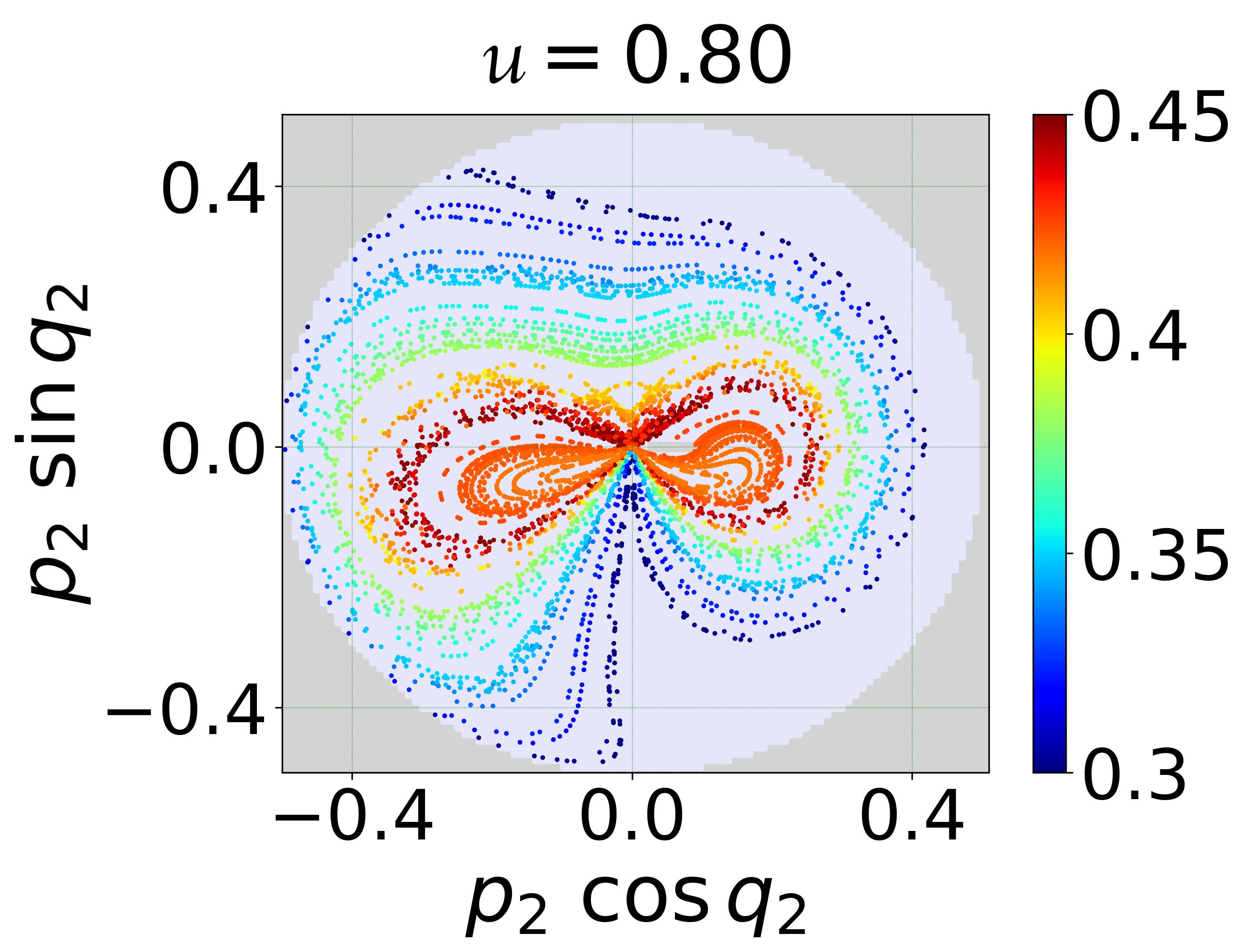}  
\includegraphics[width=3.5cm, height=3.20cm]{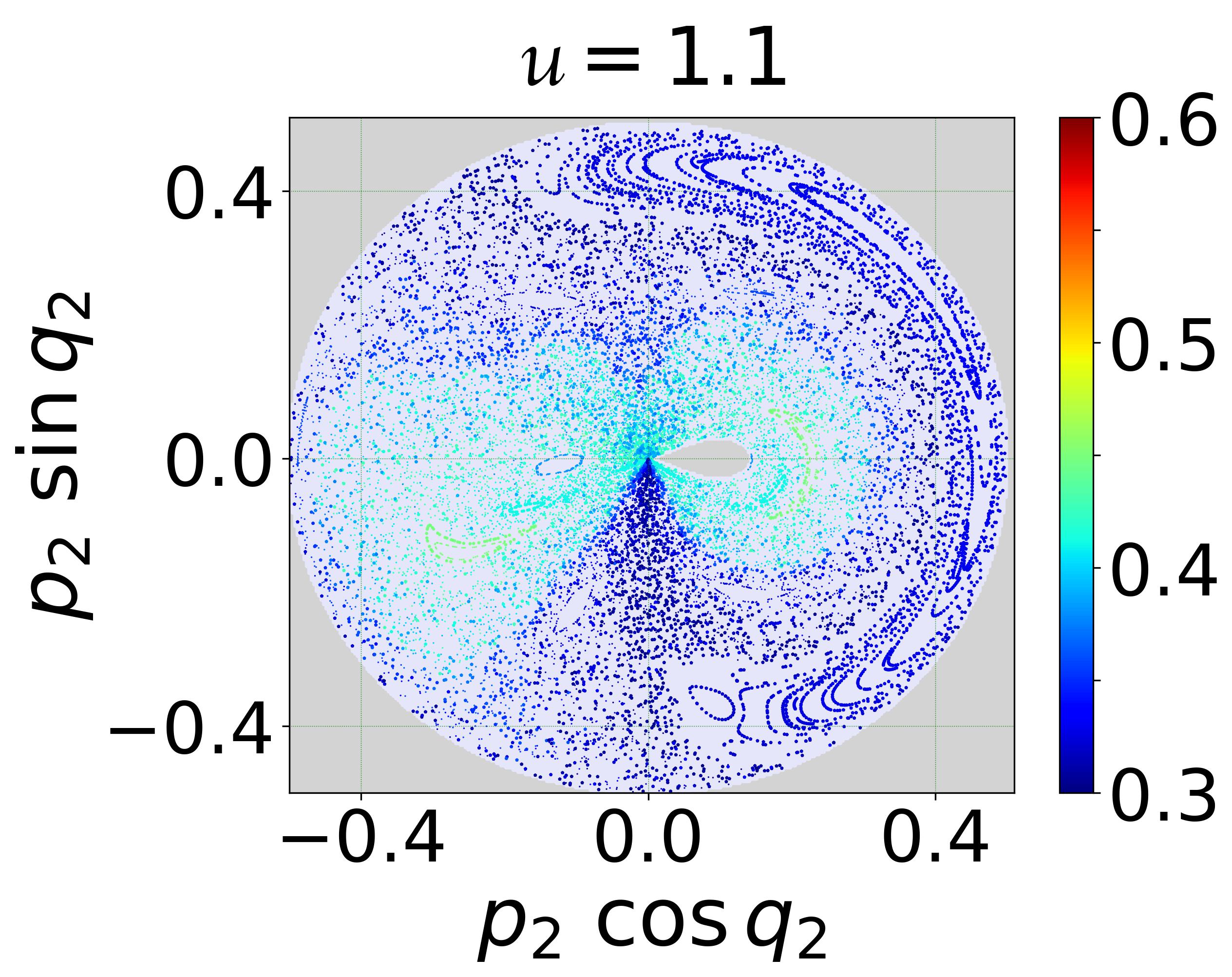}
\includegraphics[width=3.5cm, height=3.20cm]{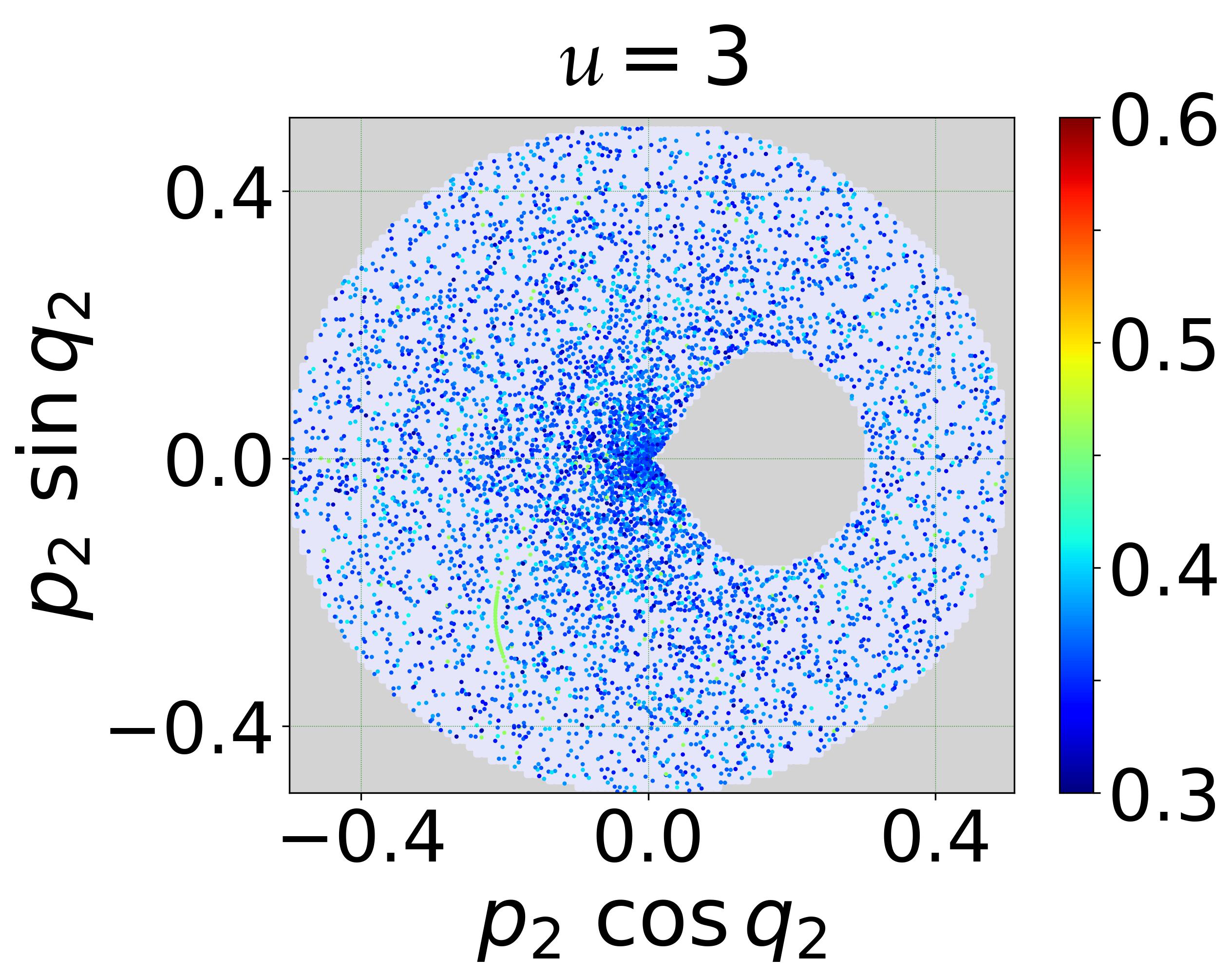}
\includegraphics[width=3.5cm, height=3.20cm]{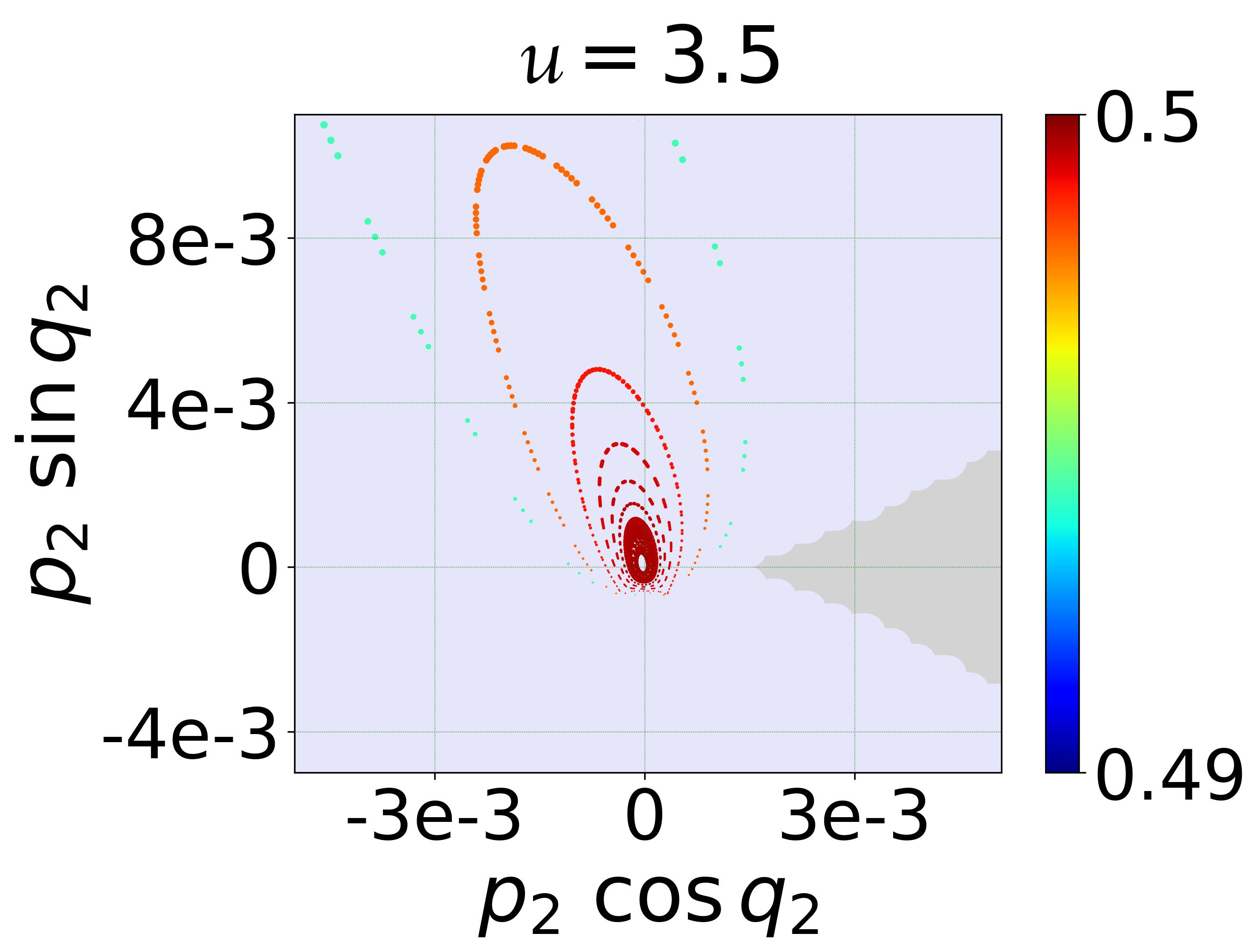}

\includegraphics[width=3.5cm, height=3.20cm]{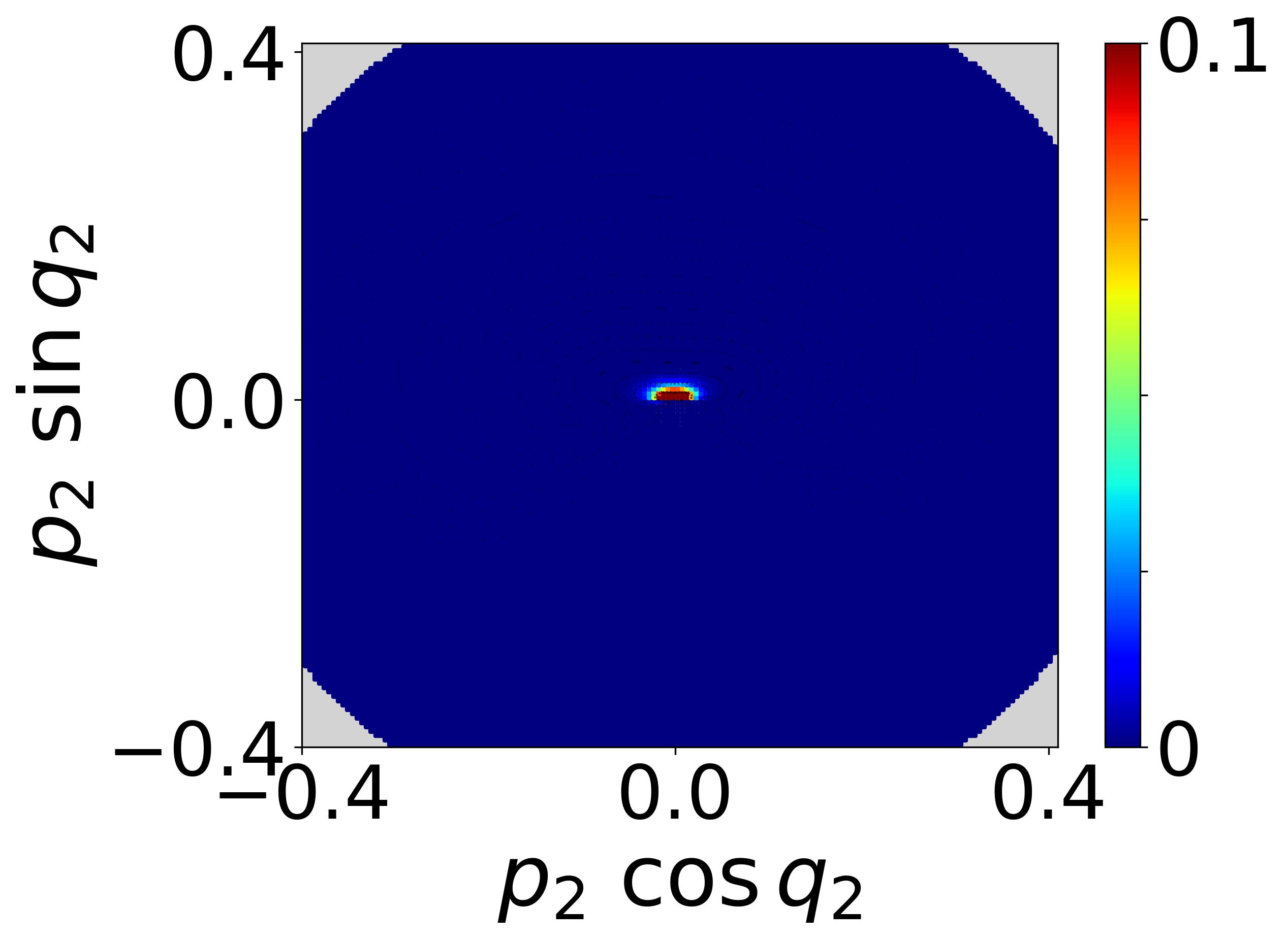}
\includegraphics[width=3.5cm, height=3.20cm]{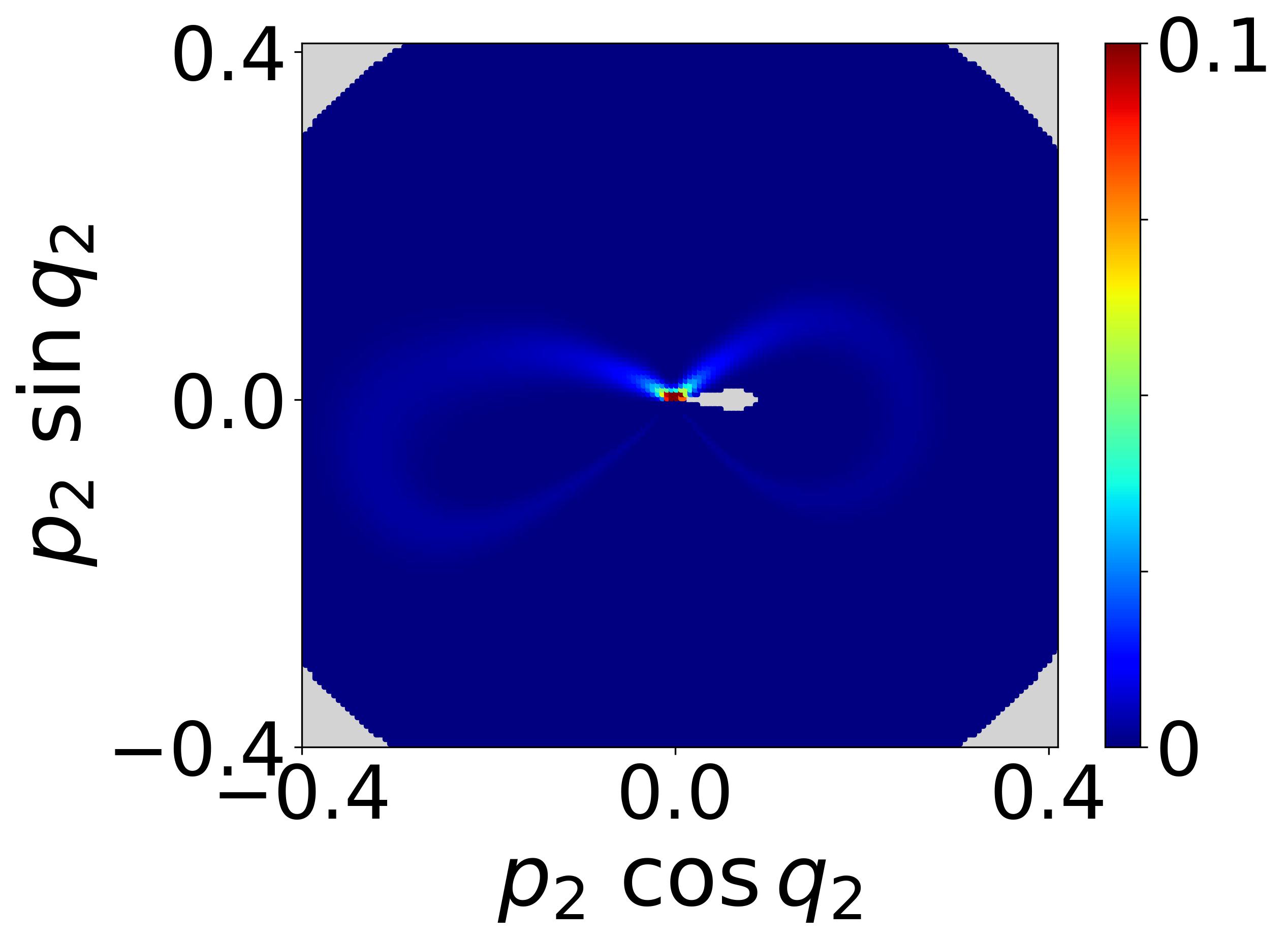}  
\includegraphics[width=3.5cm, height=3.20cm]{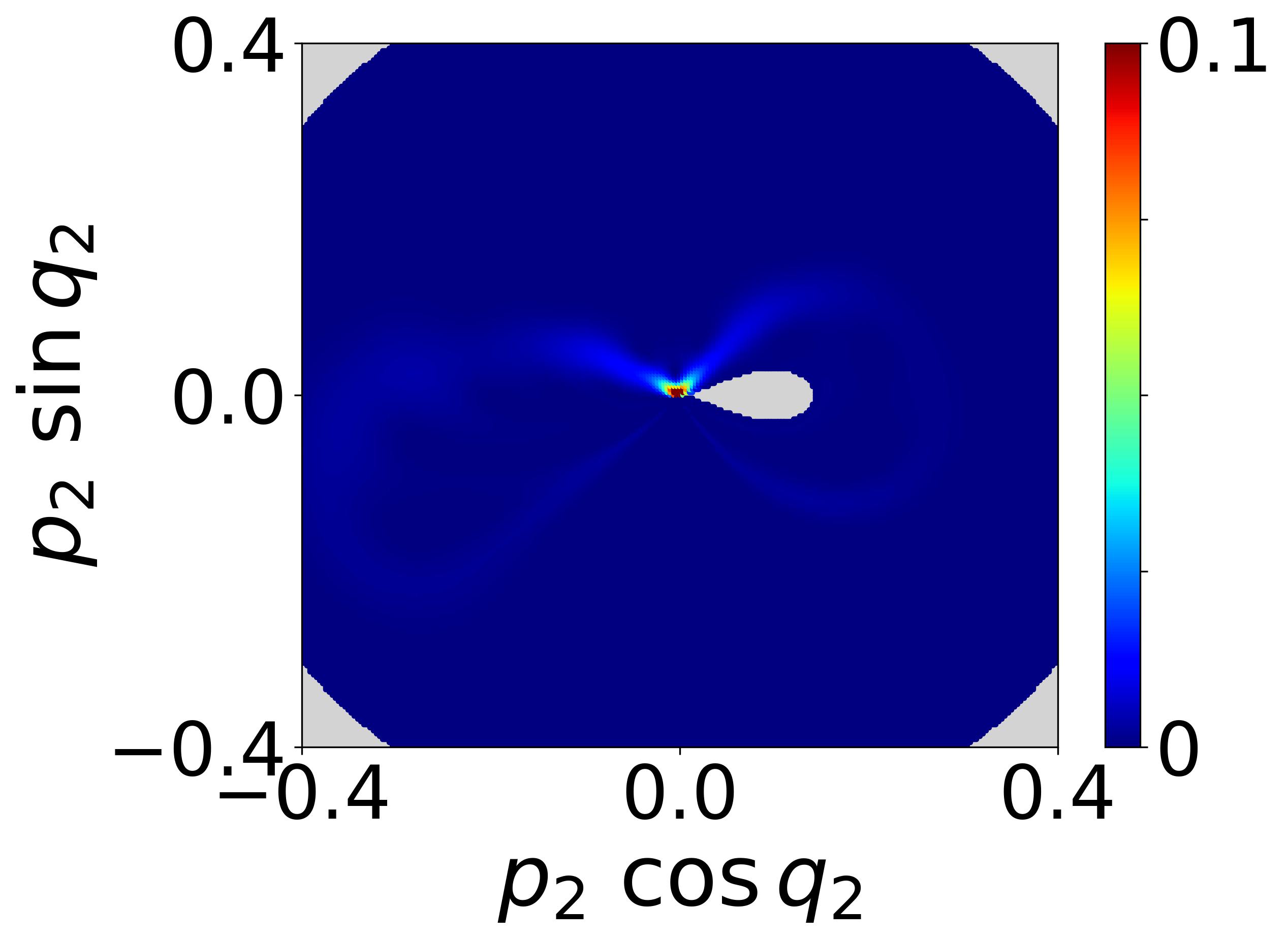}
\includegraphics[width=3.5cm, height=3.20cm]{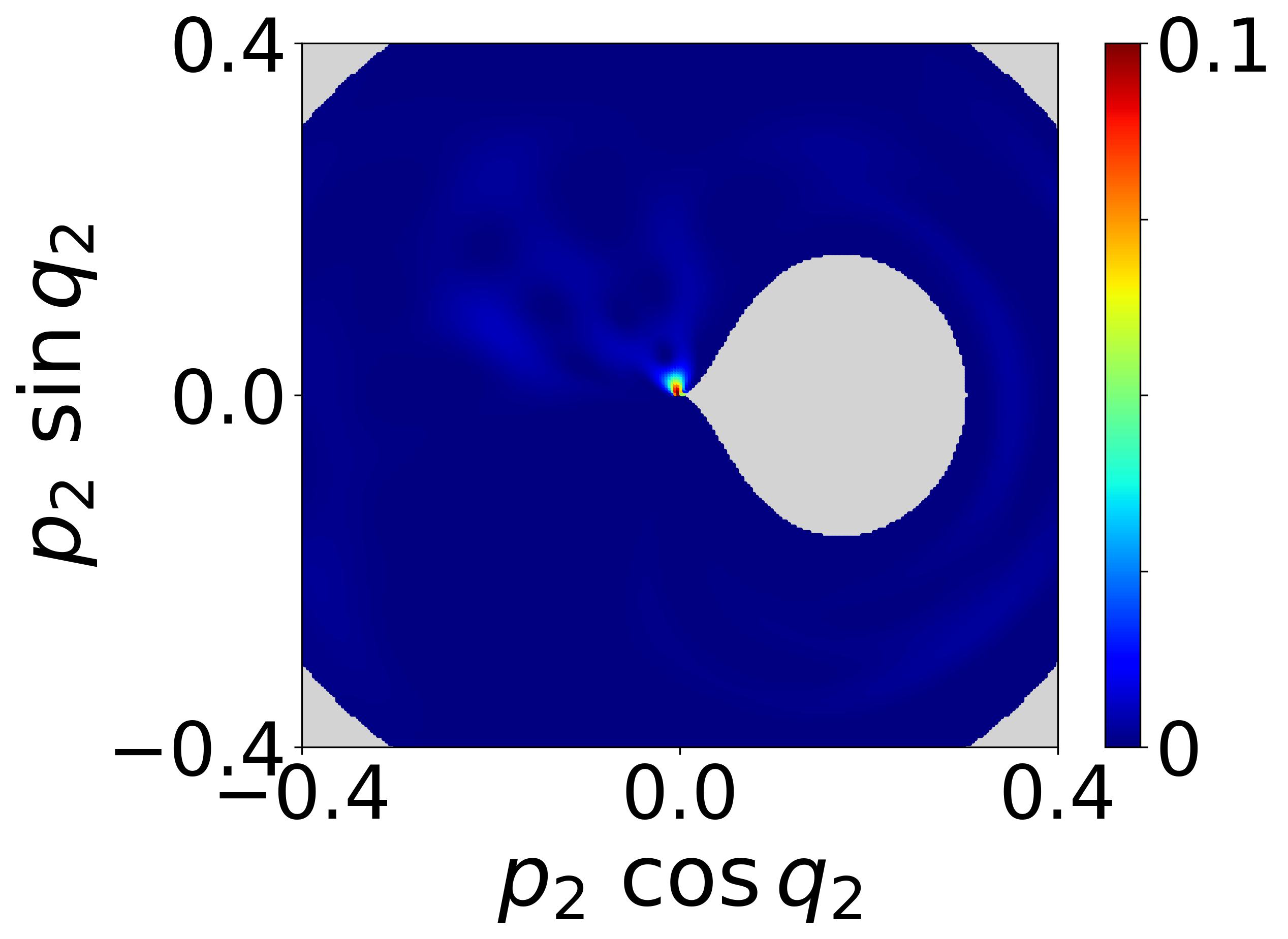}
\includegraphics[width=3.25cm, height=3.20cm]{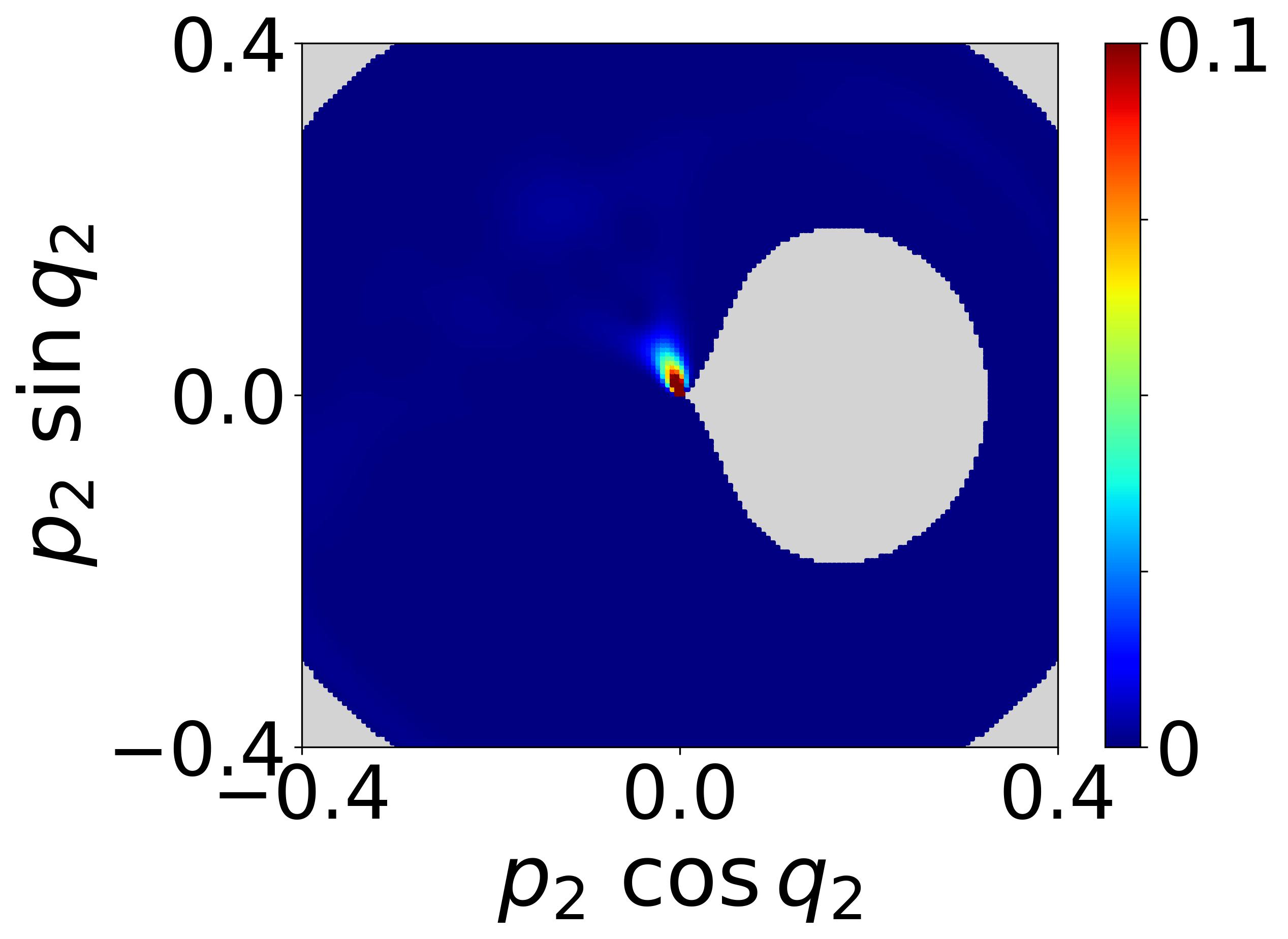}
 
\includegraphics[width=3.5cm, height=3.20cm]{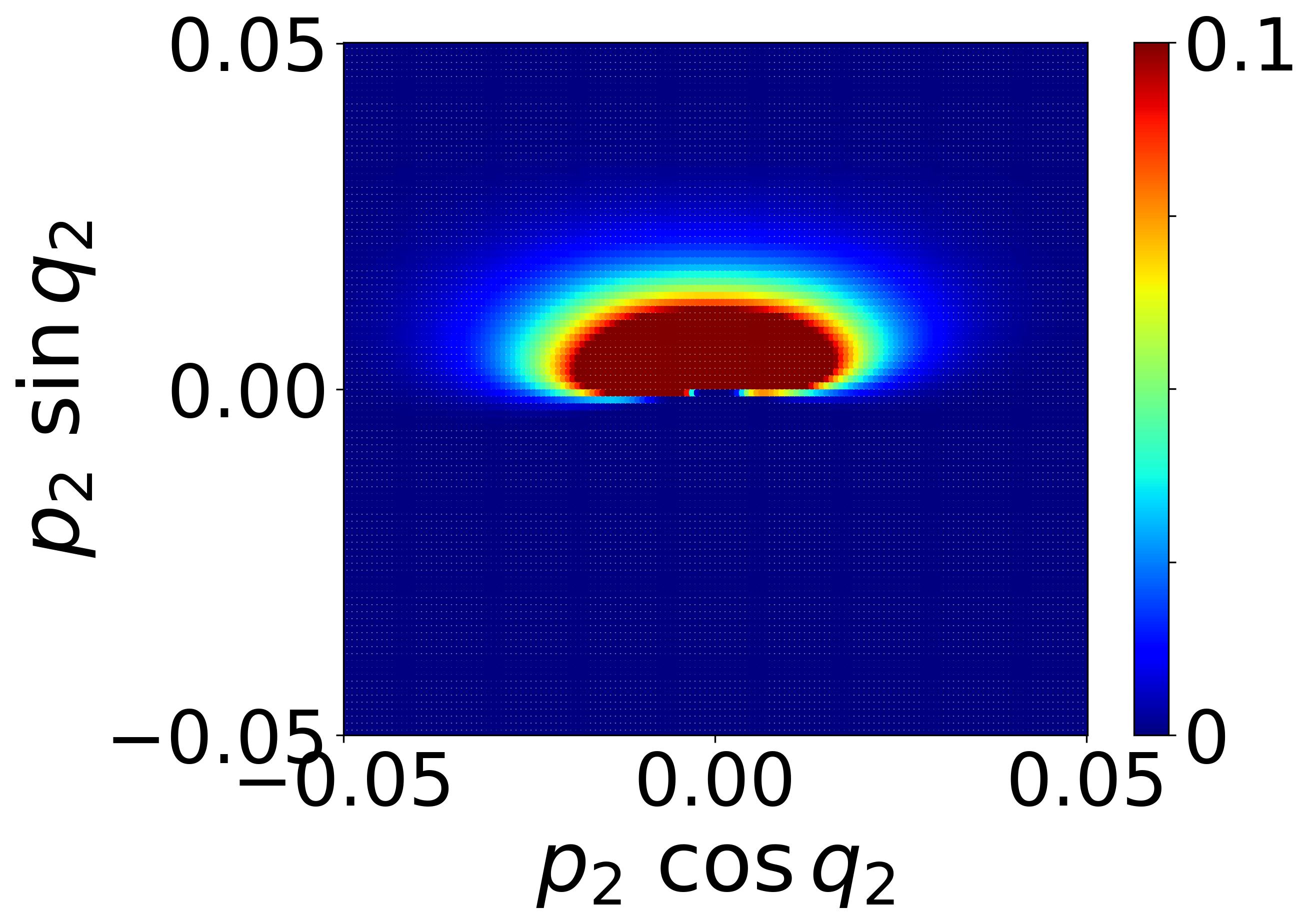}
\includegraphics[width=3.5cm, height=3.20cm]{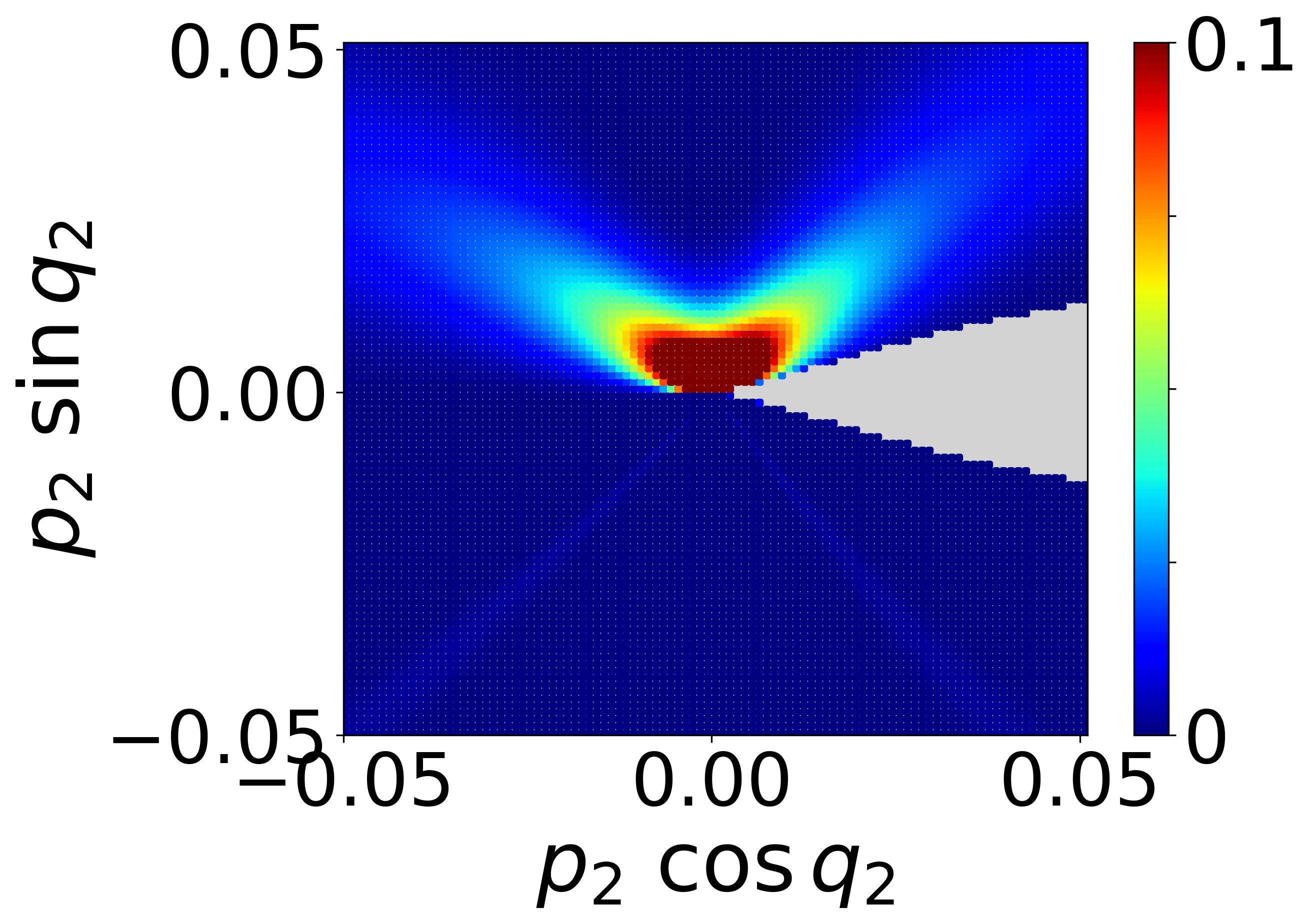} 
\includegraphics[width=3.5cm, height=3.20cm]{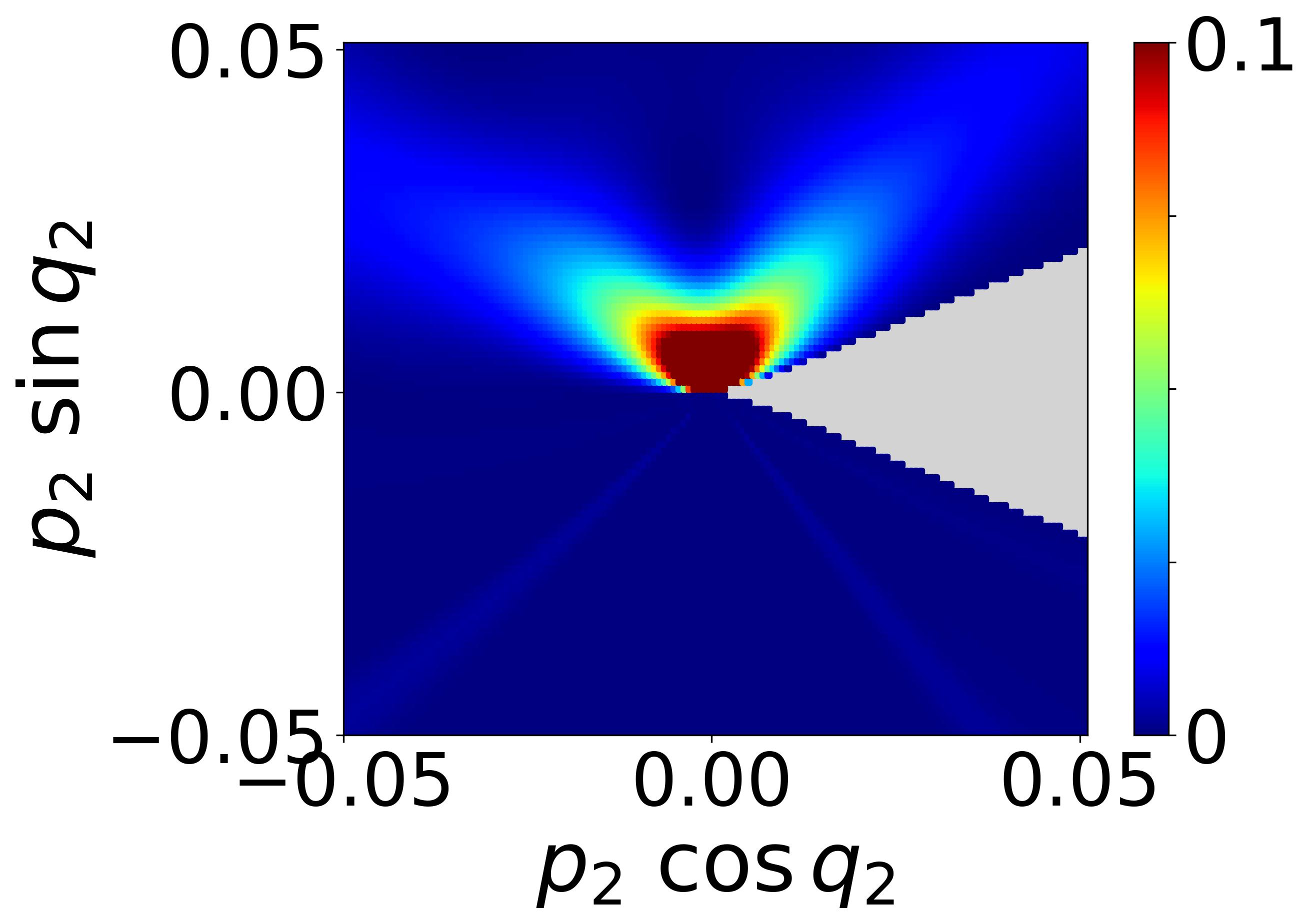}
\includegraphics[width=3.5cm, height=3.20cm]{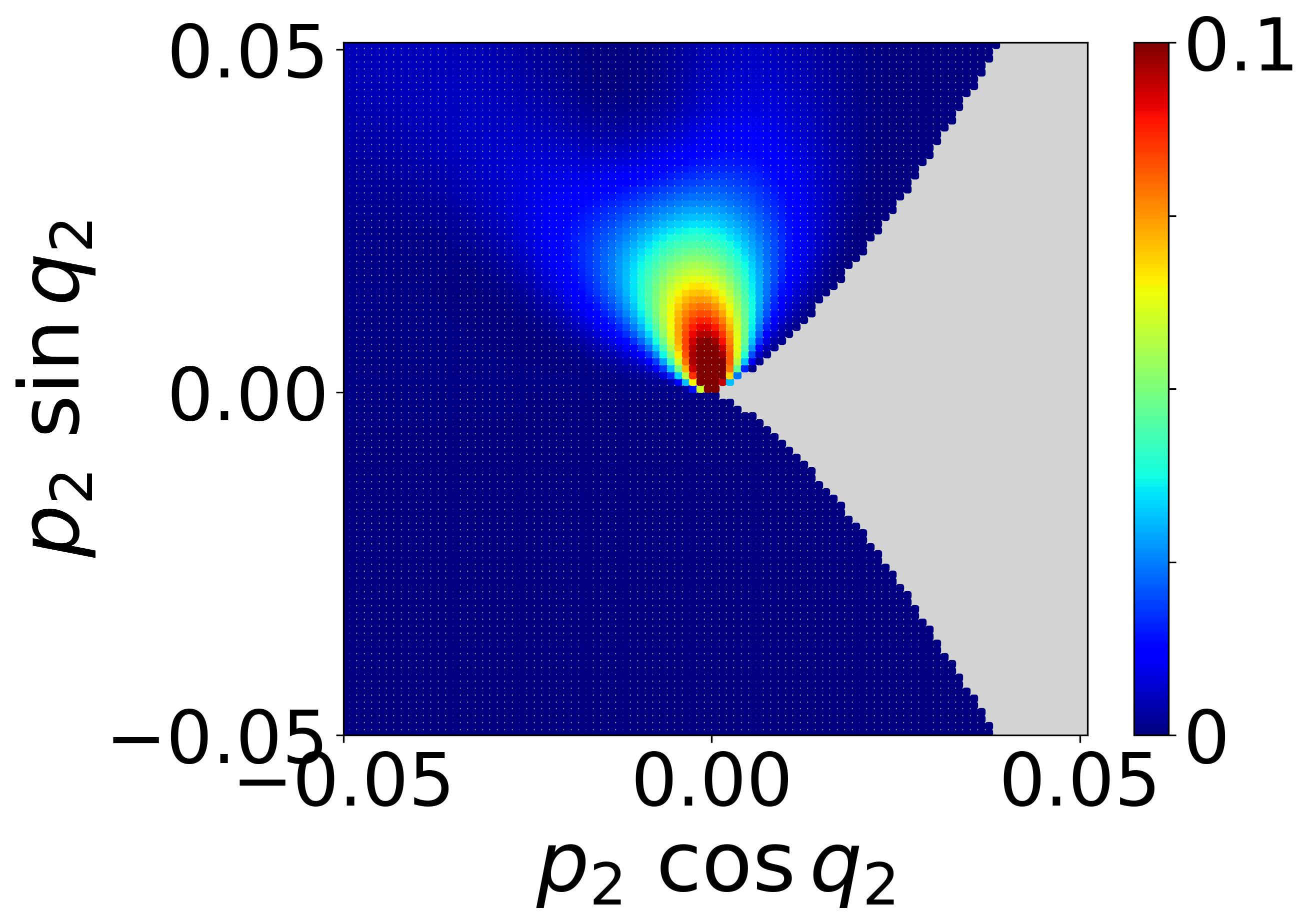}
\includegraphics[width=3.5cm, height=3.20cm]{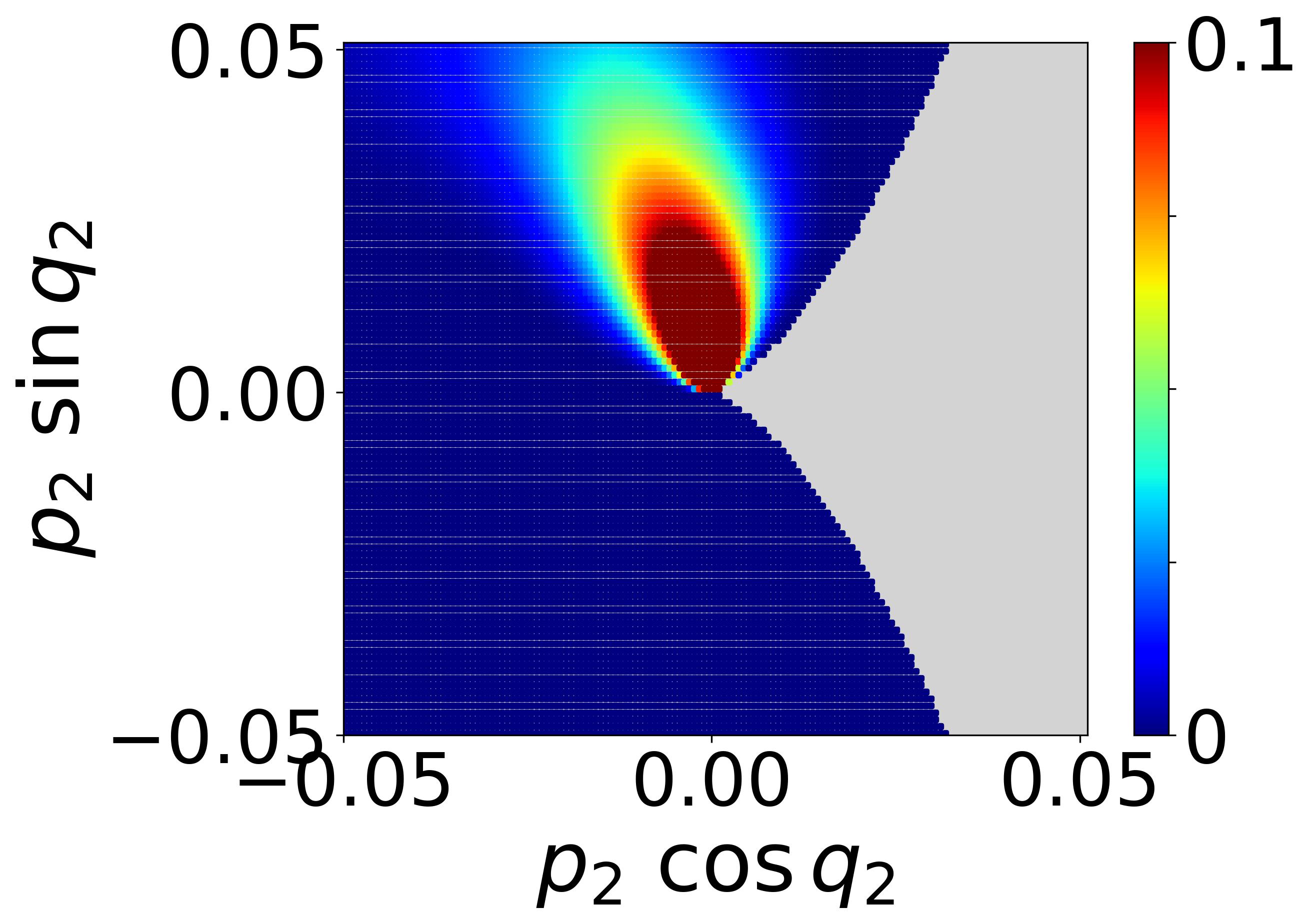}

\includegraphics[width=3.5cm, height=3.20cm]{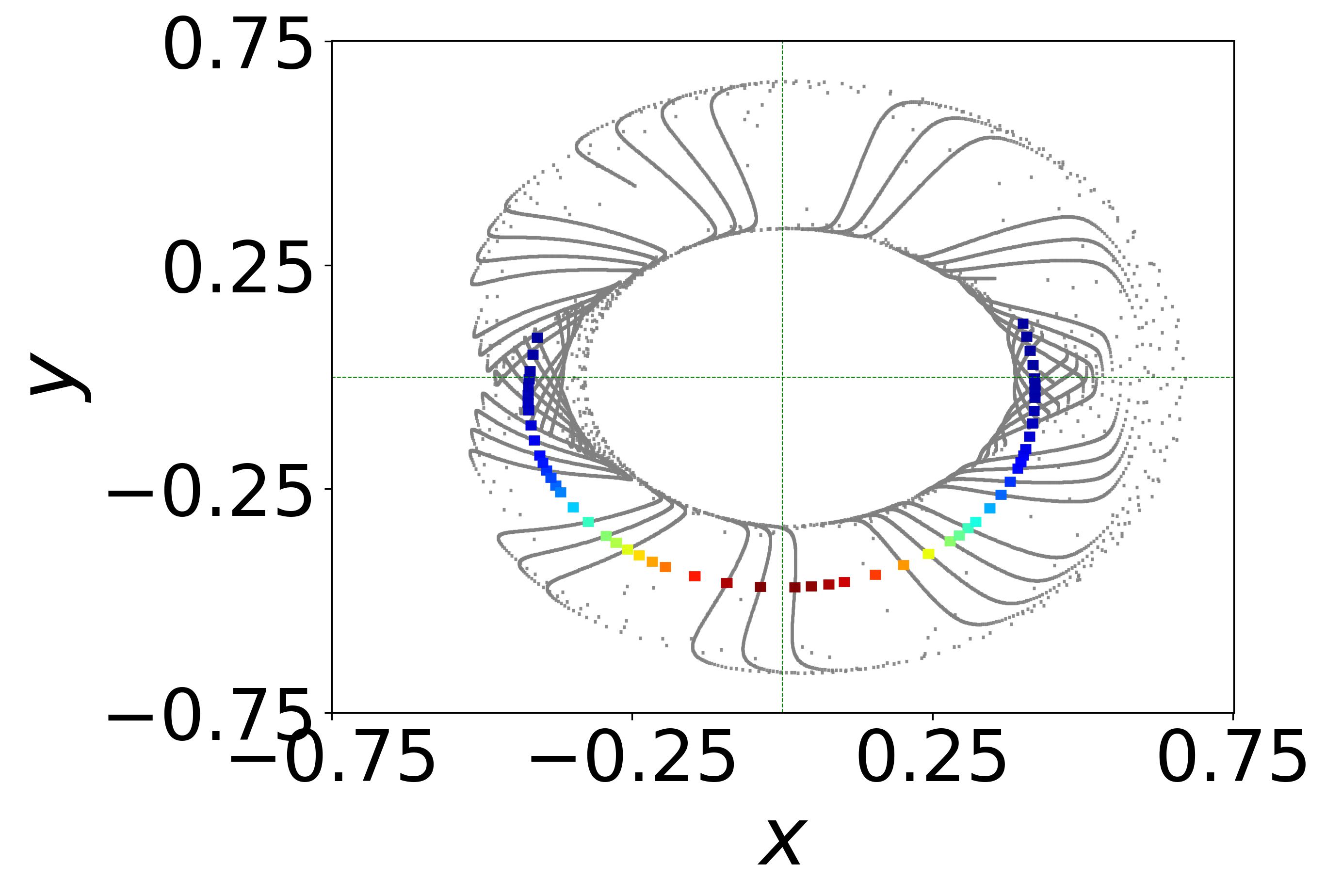}
\includegraphics[width=3.5cm, height=3.20cm]{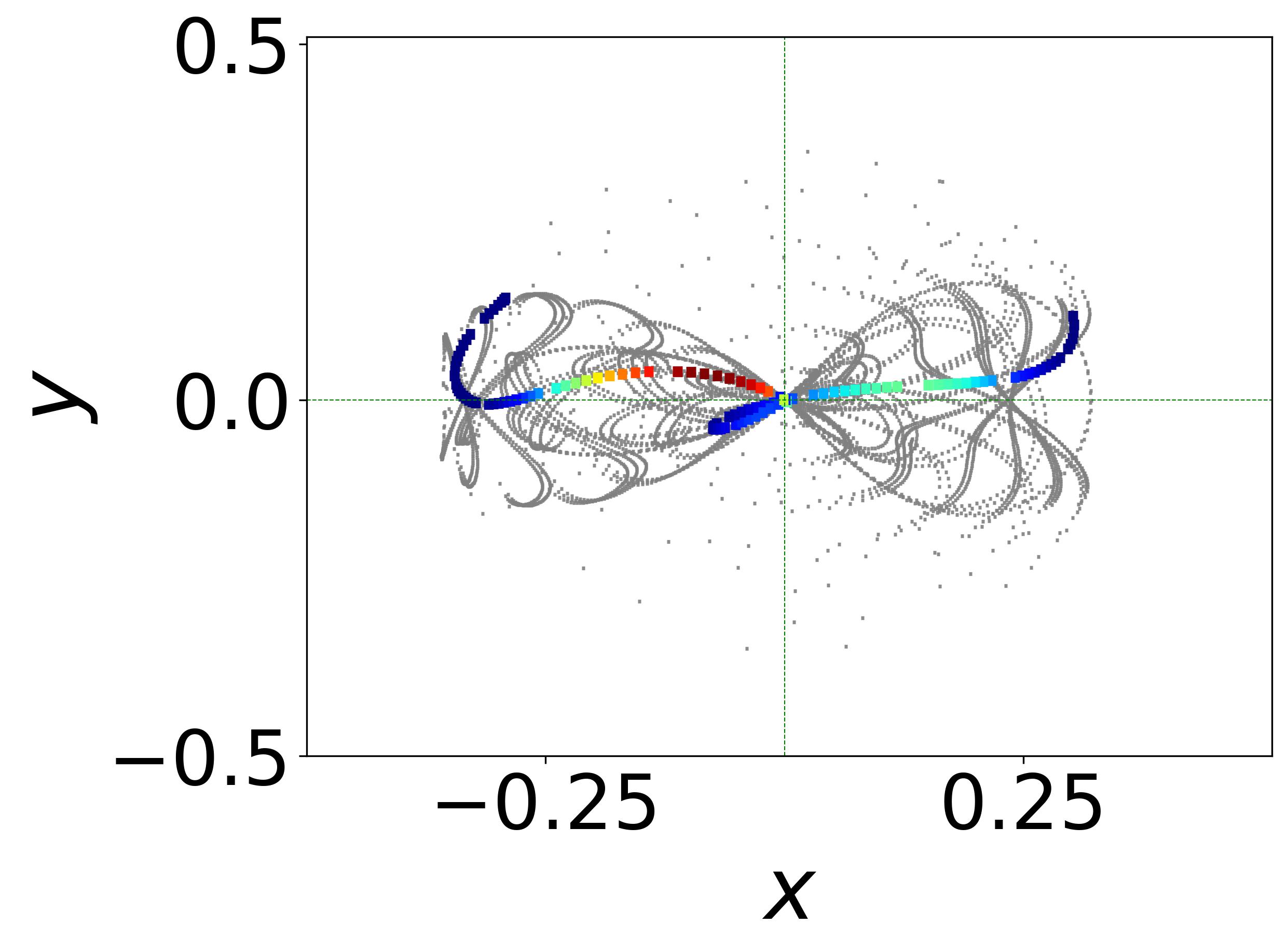}
\includegraphics[width=3.5cm, height=3.20cm]{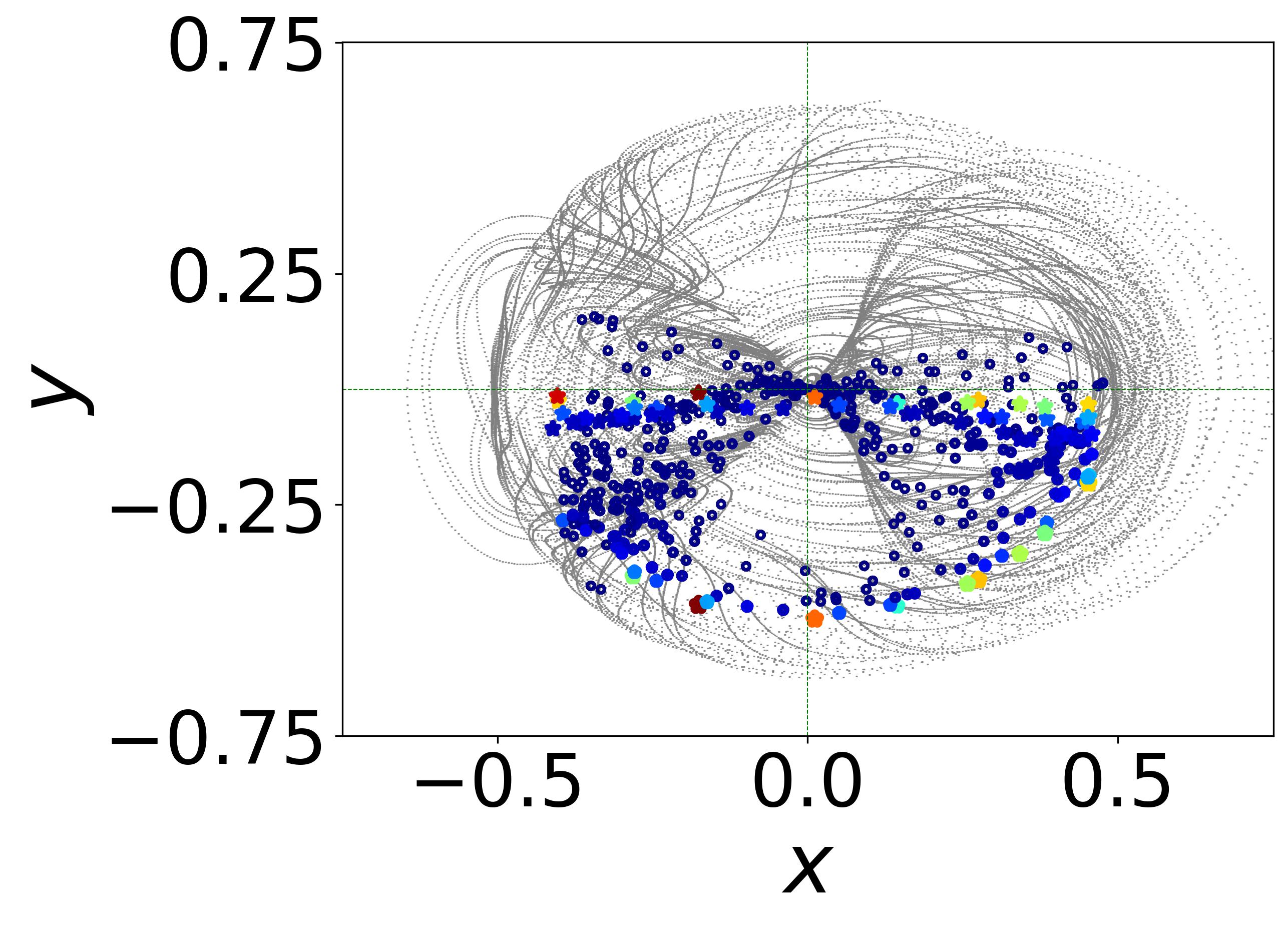}
\includegraphics[width=3.5cm, height=3.20cm]{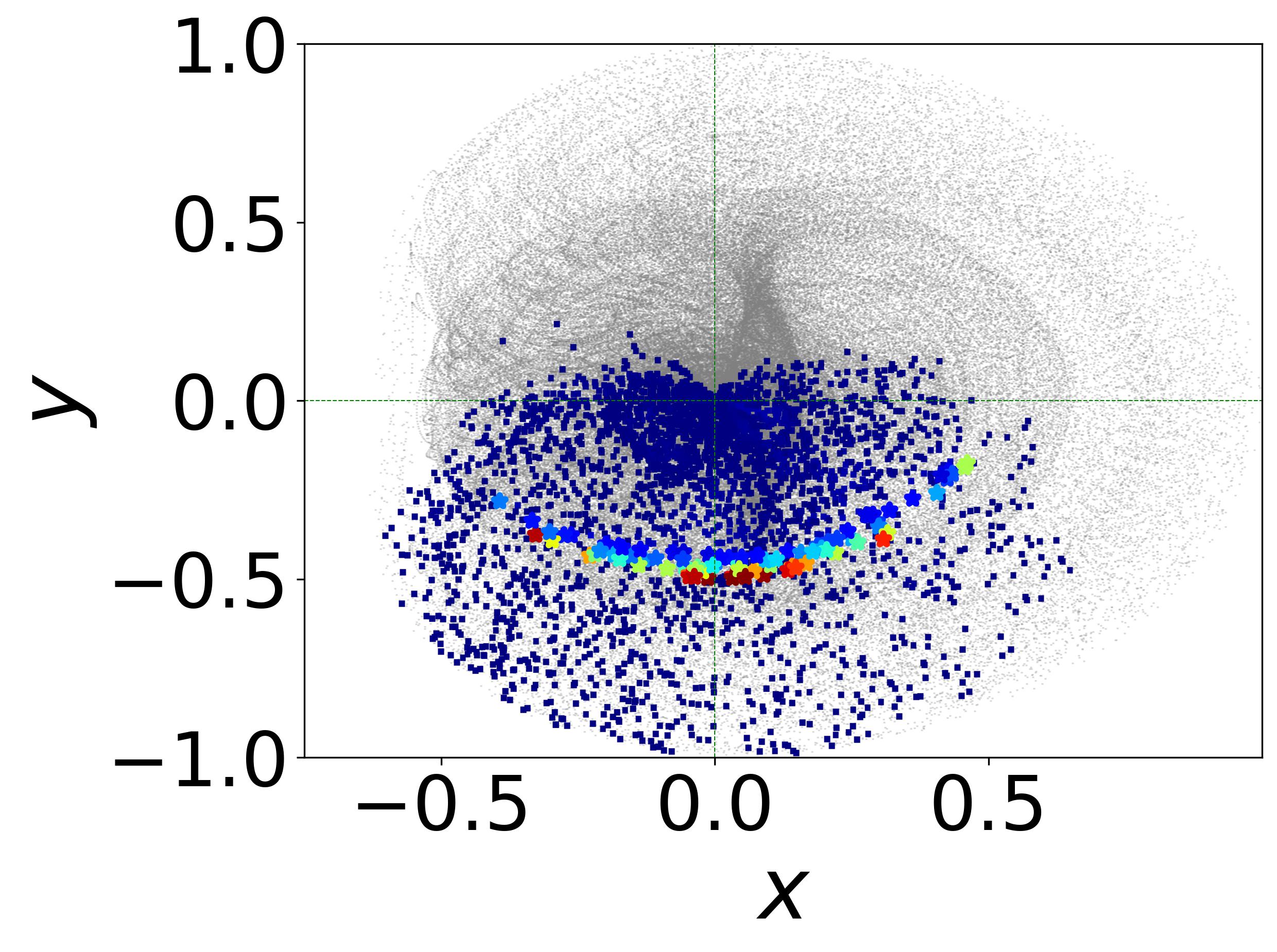}
\includegraphics[width=3.5cm, height=3.20cm]{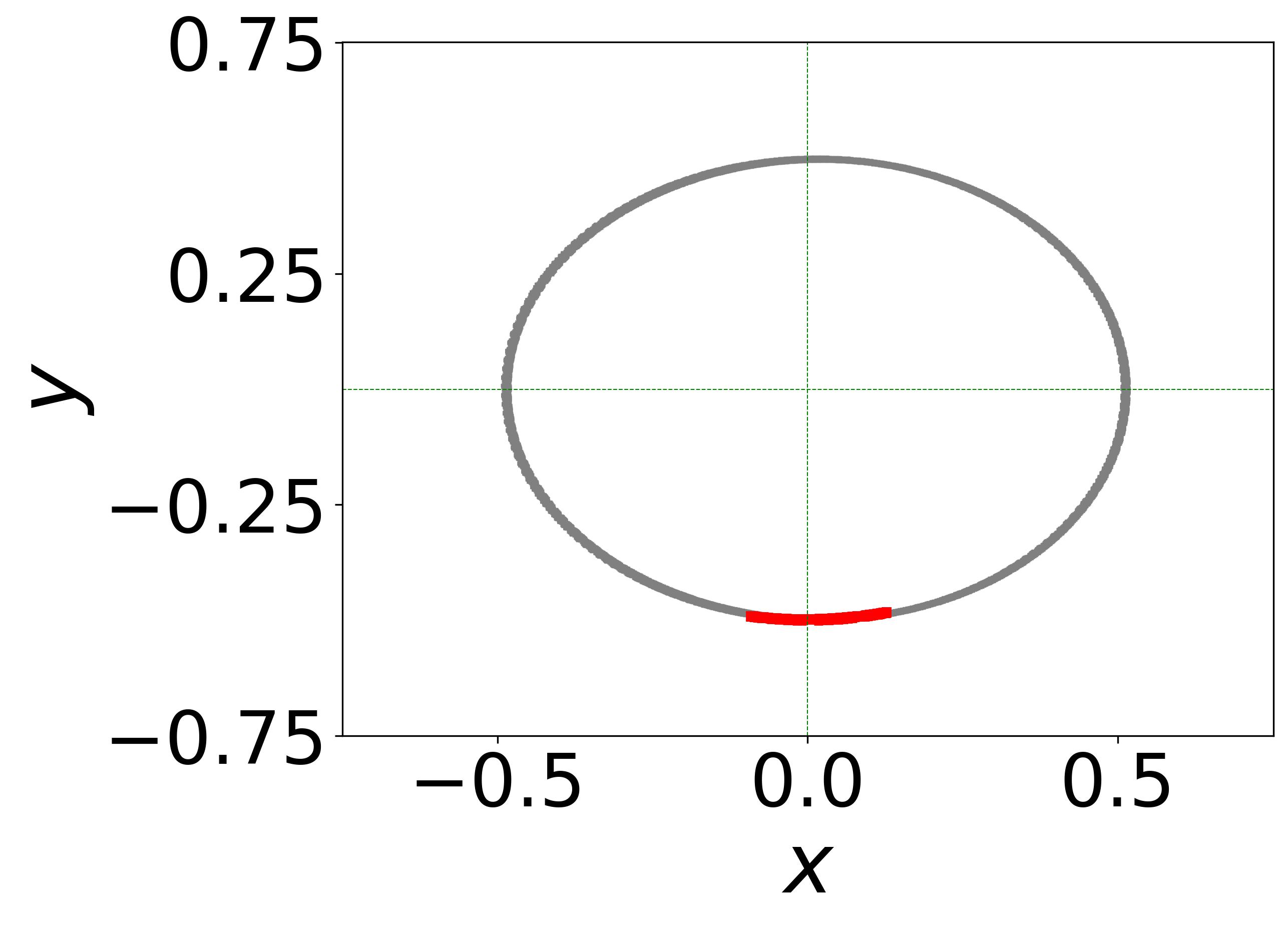}

\includegraphics[width=3.5cm, height=3.20cm]{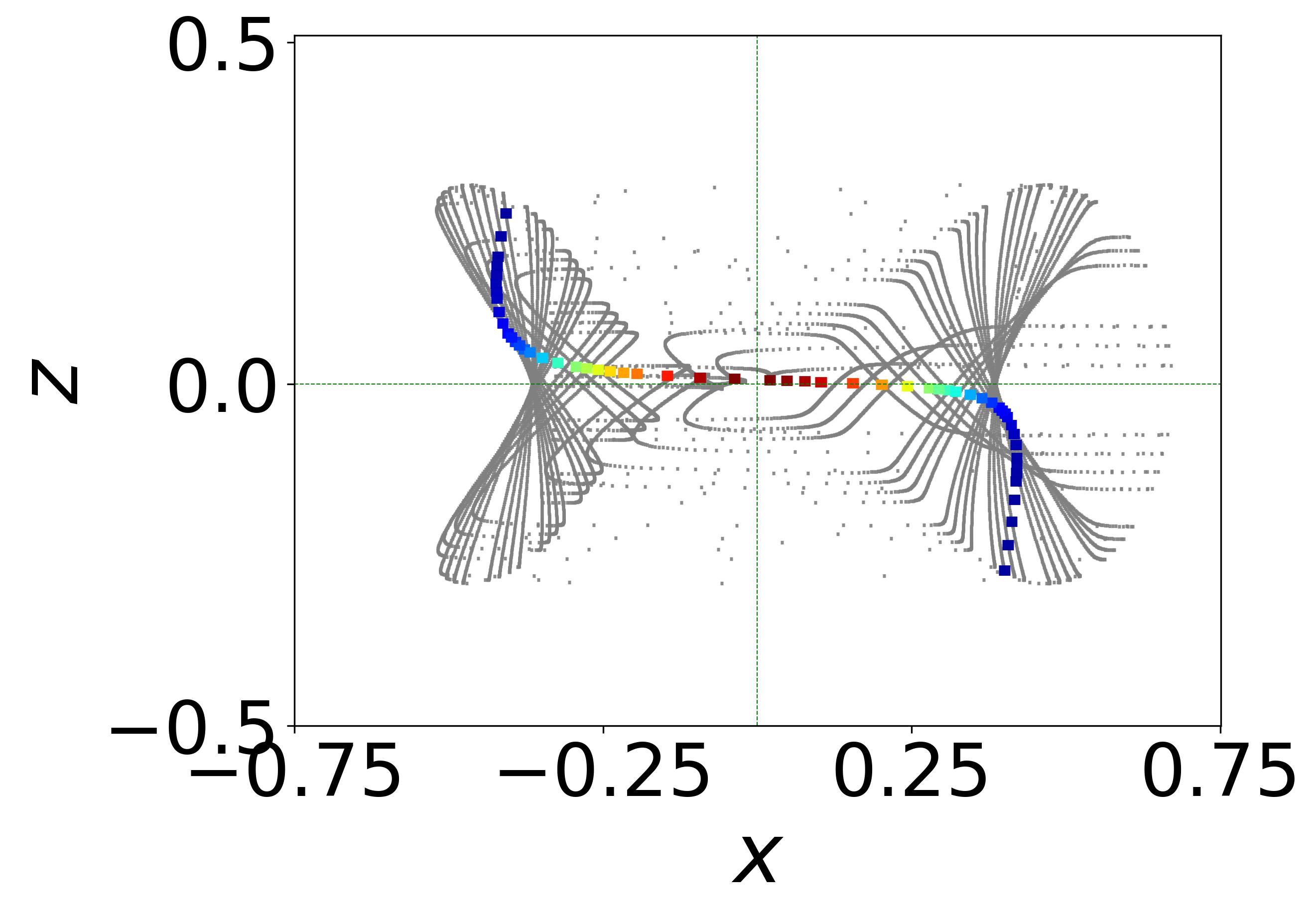}
\includegraphics[width=3.5cm, height=3.20cm]{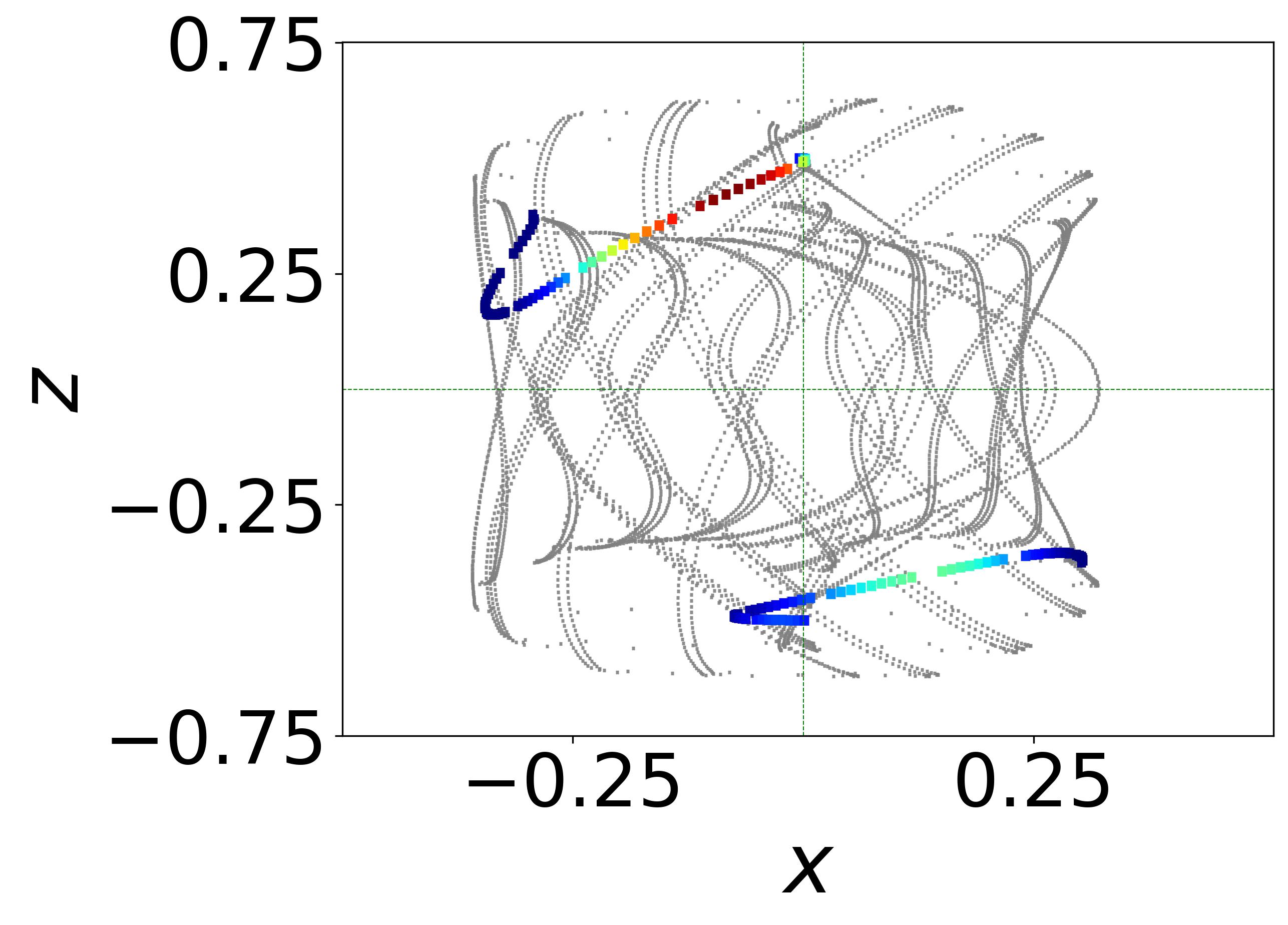}
\includegraphics[width=3.5cm, height=3.20cm]{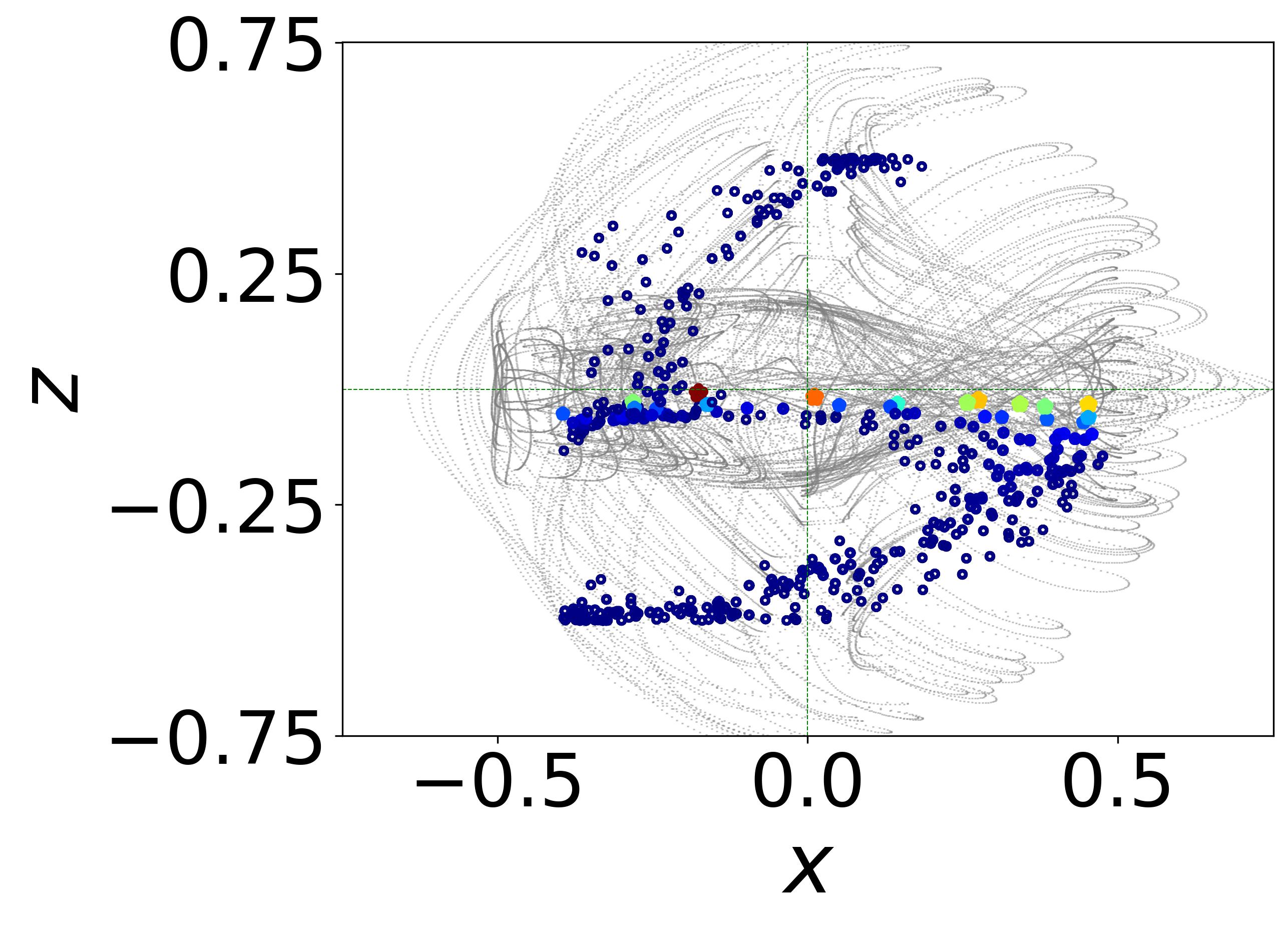}
\includegraphics[width=3.5cm, height=3.20cm]{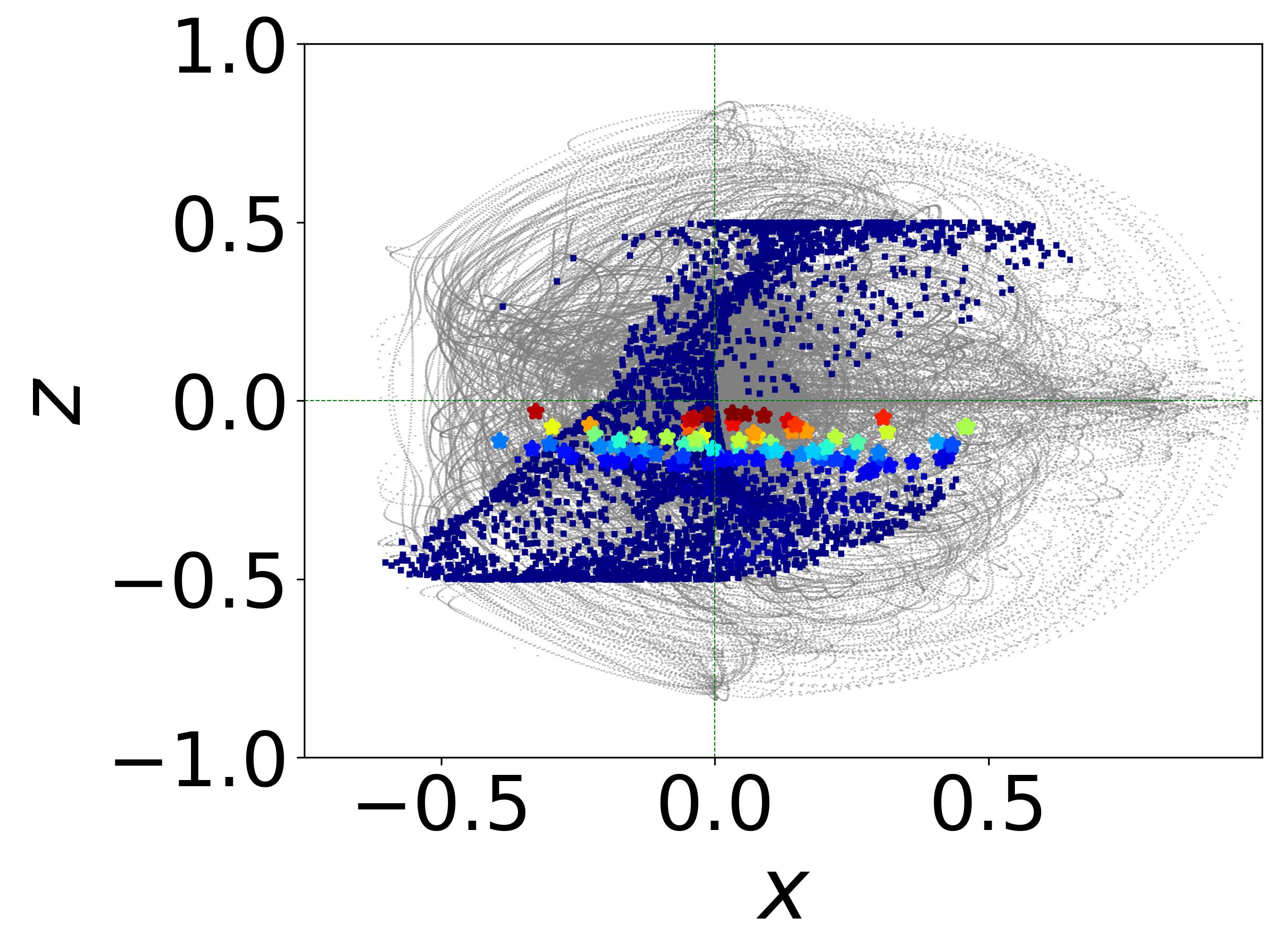}
\includegraphics[width=3.5cm, height=3.20cm]{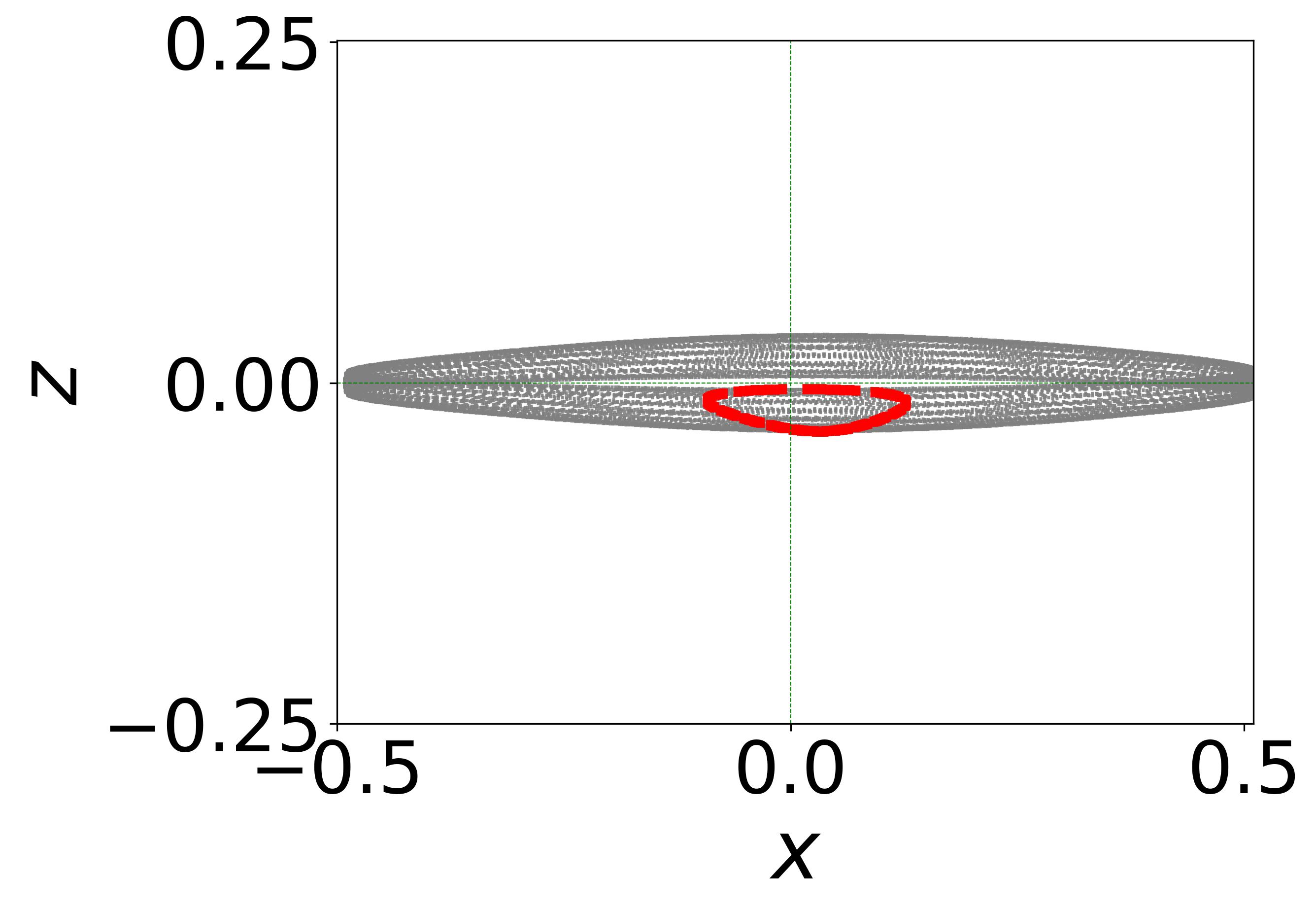}

\caption{{\bf Classical dynamics vs quantum phasespace distribution of the SP-supported eigenstate.}
Top row: Poincare sections at $p_1=1/2$, taken at $E_{\text{SP}}$, namely, along the dashed line of \Fig{fUE}. 
Five representative values of $u$ were selected, 
where the SP is stable; unstable but regular; unstable in chaos;  embedded in global chaos; and restabilized by self-trapping. 
The color code represents the average value of $p_{1}$ along the trajectory. 
Second row: The Husimi phasespace distribution of the SP-supported eigenstate for the same values of $u$. The color code represents the values $|\langle \alpha | E_{\nu}\rangle|^{2}$, where $\nu$ corresponds to the SP-supported eigenstate index, and the number of particles is $N=150$. Other parameters are the same as in the previous figures. 
Third row: zoom on the same Husimi distributions.
Last two rows: 3D plot of representative trajectory for each~$u$ from two different view aspects. Points of the Poincare sections are color-coded as in the Husimi plots. 
The torodial coordinates are: 
$x = (p_{2} + p_{1} \cos q_{1}) \cos q_{2}$, 
and $y = (p_{2} + p_{1} \cos q_{1}) \sin q_{2}$, 
and $z = p_{1} \sin q_{1}$.  
}
\label{fHSTg}
\end{figure*}

\section{The central stationary point}
\label{sV}

The classical Hamiltonian has a mid-spectrum stationary point (SP) at 
${\alpha}_{\text{SP}}=(1/\sqrt{2},0,-1/\sqrt{2})$, 
that corresponds to the dark-state orbital. The corresponding canonical variables are thus $p_1{=}1/2$ , $q_1{=}\pi$, $p_2{=}0$ while $q_2$ is ill defined.  This SP dominates the phasespace structure. Its energy is
\begin{equation} \label{eSP}
E_{SP} = \dfrac{1}{4}Nu
\end{equation}
Quantum mechanically the SP supports a coherent state where all particles occupy the dark state orbital, namely,   
\begin{equation}
|\alpha_\text{SP}\rangle = \dfrac{1}{\sqrt{2^{N} N!}} (\hat{a}^{\dagger}_{1} -\hat{a}^{\dagger}_{3})^{N} \ket{0}~.
\end{equation}

The mid-spectrum SP remains a stationary point of the classical dynamics even in the presence of interaction. It is a fixed point of the discrete nonlinear Schr\"{o}dinger equation \cite{PhysRevA.73.013617, Pethick_Smith_2008}.  By contrast, the SP-supported coherent state is an exact eigenstate of the many-body Hamiltonian only for $u=0$.  Calculating the overlap $Q_\nu(\alpha_{\text{SP}})$ for all the many-body eigenstates $|E_\nu\rangle$, we define an SP-supported eigenstate as the one having the maximal overlap. We aim to relate the properties of this many-body eigenstate to the classical stability of the underlying SP.

The classical stability analysis of the mid-spectrum SP is presented in Appendix \ref{Appendix1}. There are three degrees of freedom and hence three Bogoliubov frequencies. One of them must be ${\omega_0=0}$ due to the conservation of the number of particles.  The two other frequencies are real up to the lower instability threshold $u=2v$ where they become complex.  Then, for large enough $u$, stability is regained due to self-trapping. For  $v=0$ this upper stability threshold lies at $u=\sqrt{8}$ whereas for ${v=0.1}$ it is $u=3.2$.     
The dependence of the Boguliobov frequencies on $u$ is displayed in \Fig{fig4_1}. It should be noted that while the emergence of complex frequencies in between these thresholds indicates the loss of dynamical stability of the SP, it does not provide a way to identify the emergence of chaos in the vicinity of the SP. 

In \Fig{fig4_3a} we plot, as a function of $u$,  the purity $S$, the overlap $Q(\alpha_{\text{SP}})$, and the participation number $M_2$ of the SP-supported eigenstate. The classical instability is clearly reflected in the low coherence measures and in the high participation number. However, within this range of instability, a transition takes place at ${u=1.1}$.  In what follows, we show that this transition can be attributed to the emergence of chaos in the vicinity of the unstable SP.

The top panels of \Fig{fHSTg} show the $p_2=1/2$ Poincare sections at $E_{\text{SP}}$ throughout the $u$ parameter range.  In the quasilinear regime ($u=0.1$, left column) it is clear that the representative trajectory is supported by a Kolmogorov-Arnold-Moser (KAM) torus \cite{10.1063/1.31418,doi:10.1142/8955}. For larger interaction strengths ($u=0.8$, 2nd column) the torus is pinched, the SP becomes hyperbolic, but the classical motion remains regular.  Increasing the interaction strength further, a stochastic layer appears ($u=1.1$, 3rd column) and expands until the last KAM torus is destroyed and global chaos is attained ($u=3.0$, 4th column). Self-trapping then restores integrability in the strong interaction limit ($u=3.5$, right column). The various transitions are reflected in the measures of  \Fig{fig4_3a}  and in the shape of the Husimi distribution function of the SP-supported state. 

The $M_2$ measure in  \Fig{fig4_3a} does not provide a sharp signature for the transitions. The SP-supported state remains localized even in regimes where the classical dynamics in the vicinity of the SP is extremely unstable. This is further illustrated by indicating the localization region on top of the classical trajectories in the two bottom rows of \Fig{fHSTg}.  The conclusion is that the SP serves as a {\em pinning center} for the localization of those states, irrespective of whether it is stable or not. 

Going back to \Fig{fUE}, we mark with horizontal lines the interaction strength values ${u=0.2,1.1,3.2}$ that indicate respectively the bifurcation of the central SP, the emergence of chaos near the SP, and the transition back to stability.

\section{Classification of many-body eigenstates}
\label{mbc}

The standard classification of quantum eigenstates includes regular eigenstates supported by quasi-integrable islands in phasespace, and irregular eigenstates supported by chaos. 
This classification is already challenged by the fact that the SP-supported states are extremely localized despite being embedded in chaos. However,  our analysis below goes further and reveals substantially richer structures. Specifically, while chaotic states may display the well-known random-wave \cite{Berry_1977} or scarred phasespace distribution, they may also exhibit hybrid localization that is implied by the slow underlying mixed-chaos dynamics. We note that our model has two degrees of freedom, hence Arnold diffusion \cite{BASKO20111577,CHIRIKOV1979263} is excluded and any quantum hybridization of chaotic and integrable regions takes place across classically forbidden boundaries.

\begin{figure}[b]
\centering
\includegraphics[width=4.25cm, height=3.0cm]{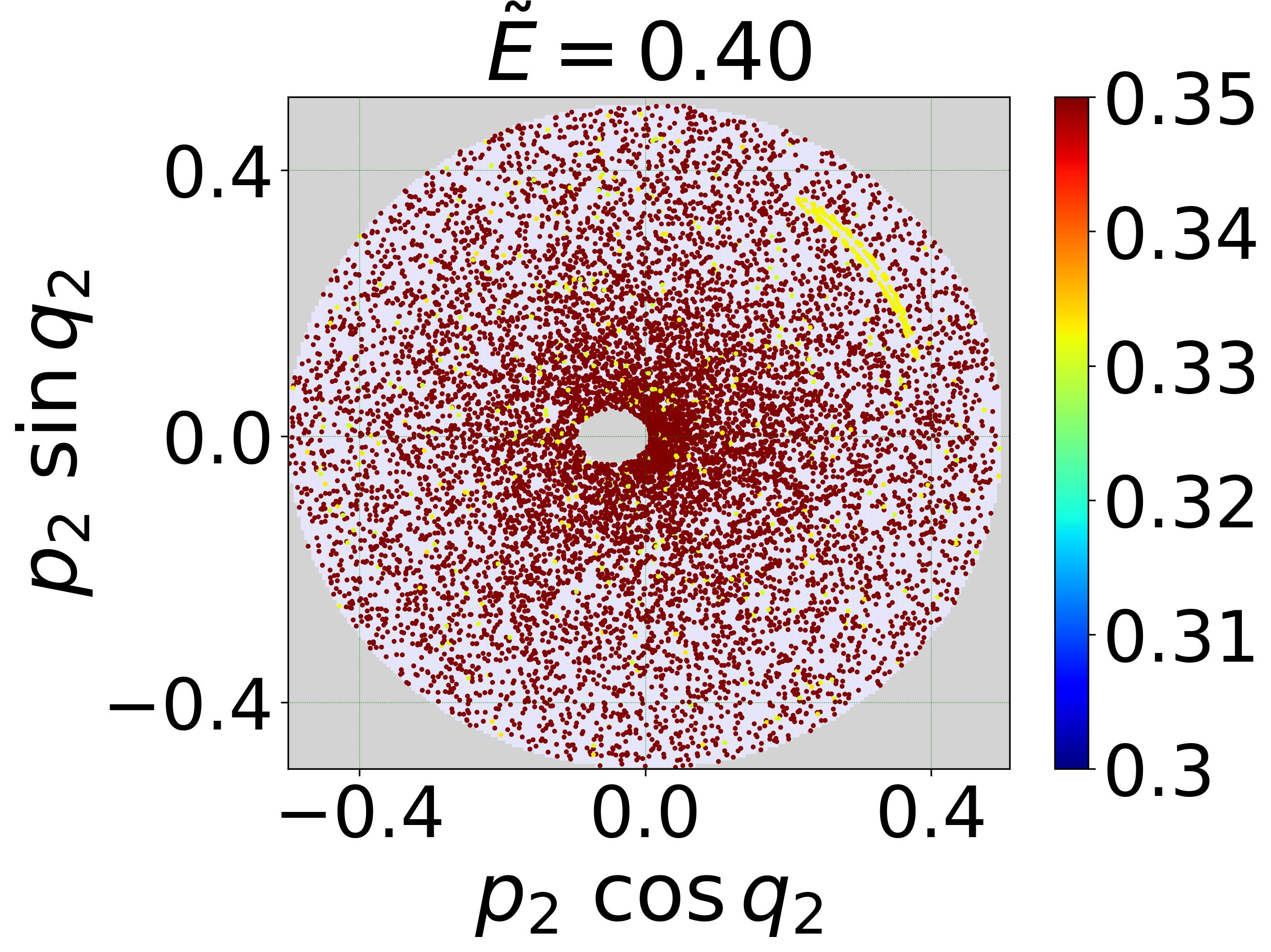}
\includegraphics[width=4.25cm, height=3.0cm]{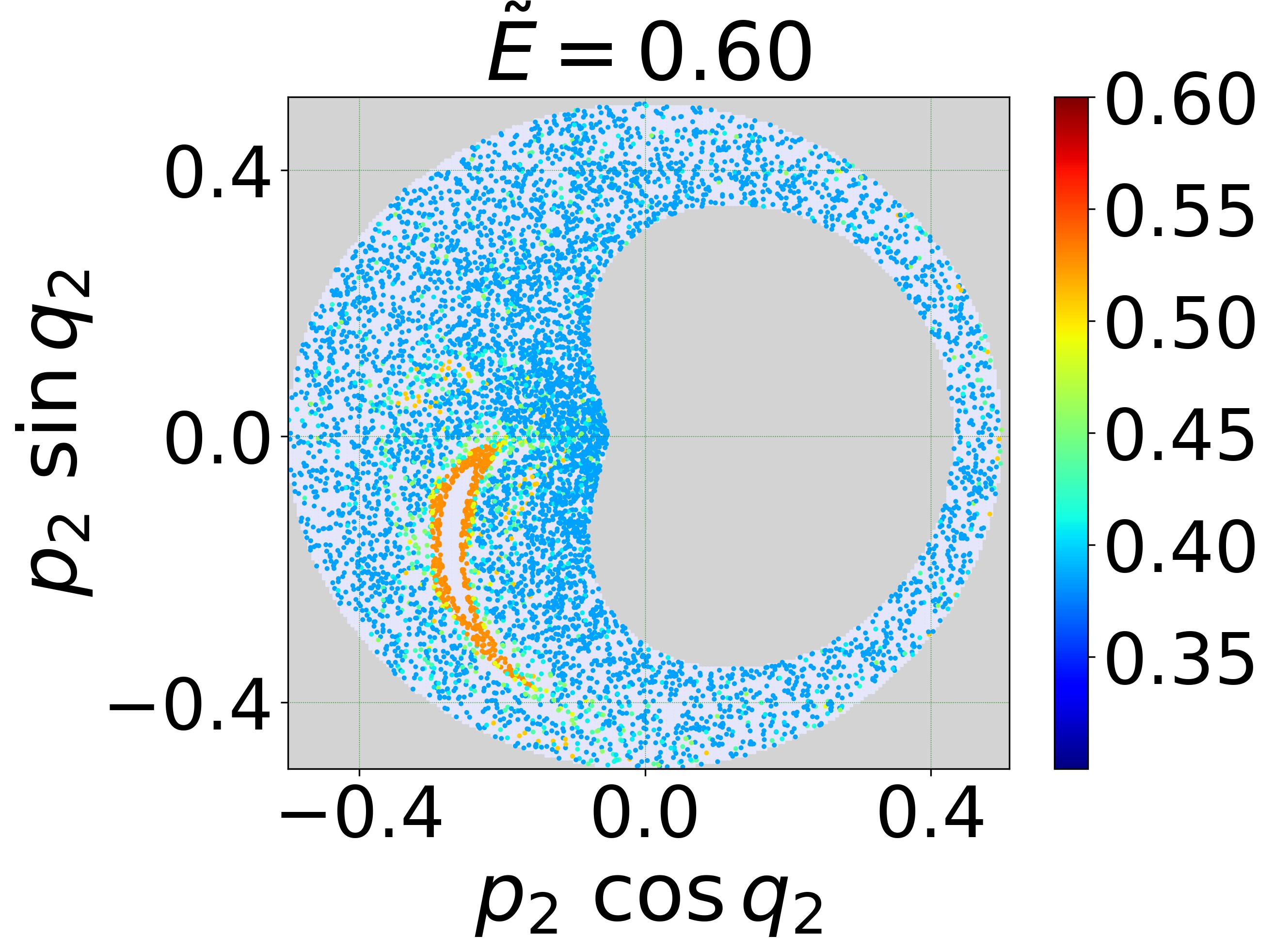}
\label{fig:Ex_Im}
\caption{{\bf Hard Chaos vs Mixed Chaos.} 
The $p_1=1/2$ Poincare sections are taken in the chaotic interaction regime at energies (a) below and (b) above the stationary point's energy $E_{\text{SP}}$. Color code is the average value of $p_{1}$ over a given trajectory.
}
\label{fPS}
%
%
\ \\
\centering 
\includegraphics[width=4.25cm, height=3cm] {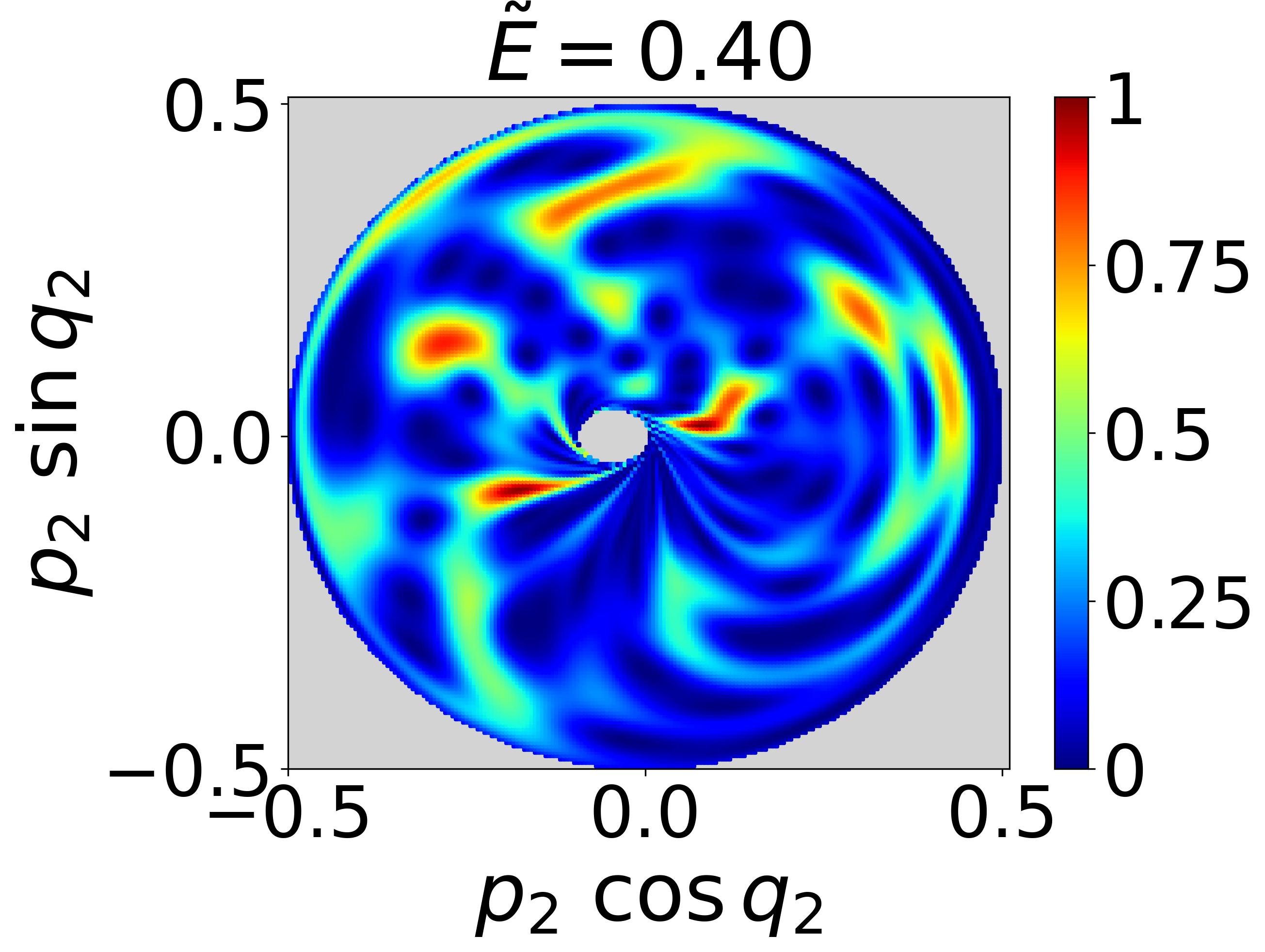}
\includegraphics[width=4.25cm, height=3.0cm]{figure6_row3_4.jpeg} \\
\includegraphics[width=4.25cm, height=3cm] {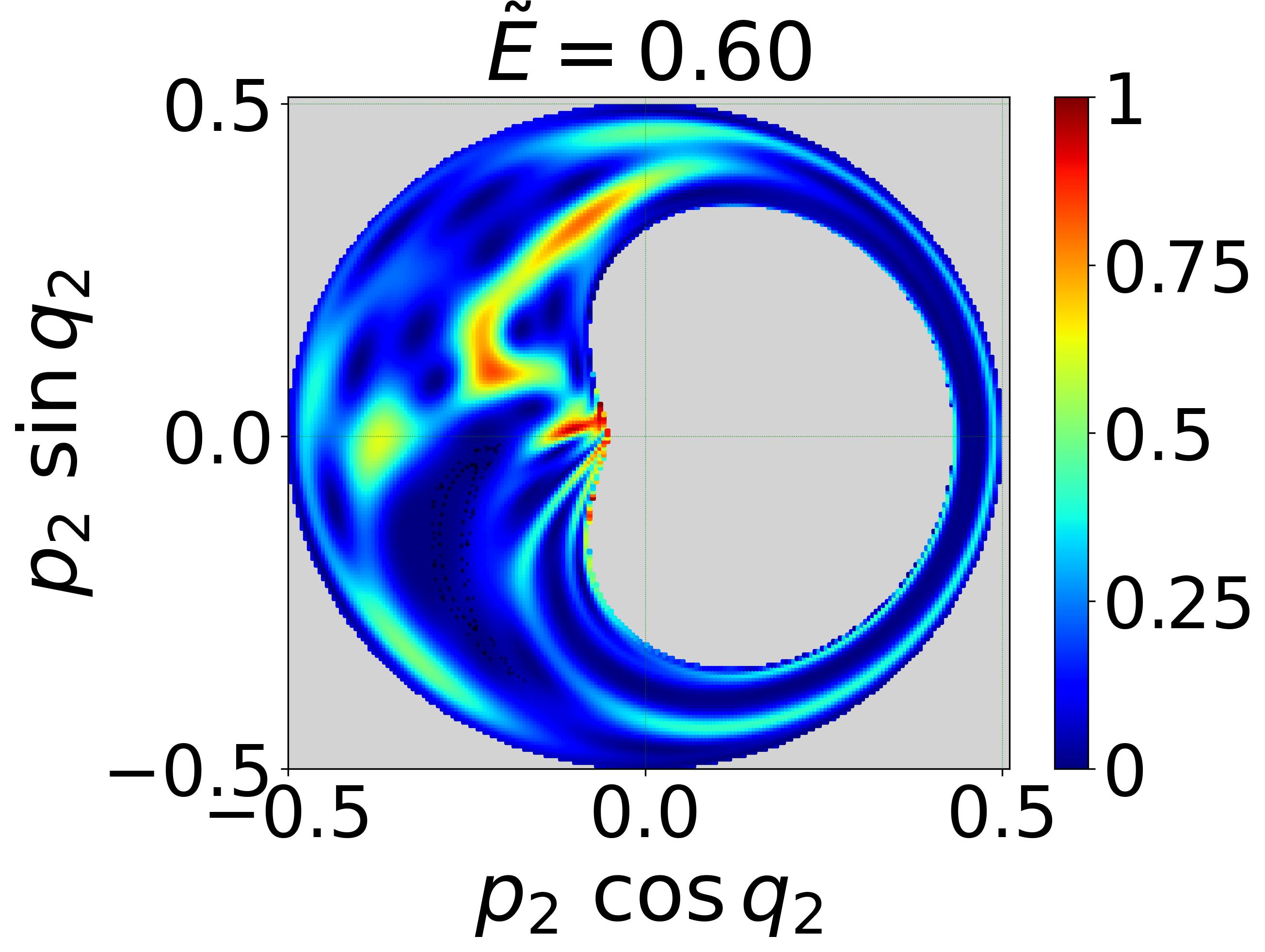}
\includegraphics[width=4.25cm, height=3cm] {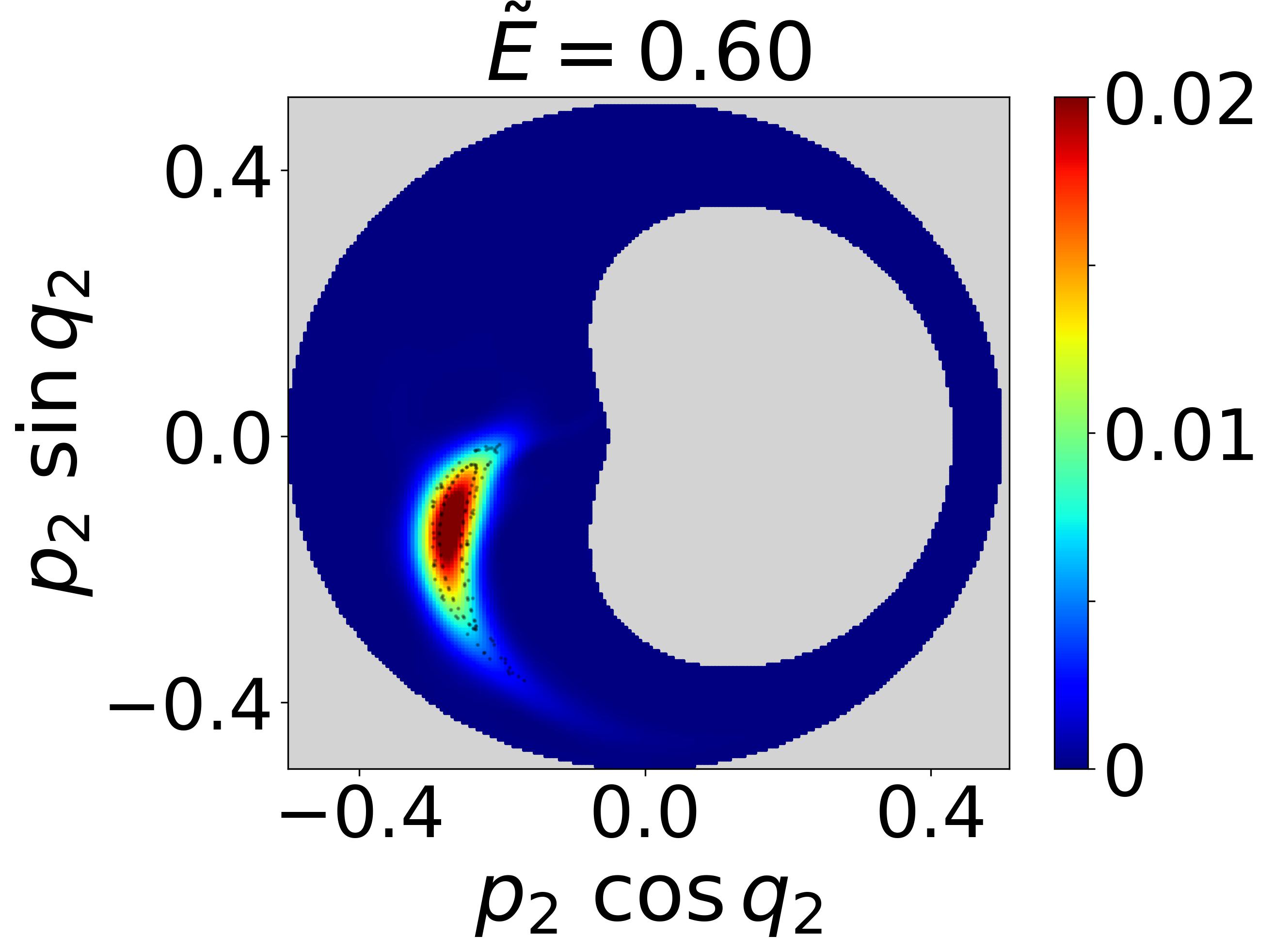} 
\caption{{\bf Husimi functions for representative states.} 
(a)~Representative chaotic state in the hard chaos regime $\tilde{E} \sim 0.4$;
(b)~The SP-supported eigenstate (zoomed) ;
(c) Representative chaotic state in the mixed-chaos regime $\tilde{E} \sim 0.6$;
(d)~An island state at the same energy. Color codes in (a) and (c) are rescaled with maximum overlap $|\langle \alpha | E_{\nu}\rangle|^{2}$ in the section, while in (b) \& (d) color code represents $|\langle \alpha | E_{\nu}\rangle|^{2}$.
}
\label{fHST}
\end{figure}

In order to demonstrate the above statement, we contrast two regimes of the classical dynamics.
For $u=3.0$, the $E_{\text{SP}}$ energy shell is globally chaotic. Nevertheless, the $E>E_{\text{SP}}$ range, as opposed to the $E<E_{\text{SP}}$ range, features a mixed classical phasespace that contains a large integrable island. This is illustrated in \Fig{fPS}, where we plot Poincare sections at representative energies. The tiny island that appears for $E<E_{\text{SP}}$ cannot be resolved quantum mechanically, as opposed to the relatively large island that dominates in the $E>E_{\text{SP}}$ range. This observation will be further discussed and established below. 

The Percival paradigm suggests that the existence of a large island will split the quantum many-body spectrum into regular and irregular groups of eigenstates. Contrary to that, we argue that {\em hybridized states} are more prevalent. These states have unique statistical properties that distinguish them from regular or fully chaotic eigenstates, and they are responsible for the large spread in $M_{q}$ values observed in \Fig{2_1a}. Their dominance is reflected in the intensity statistics discussed in the following section.

The Husimi phasespace distribution of four representative quantum eigenstates is shown in \Fig{fHST}. The top panels are obtained for parameters where the underlying classical dynamics is globally chaotic throughout the pertinent energy surfaces. While some eigenstates in this regime are delocalized irregular eigenstate as the one shown in \Fig{fHST}a, there are also highly localized SP-supported eigenstates like in \Fig{fHST}b.  The bottom panels contrast a 'chaotic' eigenstate (\Fig{fHST}c) and an 'island'  eigenstates (\Fig{fHST}d) lying on the same classically mixed energy surface. As seen below, the distinction between the two is blurred and most eigenstates on this surface lie in between these extreme examples and display hybridization between chaotic and integrable regions.

\begin{figure}
\centering
\includegraphics[width=8.5cm, height=4cm] {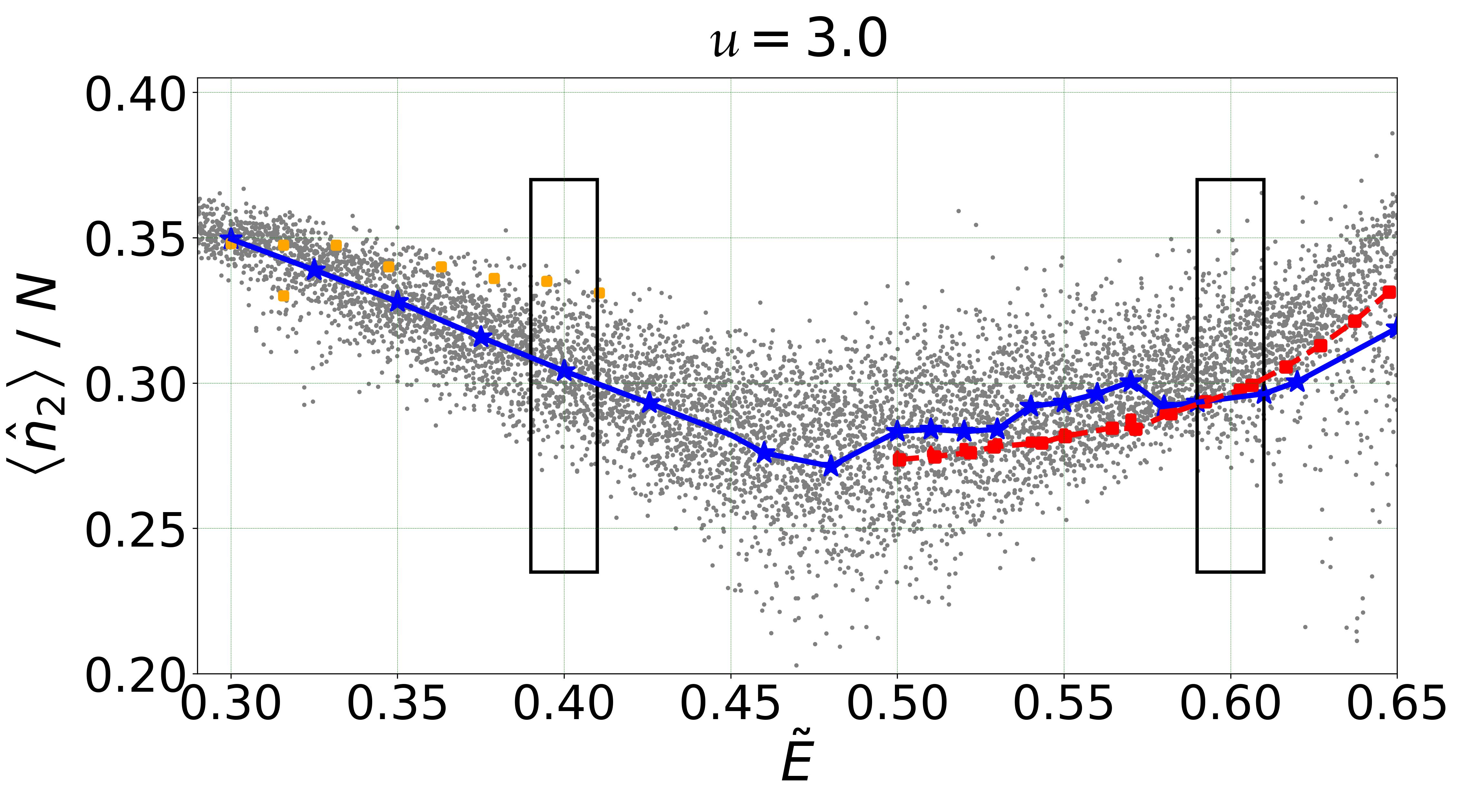}
\caption{{\bf The classical skeleton of the spectrum.}
The blue solid line marks the time-averaged value of $n_{2}$ for chaotic trajectories. The red dashed line is the time-averaged value of $n_{2}$ for regular trajectories corresponding to 'island states'. Rectangles mark the energy windows in which all the states have been considered for scaling and comparison analysis of chaotic states of two kinds namely 'hard chaotic' and 'mixed chaotic'. }
\label{5aa_2}
\end{figure}

\begin{figure}
\centering 
\includegraphics[width=8.5cm, height=4cm] {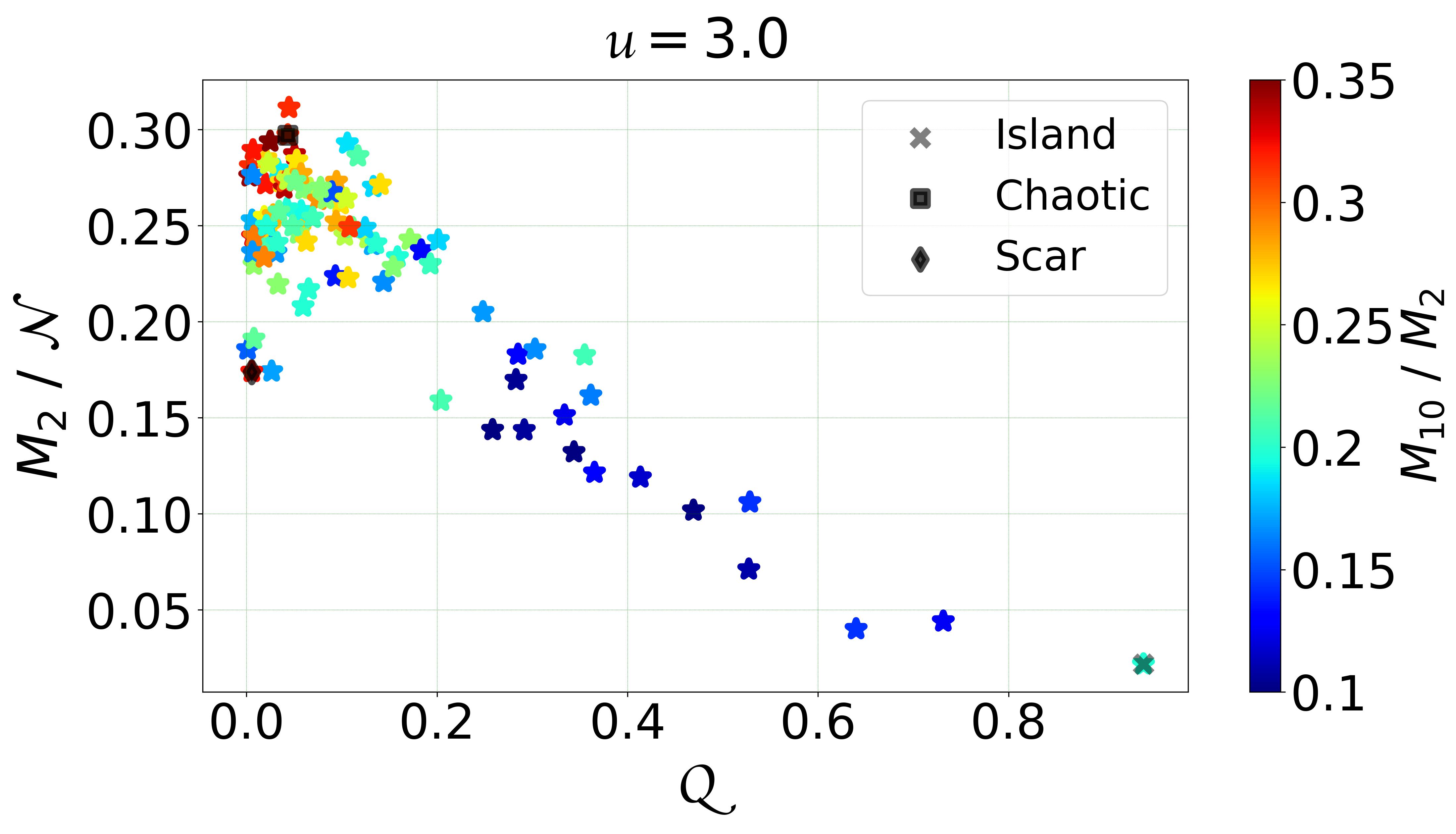}
\caption{{\bf Hybridization of chaotic and island states.}
$M_{2}$ vs $Q$, color-coded with $M_{10}/M_{2}$. Here $Q$ is the sum of the overlaps (normalized) between the coherent states on the island and energy eigenstates within the energy shell $\tilde{E}=0.60$.} 
\label{fMvQ}
\end{figure}

In the mixed chaos regime, we construct measures that distinguish  between chaotic and regular states. Classically, each trajectory is either regular or irregular. The time-average of  $n_2$ for all irregular trajectories (blue points in \Fig{5aa_2}) is the same and equals its mean over the chaotic sea. Doing the same for regular trajectories gives different values associated with the mean over the pertinent invariant tori  (red points in \Fig{5aa_2}). This procedure thus generates a classical 'skeleton' for the many-body spectrum.

Identifying the regular trajectories associated with the red points, we select a subset of coherent states that are located at their Poincare sections (the intersection of the pertinent torus with the $p_1=1/2$ plane). The projection $Q$ of the many-body eigenstates onto this subset quantifies their regularity.  The results are displayed in the scatter diagram of \Fig{fMvQ} where the quantum eigenstates are classified according to their $M_2$ and $Q$ values. The states presented in \Fig{fHST}c,d are respectively one that has a very large value of $M_2$ (with low $Q$) and one that has the maximal $Q$ (with low $M_2$). As such, they are distinctly irregular (chaotic) and regular (island) examples. However, the scatter of the points in \Fig{fMvQ} suggests that such binary classification is inappropriate, and that typically the many body eigenstates can not be associated with one or another classical region.

\begin{figure}
\centering
\includegraphics[width=8.5cm, height=4cm]{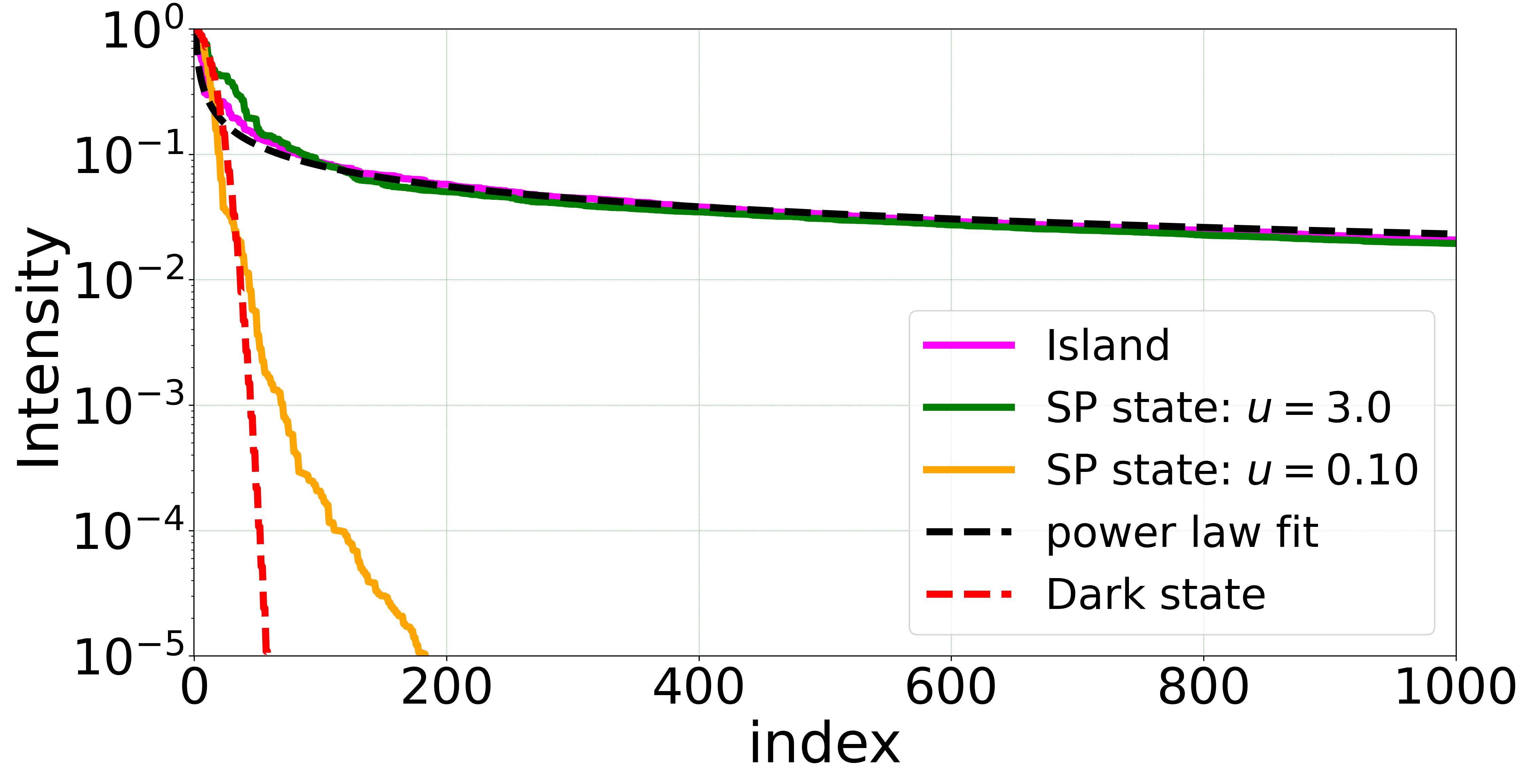}
\caption{{\bf Lineshape analysis.} 
The sorted values of $X_{\nu,n}$ normalized as: 
Intensity =$X_{\nu,n}/\text{max}(X_{\nu,n})$, for the selected island and SP-supported states of \Fig{fHST}. 
For reference, we show the lineshape of a coherent dark state.
The dashed curve is $\sim 1/\textbf{n}^{0.55}$. 
Vanishingly small intensities that correspond 
to forbidden regions have been excluded.}
\label{shape_scaling}
%
\ \\
\centering 
\includegraphics[width=8.5cm, height=4cm]  {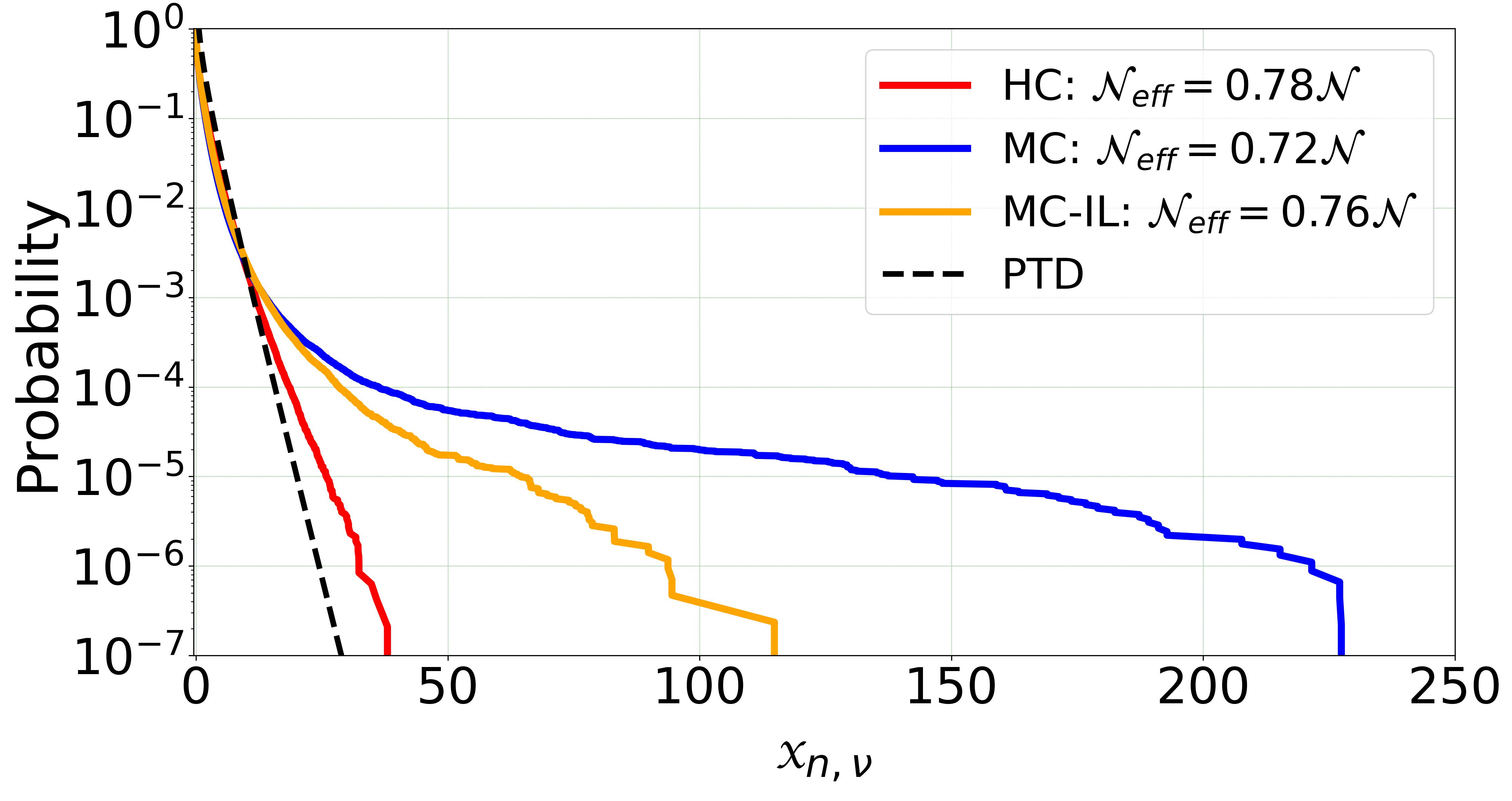}
\caption{{\bf Intensity statistics for chaotic states.}
Comparison of Porter-Thomas distribution (PTD) with the intensity distributions of the three different kinds of chaotic groups namely, Hard chaotic (HC); Mixed chaotic (MC); and Mixed chaotic with island region excluded (MC-IL). Probability on the y-axis implies same as $ P(x)$ in \Eq{ePT}.} 
\label{PTD}
\end{figure}

\section{Intensity Statistics}

Having identified different families of many-body eigenstates, we now turn to the detailed analysis and characterization of their Fock-basis intensity distribution, namely the distribution of the $X_{\nu,n}$, as defined in \Eq{eX}.  The ratio between the different $M_q$ moments  of this distribution are set either by its overall envelope, or by the statistical noise within it. In the numerical results below, we focus on the particular ratio $M_{10}/M_2$.

\begin{figure}

[Hard Chaos]

\includegraphics[width=4.25cm, height=3.0cm]{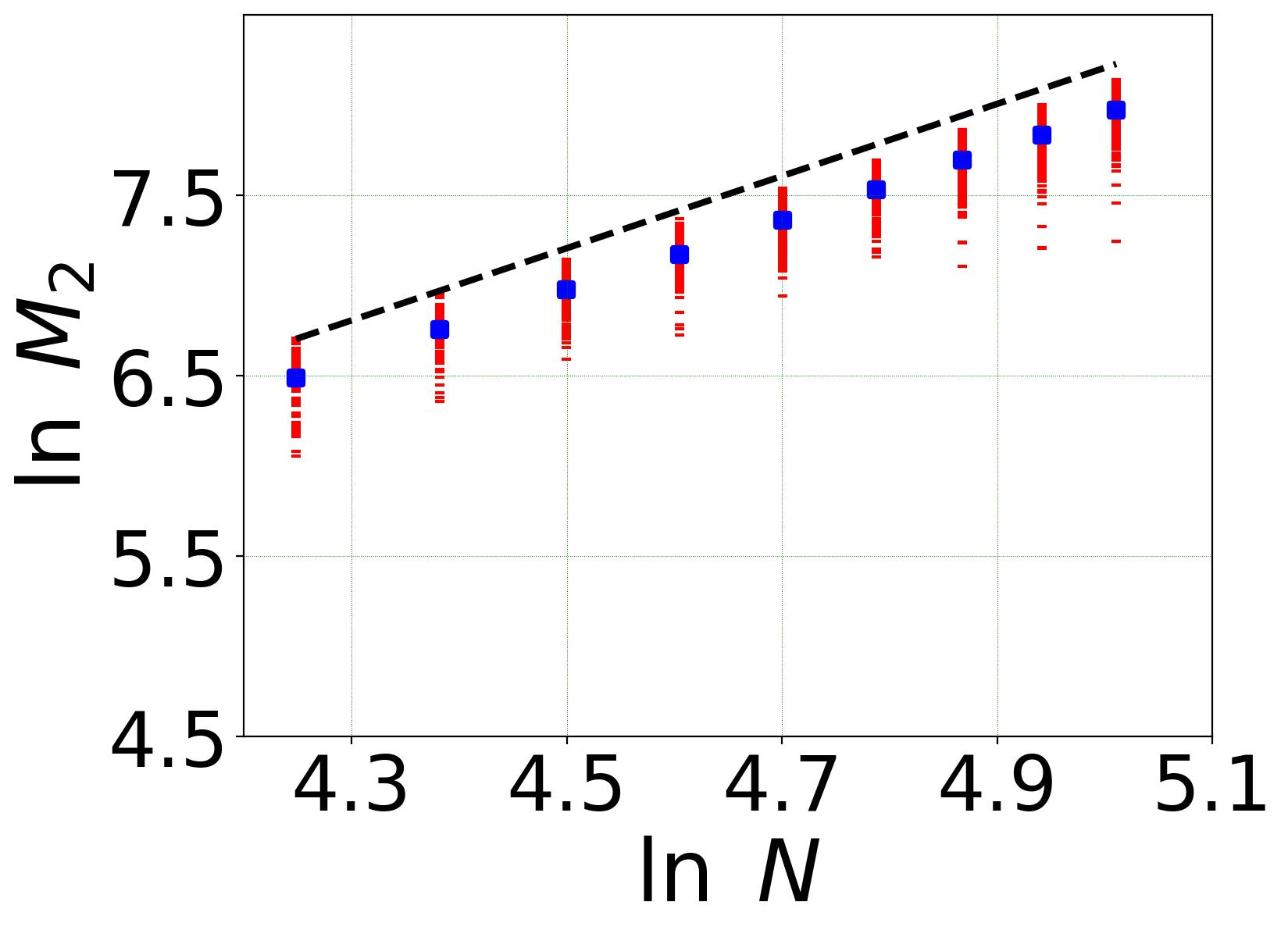}
\includegraphics[width=4.25cm, height=3.0cm]{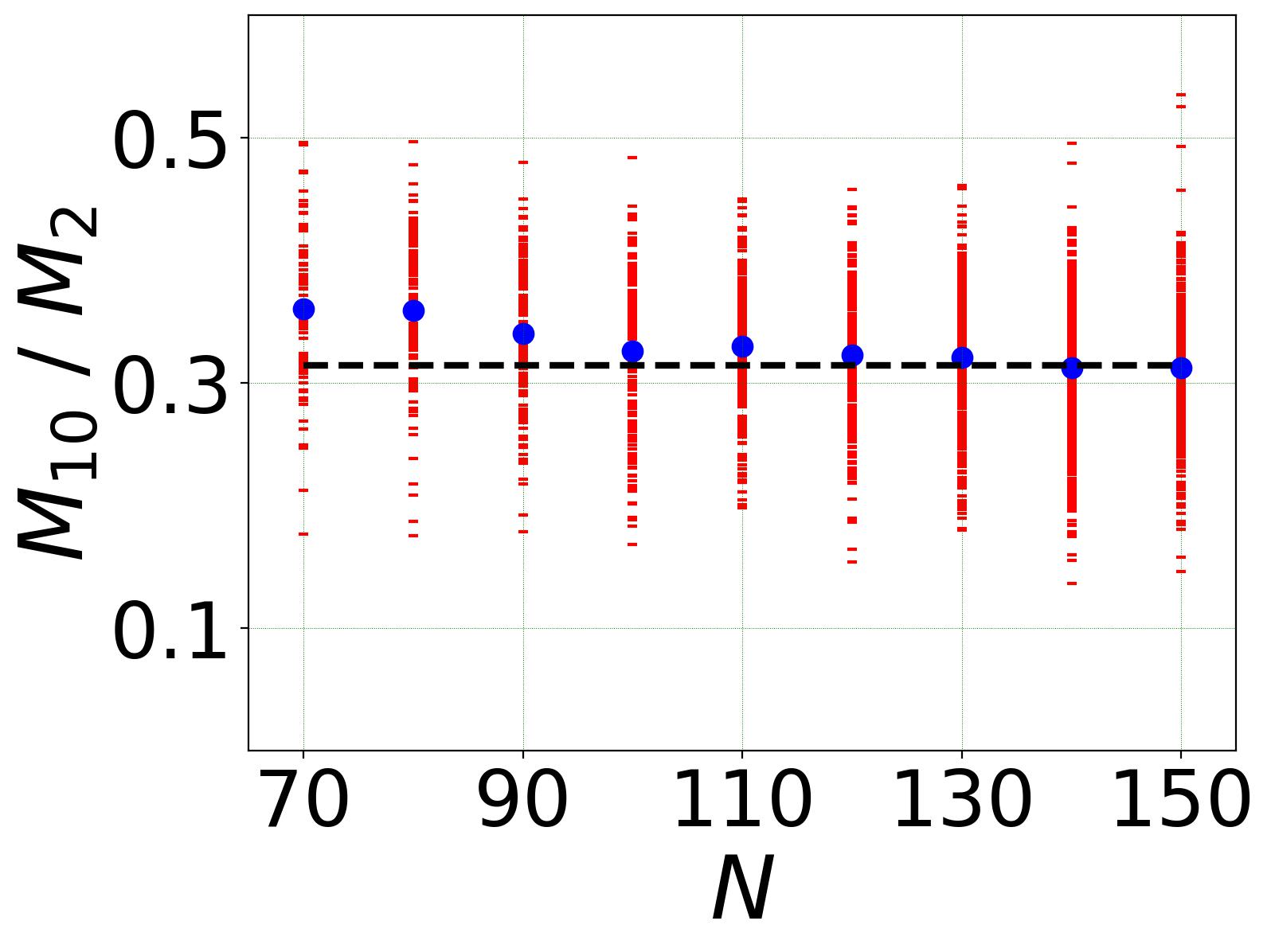}

[Mixed Chaos]

\includegraphics[width=4.25cm, height=3.0cm]{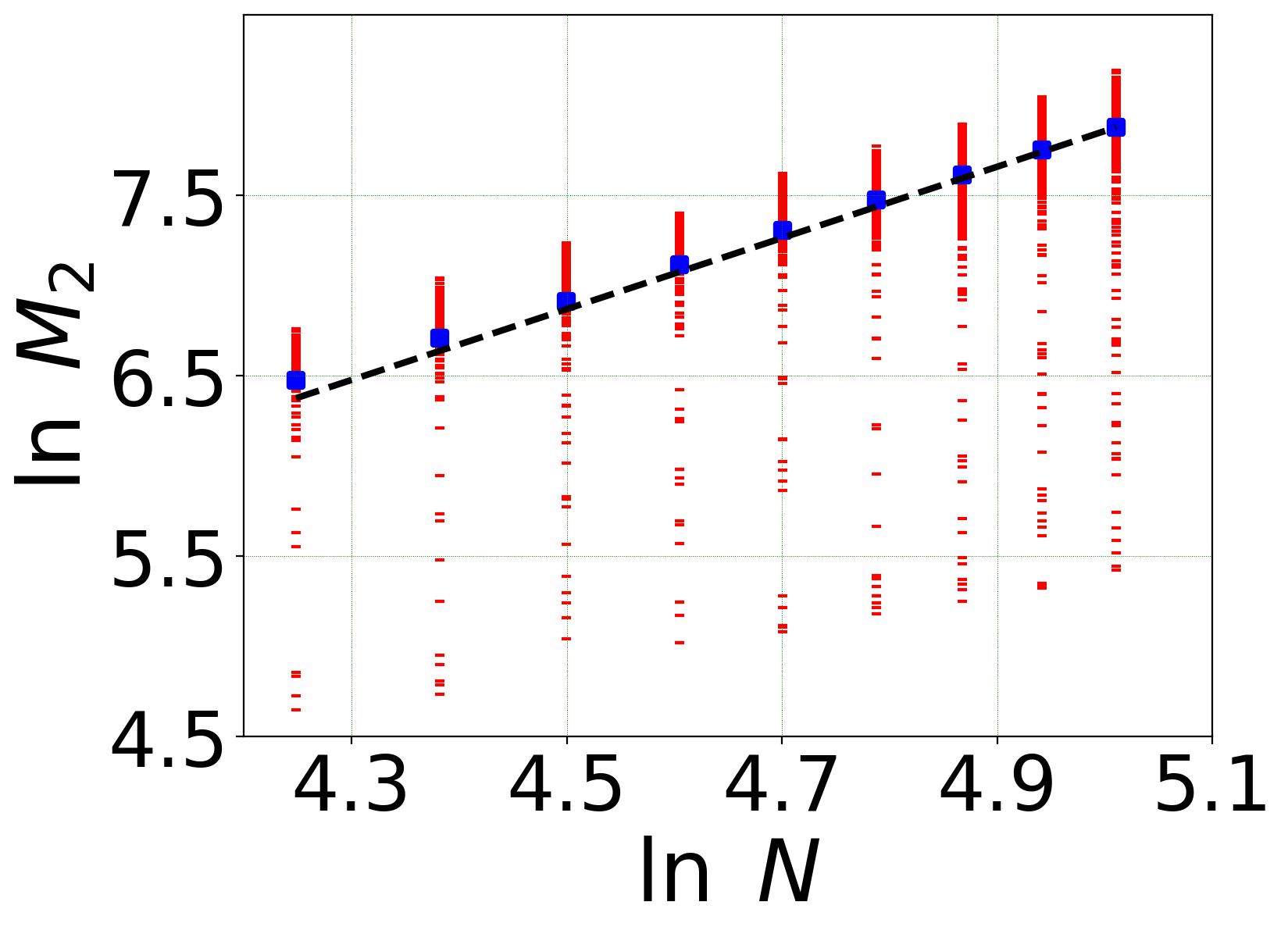}
\includegraphics[width=4.25cm, height=3.0cm]{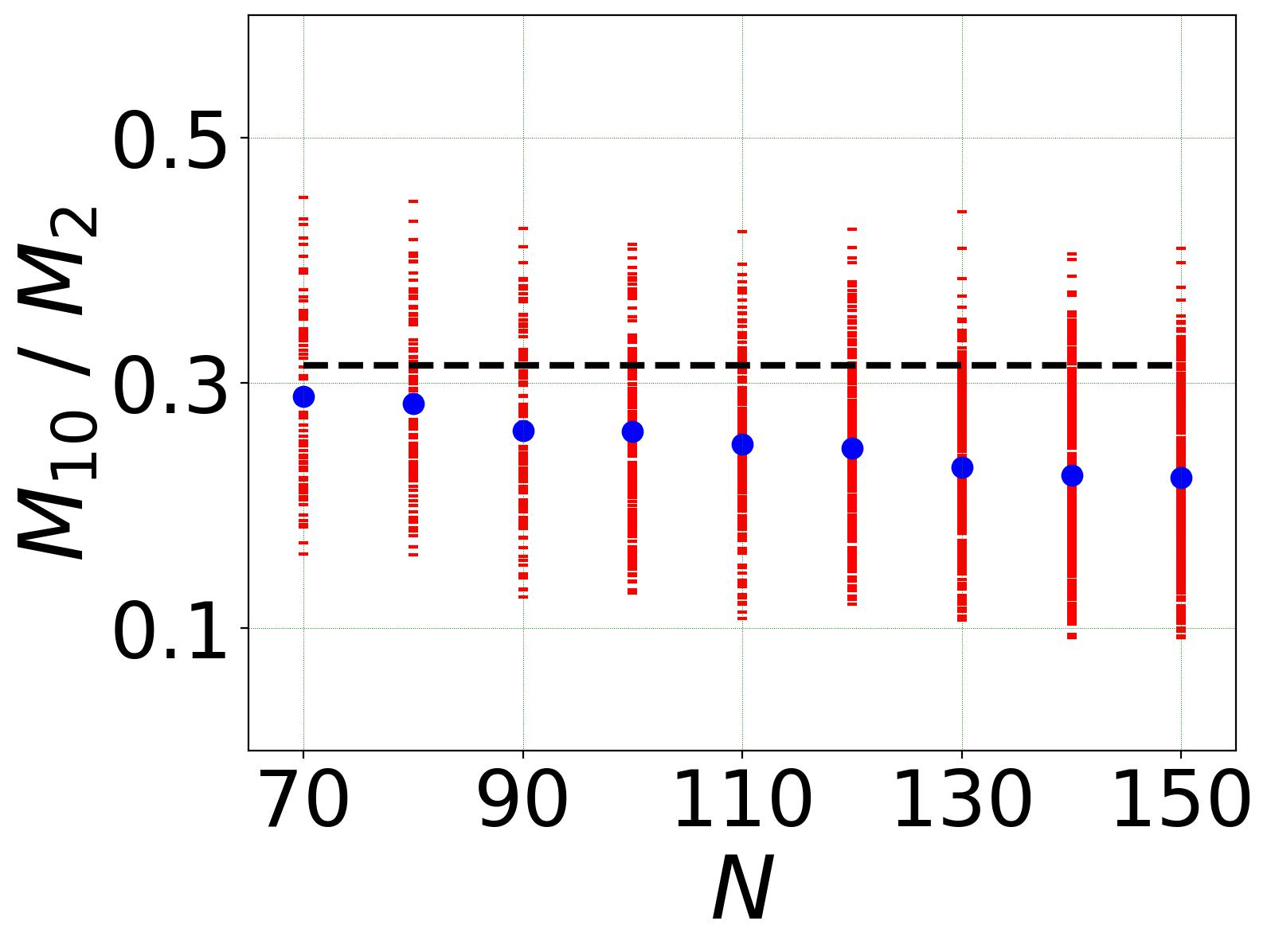}

[Island states]

\includegraphics[width=4.25cm, height=3.0cm]{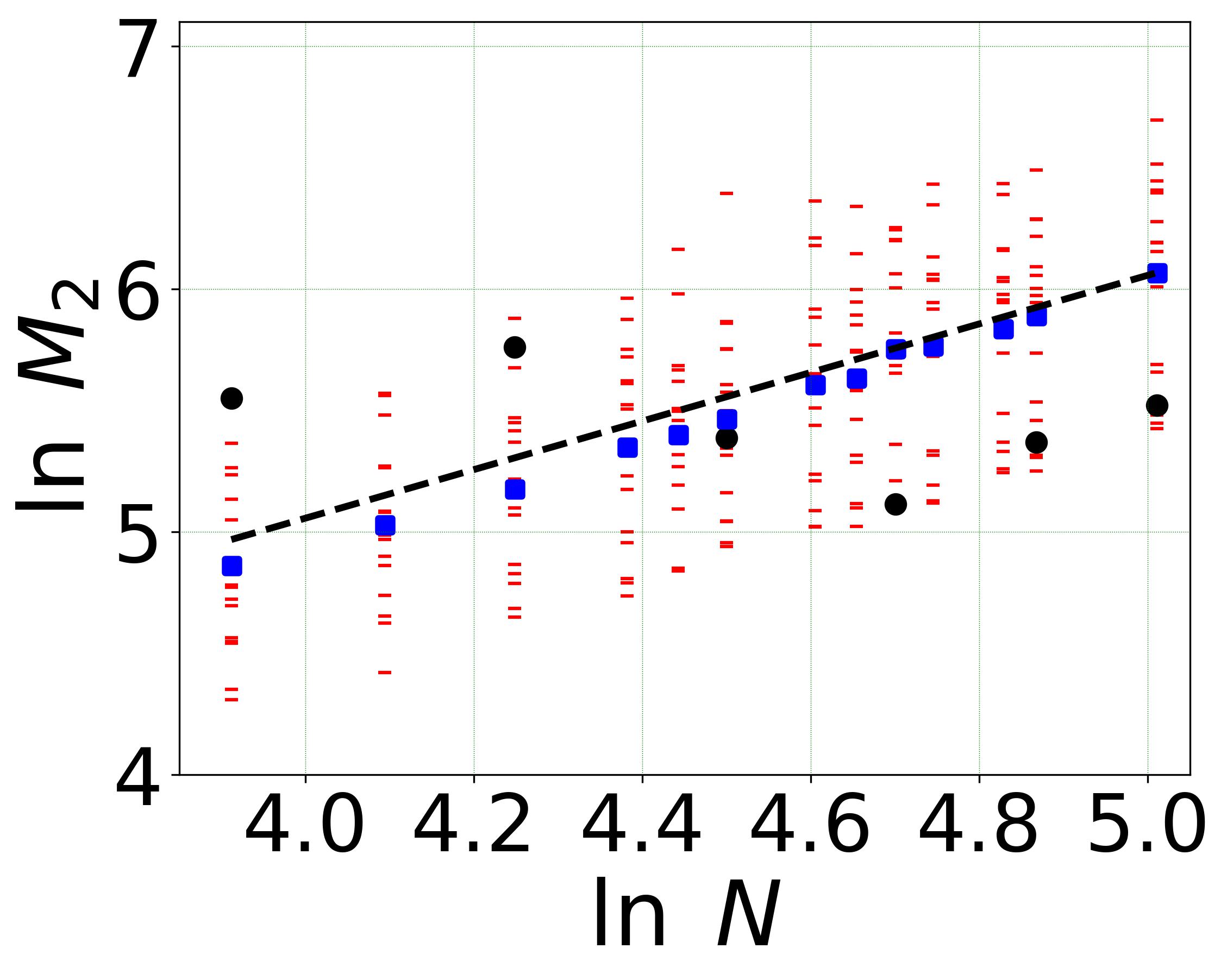}
\includegraphics[width=4.25cm, height=3.0cm]{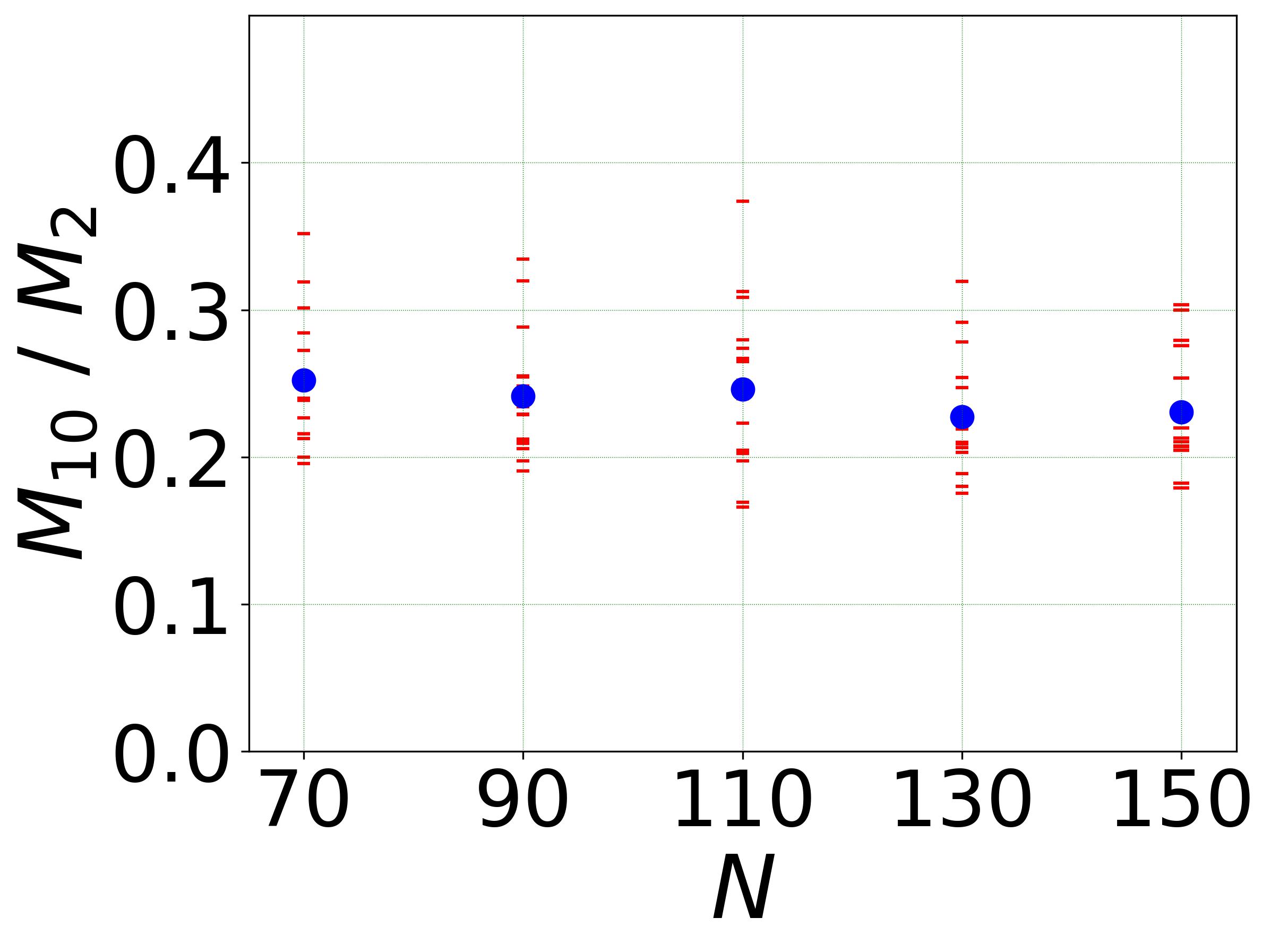}

[SP supported states]

\includegraphics[width=4.25cm, height=3.0cm]{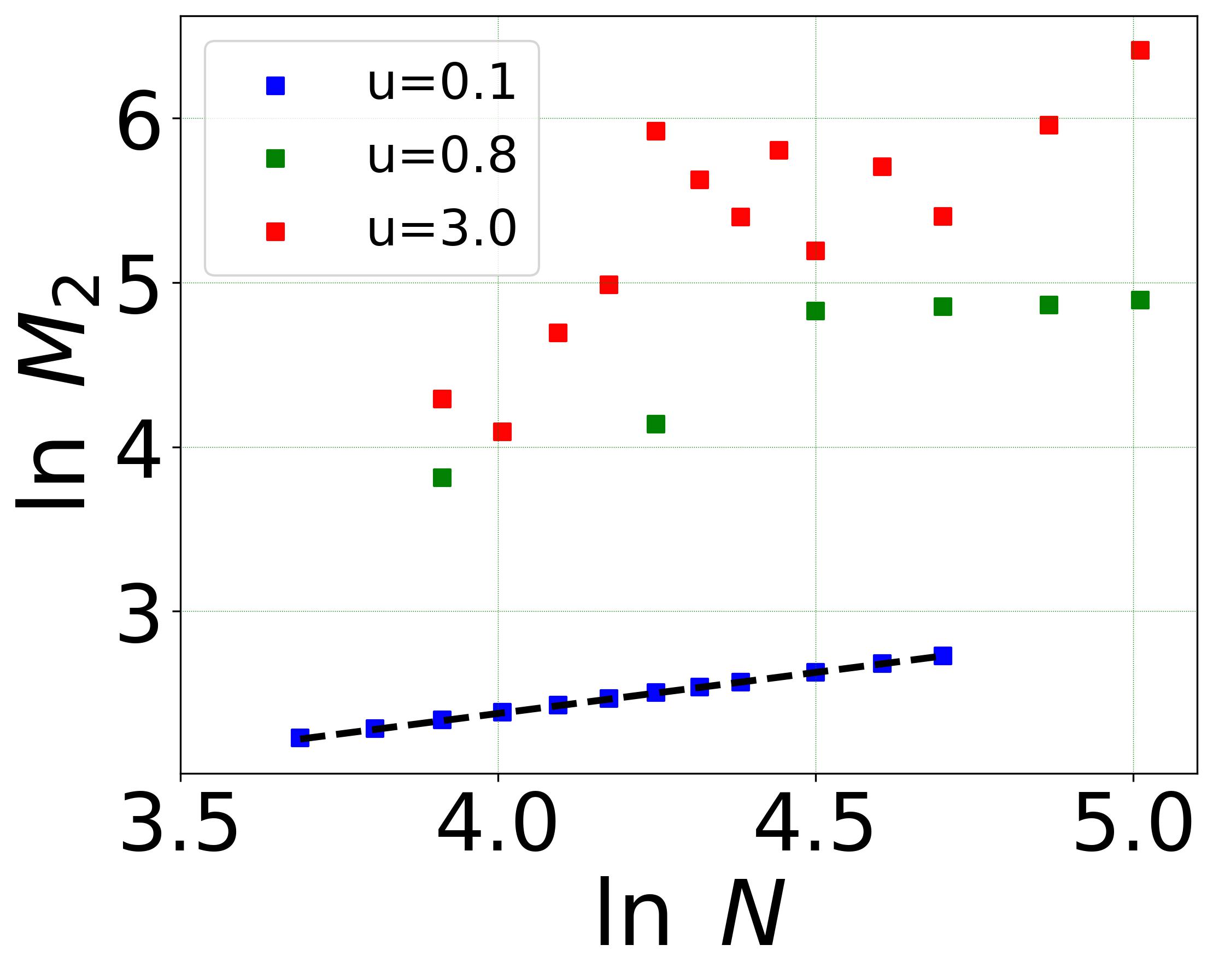}
\includegraphics[width=4.25cm, height=3.0cm]{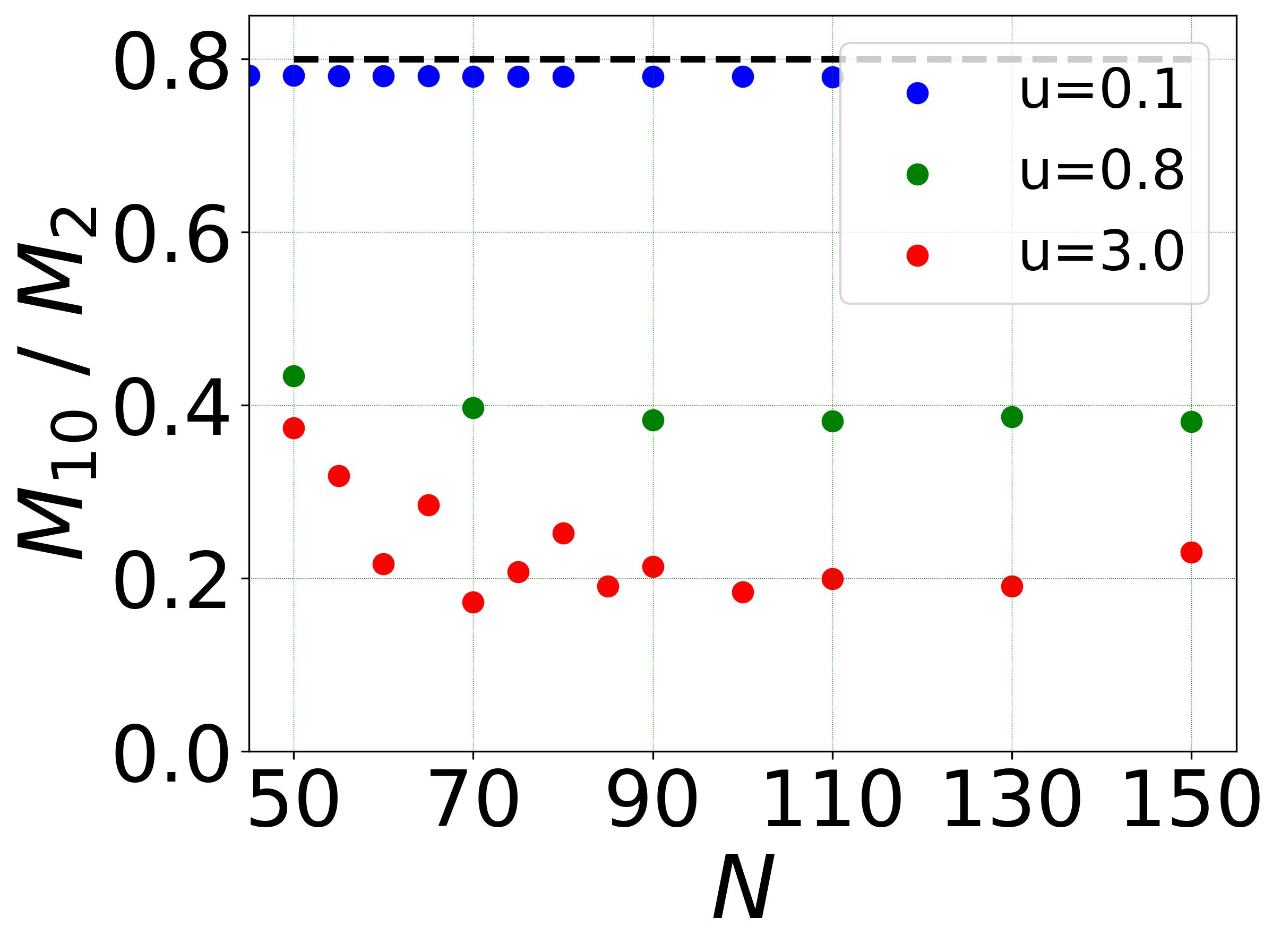}

\caption{{\bf The $N$ dependence of the statistics.} 
(a) Scaling of the states in the hard chaos regime (left), with the slope=1.94. 
(b) Scaling of the states in the mixed chaos regime (right), with the slope=1.85. 
(c) Scaling of the 'island state' with the slope=1. 
Black dots in the $\ln M_{2}$ vs $\ln N$ plot are the $M_{2}$ values of a single eigenstate in the energy shell $\tilde{E} =0.60$ and fluctuates with $N$. 
(d) Scaling of the SP-supported state with the slope=1/2. 
Slope in all the figures refers to the slope of the curve $\ln M_{2}$ vs $\ln N$.} 

\label{fig7}
\end{figure}

If the variation of $X_{\nu,n}$ as a function of $n$ follows a smooth envelope, the $M_q$ ratios are determined by the  lineshape.  For a uniform distribution $X_{\nu,n}=1/{\cal N}$, we have $M_q={\cal N}$ for all $q$. By contrast, a power-law lineshape results in a rapid drop of $M_q$ as $q$ is increases, leading to very low non-universal $M_{10}/M_2$ ratio.

\begin{table}
\begin{tabular}[c]{ |p{4.5cm}||p{1.75cm}|p{1.75cm}|  }

\hline
\multicolumn{3}{|c|}{Standard states } \\

\hline

State type & $M_{10}/M_{2}$ & slope \\

\hline  
 
\vspace{0.05cm} Dark state \  $\alpha = \frac{1}{\sqrt{2}} (1,0,-1)$ & \vspace{0.05cm} 0.80  & \vspace{0.05cm}  1/2 \\

\vspace{0.05cm} Generic Coherent State  
& \vspace{0.05cm}  0.65 & \vspace{0.05cm} 1 \\

\vspace{0.05cm} Rectangular:  $X_{n}= 1/\mathcal{N}$ & \vspace{0.05cm} $1$ & \vspace{0.05cm} $2$ \\

\vspace{0.05cm} GOE:  $X_{n}\sim 1/\mathcal{N}$+fluctuations & \vspace{0.05cm} $0.3145$ & \vspace{0.05cm} $2$ \\

\vspace{0.05cm}Power law:  $X_{n} \propto 1/n^{0.5}$ & \vspace{0.05cm} $\sim$ 0.22  & \vspace{0.05cm} $\sim 0.45$ \\

\hline 
 
\vspace{0.05cm} SP state (stable)  &  \vspace{0.05cm} $\sim 0.78$  &  \vspace{0.05cm} 1/2 \\
\vspace{0.05cm} SP state (unstable regular) &  \vspace{0.05cm} $\sim 0.4$  &  \vspace{0.05cm} noisy \\
\vspace{0.05cm} SP state (unstable chaotic) &  \vspace{0.05cm} $\sim 0.22$  &  \vspace{0.05cm} noisy \\
\vspace{0.05cm} Island state:  $u=3.0$ & \vspace{0.05cm} $\sim 0.22$ & \vspace{0.05cm} 1 \\
\vspace{0.05cm} Mixed chaos:  $u=3.0$ & \vspace{0.05cm} $\sim 0.22$ & \vspace{0.05cm} $1.85$ \\
\vspace{0.05cm} Hard chaos:  $u=3.0$ & \vspace{0.05cm} $\sim 0.313$ & \vspace{0.05cm} $\sim 1.94$ \\

\hline

\end{tabular}
\caption{{\bf Characterization of quantum states.}  
Summary of the expected $M_{10}/M_2$ ratio and the power-law dependence of the participation number $M_2$ on $N$ for several reference states (top block) and various classes of trimer eigenstates (bottom block)}
\label{TB}
\end{table}

On the other extreme, the ratio between different $M_{q}$ can be affected by statistical fluctuations within an otherwise uniform envelope, as in the case of a Billiard system \cite{PhysRevLett.74.62}. The GOE statistics typical to fully irregular states in the presence of time-reversal symmetry, yield a universal ratio $M_{10}/M_{2}\approx 0.3$. For the hybridized eigenstates that dominate the trimer spectrum, the challenge is to identify what feature of their intensity distribution is responsible for the numerically observed value of $M_{10}/M_{2}$. 

The values of $X_{\nu,n}$ sorted according to their size from the largest to the smallest, are plotted in \Fig{shape_scaling} for the island and SP-supported states. The curves of these eigenstates simply reflect the lineshape of their smooth envelope. In the stable regime ($u=0.1$) the SP-supported eigenstates have an envelope that is similar to a reference coherent dark state. As the interaction strength is increased, stability is lost and chaos emerges ($u=3$), the SP-supported eigenstates develop long power-law decaying tails, identical to those of the regular island-supported state. We conclude that both the island and the SP-supported eigenstates are pinned down by their classical (island and fixed-point, respectively) localization centers: The tails that extends into the mixed chaotic region reflect power-law localization.      

The same procedure is used for chaotic states. In this case, the ordering does not reflect an overall lineshape but merely the characteristics of the statistical noise within a rather uniform envelope. Swapping the axes, we obtain a count of the number of intensities that satisfy $X_{\nu,n} > X$ or upon normalization by ${\cal N}$ the inverse cumulative histogram $\text{Prob}(X_{\nu,n}>X)$. The latter is displayed in \Fig{PTD}.  Chaotic GOE states are characterized by the Porter-Thomas statistics \Eq{ePT}. The inverse cumulative histogram of the trimer's eigenstates in the hard chaos regime follows roughly the Porter-Thomas exponential decay with a deviation that may be attributed to an envelope effect that has not been eliminated. By contrast, the inverse cumulative histograms of the eigenstates in the mixed chaos regime have much longer tails. In order to rule out the possibility that these tails are due to regular states localized in the relatively large island, we eliminate the island from the statistics and obtain a distribution that still deviates substantially from the Porter-Thomas form. We conjecture that there is a hierarchy of smaller and smaller islands that affect the statistics. Thus the tails of the distribution reflect the non-uniformity of the mixed landscape.

The different classes of many-body eigenstates discussed above have fundamentally different dependence on the effective Planck constant $\hbar\propto 1/N$. In \Fig{fig7} we plot the dependence of the participation number $M_2$ and of the ratio $M_{10}/M_2$ on the total number of particles $N$.  
The participation number of the ergodic eigenstates in the hard chaos regime scales as the Hilbert space dimension of  the 2 DoF system ${M_2 \propto \mathcal{N} \sim N^2}$, 
while the $M_{10}/M_2$ ratio approaches the expected GOE value ${M_{10}/M_{2}=3/9.54=0.314}$ as $N$ is increased. 
In contrast, in the mixed chaos regime, the participation number $M_2$ scales as $N^{1.85}$, while the ratio $M_{10}/M_2$ drops below the GOE expectation, indicating hybrid localization.

Both island-states and SP-supported states that are immersed in chaos exhibit $M_{10}/M_2\approx 0.22$. This finding supports the claim that the underlying island or the SP are merely pinning centers for a hybrid localized state.  The dependence of $M_2$ on $N$ in both cases is erratic if we follow an individual state that is selected by a maximum overlap criterion. However, for island states, we can accumulate statistics and consider the dependence of the {\em mean} participation on particle number, obtaining ${\bar{M}_2 \propto [\sqrt{N}]^2 }$  as expected for a minimal wavepacket in two-degree of system system. This dependence should be contrasted with the quasi one-degree-of-freedom result observed for an SP-supported state in the regular region, namely ${ M_2 \propto \sqrt{N} }$,  same as for a dark state. In the latter case $M_{10}/M_2\approx 0.8$ as expected. 

Our findings are summarized in table~\ref{TB} where the obtained dependence of $M_2$ on $N$ and the ratio $M_{10}/M_2$ for the different classes of many-body eigenstates, are compared to the expected behavior of states that possess various lineshapes and statistical fluctuations.

\section{Summary}

Considering a generic many-body Hamiltonian system that features a mixed classical phasespace with chaotic and quasi-regular motion,  one can identify in the spectrum irregular and regular eigenstates. However, these are idealizations and most quantum eigenstates do not adhere to the traditional paradigm. In this work, we have highlighted two notable deviations from the binary regular-irregular classification: strongly localized eigenstates in a fully chaotic classical phasespace and hybrid eigenstates that extend across chaos-integrability borders in a mixed phasespce.

In the first case, eigenstates remain localized in regions where the classical chaotic dynamics is slow with respect to some characteristic quantum timescale (e.g. the Heisenberg time). This type of localization, known as 'dynamical localization' is related to the theory of Anderson localization in disordered systems \cite{Anderson,RevModPhys.80.1355,Modugno_2010}. Slow regions in phasespace are found near the boundaries of the chaotic sea as discussed in \cite{PhysRevA.101.043603}, or in the vicinity of unstable stationary points, as illustrated in this work. 

The majority of the eigenstates in mixed phasespace regions are hybrid. They are pinned by the underlying rugged phasespace structure. Past literature has emphasized localization due to the last KAM torus that is destroyed in the Chirikov scenario, or due to the  remnants of the last KAM torus, aka cantori, or by remnants of unstable manifolds \cite{Heller,10.1063/5.0130682,PhysRevLett.132.047201}. Such localization mechanisms are highly specific and occur only if the model parameters are carefully tuned. A related scenario can be seen  in \Fig{fHSTg} for ${u=1.1}$, where the last KAM still survives, but it does not signify a dramatic crossover in the global statistics. 

Contrasting with previous publications, we realize here that the signature of the mixed phasespace persists for a wide range of~$u$ values. On the practical side we have utilized $M_{q}$ ratios as a measure for the identification of new families of states that do not fall under standard categorization. Such ``generalized entropy measures" should be employed in the spectral analysis of any system with a mixed phasespace.

\rmrk{The effects of scarring, localization, and hybridization are all related to the underlying classical phase-space, and it might be useful to summarize what are the ``classical" ingredients that are required for the analysis. It is possibly natural to start with Lyapunov exponent analysis as in \cite{PhysRevE.107.024210}. Local dispersion in the value of the largest exponent may indicate mixed regions of regular and chaotic motion, as opposed to ergodic regions where it has a well defined value. In a fully chaotic region, an idealized theory \cite{LKaplan_1999} provides a direct relation between the scar ``intensity" and the instability exponents of the underlying periodic orbits. Stationary points at the {\em corners} of phase space are somewhat special and possibly can be regraded as the upper unstable fixed-point of a mathematical pendulum. More generally, near the {\em boundaries} that separate chaotic from quasi-regular regions, the distribution of the Lyapunov exponents becomes fragmented, and dynamical localization is related to the transport coefficients of the slow dynamics. The latter are not determined merely by the Lyapunov spectrum, and in some cases are related to cantori or to remnants of unstable manifolds \cite{Heller,10.1063/5.0130682,PhysRevLett.132.047201}. Irrespective of the localization mechanism, dynamical tunneling allows hybridization that blurs the classification of eigenstates.}

\appendix

\section{Stability analysis for the central SP}\label{Appendix1}

The classical SPs of the Bose-Hubbard trimer model are found by solving:
\begin{equation}
i \dot{\textbf{a}} = (H_{0} + u \mathcal{P}) \bm{a} = \mu \bm{a},
\label{A1}
\end{equation}
where $\bm{a} = (1/\sqrt{N})( a_{1},a_{2},a_{3} )$ represents re-scaled classical amplitudes and the operators $H_{0}$ and $\mathcal{P}$ are:
\begin{equation}
H_{0} = \begin{pmatrix}
 0 & -\frac{\Omega}{2}  & 0 \\ 
-\frac{\Omega}{2} & v &  -\frac{\Omega}{2} \\ 
0  & -\frac{\Omega}{2} &  0
\end{pmatrix}, \mathcal{P} = \begin{pmatrix}
 P_{1} & 0 & 0\\ 
0 &  P_{2} & 0\\ 
0 & 0 &  P_{3}
\end{pmatrix},
\label{A2}
\end{equation}
where $P_{i}= |a_{i}|^{2}/N $. The dark-state  SP  
is given by $\bm{a} = \alpha_{\text{SP}}=(1/\sqrt{2},0,-1/\sqrt{2})$.
The dynamical stability analysis of this SP is carried out via diagonalization of the Bogoliubov matrix:
\begin{equation}
\begin{pmatrix}
H_{0} + 2 u \mathcal{P}- \mu & -u \mathcal{P}   \\ 
u \mathcal{P}  & -(H_{0} + 2 u \mathcal{P}- \mu)  
\end{pmatrix}
\label{A3}
\end{equation}
resulting in 3 pairs of characteristic frequencies, 
namely $\pm\omega_{q}$ indexed by ${q=\{0,+,-\}}$. 
The trivial frequency $\omega_{0}=0$ is implied by conservation of particles, while
\beq 
&& \textstyle \nonumber
\omega_{+} = \frac{\sqrt{\sqrt{\left((u-2 v )^2+4\right)^2-16 \left(u^2-2 u v +1\right)}+u^2-4 u v +4 v ^2+4}}{2 \sqrt{2}}  
\\ && \textstyle \nonumber
\omega_{-} = \frac{\sqrt{-\sqrt{(u-2 v ) \left(u^3-6 u^2 v +4 u \left(3 v ^2-2\right)-8 v  \left(v ^2+2\right)\right)}+(u-2 v )^2+4}}{2 \sqrt{2}}
\eeq
For $v=0$ the non-vanishing frequencies coalesce to give,
\beq
\omega_{\pm} = \pm \frac{1}{2\sqrt{2}} 
\left[ (4+u^2) \pm u\sqrt{u^2-8} \right]^{1/2}
\label{A4}
\eeq

\vspace{1cm}

{\bf Acknowledgment.-- } 
This research was supported by the Israel Science Foundation (Grant  No.518/22).
We thank Maxim Olshanii for valuable discussions.



%

\end{document}